%% file: main.tex
\documentclass[11pt]{article}

\usepackage[table]{xcolor}
\usepackage{soul}
\usepackage[normalem]{ulem}
\usepackage{graphicx} 
\usepackage{amsmath}
\usepackage{multirow}
\usepackage{xspace}
\usepackage{listings}
\usepackage{algorithm}
\usepackage{algpseudocode}
\usepackage{stackengine}
\usepackage{tcolorbox}
\usepackage{subcaption}
\usepackage{wrapfig}
\usepackage{makecell}

\usepackage{url} 
\usepackage{booktabs} 
\usepackage[bookmarks=false]{hyperref}
\hypersetup{hidelinks}
\usepackage{natbib}
\usepackage{orcidlink} 
\usepackage{authblk}

\definecolor{codegreen}{rgb}{0,0.6,0}
\definecolor{codegray}{rgb}{0.5,0.5,0.5}
\definecolor{codepurple}{rgb}{0.58,0,0.82}
\definecolor{backcolour}{rgb}{0.95,0.95,0.92}

\lstdefinestyle{alloystyle}{
    backgroundcolor=\color{white},
    commentstyle=\color{codegreen},
    keywordstyle=\color{blue},
    numberstyle=\tiny\color{codegray},
    stringstyle=\color{codeblue},
    basicstyle=\ttfamily\footnotesize,
    breakatwhitespace=false,
    morecomment=[l]{//}, 
    morecomment=[s]{/*}{*/}, 
    morekeywords={abstract, all, and, as, assert, but, check, disj, else, enum, exactly, extends, fact, for, fun, if, implies, in, int, let, lone, module, no, none, not, one, open, or, part, pred, run, seq, set, sig, some, sum, then, this, univ, with}, 
    morestring=[b]", 
    breaklines=true,
    captionpos=b,
    keepspaces=true,
    numbers=left,
    numbersep=5pt,
    showspaces=false,
    showstringspaces=false,
    showtabs=false,
    tabsize=2
}

\lstdefinestyle{pythonstyle}{
    backgroundcolor=\color{gray!20},   
    frame=single,                      
    rulecolor=\color{black},           
    commentstyle=\color{teal},         
    keywordstyle=\color{blue},      
    numberstyle=\tiny\color{black},    
    stringstyle=\color{blue},        
    basicstyle=\ttfamily\tiny,         
    breakatwhitespace=false,           
    breaklines=true,
    captionpos=b,                      
    keepspaces=true,                   
    numbers=left,                      
    numbersep=5pt,                     
    showspaces=false,                  
    showstringspaces=false,            
    showtabs=false,                    
    tabsize=2,                         
    language=Python,                   
    xleftmargin=8pt,                   
    framexleftmargin=5pt,              
}

\usepackage[normalem]{ulem} 

\newcommand{\tool}[0]{\textsc{LLM-FLAR}\xspace}

\usepackage{pifont}

\newcommand{\cmark}{\ding{51}}
\newcommand{\xmark}{\ding{55}}

\newcommand{\no}{\textcolor{red}{\xmark}}
\newcommand{\yes}{\textcolor{black}{\cmark}}
\newcommand{\ma}[1]{\color{black}#1}

\newcommand{\rashed}[1]{\textcolor{black}{#1}}

\newcommand{\orcid}[1]{\href{https://orcid.org/#1}{\includegraphics[width=10pt]{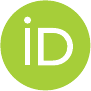}}}
\usepackage{geometry}
\usepackage[misc,geometry]{ifsym}
\renewcommand{\cite}[1]{\citep{#1}}

\newcommand{\moh}[1]{{\color{black}#1}} 
\newcommand{\hamid}[1]{{\color{black}#1}}

\begin{document}

\title{An Empirical Evaluation of Pre-trained Large Language Models for Repairing Declarative Formal Specifications}

\author[1]{Mohannad Alhanahnah\orcidlink{0000-0001-7108-3809}}
\author[2]{Md Rashedul Hasan\orcidlink{0009-0009-3417-4352}}
\author[2]{Lisong Xu\orcidlink{0000-0003-3465-4056}}
\author[2]{Hamid Bagheri\orcidlink{0000-0001-6686-466X}}

\affil[1]{Chalmers University of Technology, Gothenburg, Sweden}
\affil[2]{University of Nebraska–Lincoln, Lincoln, NE, USA}
\date{}



\maketitle
\input{abstract}
\input{introduction}

\input{background}

\input{approachV2}
\input{experimentDesign}
\input{experimentResults}

\input{threatsToValidity}
\input{relatedwork}

\input{conclusion}

\subsection{Data Availability Statement}
All data used to write this paper is open-source and publicly available at \url{https://github.com/Mohannadcse/AlloySpecRepair}.

\newpage
\appendix
\input{appendix}

\newpage

\clearpage
\bibliographystyle{plainnat}
\bibliography{alloy,apr,llm}

\end{document}

%% file: abstract.tex
\begin{abstract}
Automatic Program Repair (APR) has garnered significant attention as a practical research domain focused on automatically fixing bugs in programs. While existing APR techniques primarily target imperative programming languages like C and Java, there is a growing need for effective solutions applicable to declarative software specification languages. 
This paper systematically investigates the capacity of Large Language Models (LLMs) to repair declarative specifications in Alloy, a declarative formal language used for software specification.
\ma{We designed 12 different repair settings, encompassing single-agent and dual-agent paradigms, utilizing various LLMs. These configurations also incorporate different levels of feedback, including an auto-prompting mechanism for generating prompts autonomously using LLMs.}
\ma{Our study reveals that dual-agent with auto-prompting setup outperforms the other settings, albeit with a marginal increase in the number of iterations and token usage. This dual-agent setup demonstrated superior effectiveness compared to state-of-the-art Alloy APR techniques when evaluated on a comprehensive set of benchmarks.}
\ma{This work is the first to empirically evaluate LLM capabilities to repair declarative specifications, while taking into account recent trending LLM concepts such as LLM-based agents, feedback, auto-prompting, and tools, thus paving the way for future agent-based techniques in software engineering.}
\end{abstract}
 

%% file: introduction.tex
\section{Introduction} \label{sec:intro}

Declarative specification languages have become instrumental in addressing a multitude of software engineering challenges. Among these languages, the Alloy specification language~\cite{Jackson2006} has emerged as a powerful tool, leveraging relational algebra and first-order logic to tackle a diverse array of tasks within the software engineering domain. Its application spans across 
software verification~\cite{DBLP:conf/icse/BagheriKM20}, the security analysis of cutting-edge platforms such as the IoT and Android systems~\cite{DSN2016,DBLP:journals/ese/BagheriWAGM21,ISSTA2020}, and the creation of test cases~\cite{Khurshid04,TrimDroid}.
Integrating Alloy with the Alloy Analyzer has facilitated automated property verification, simplifying the process of checking whether specifications adhere to desired properties, and seamlessly integrating it into the Alloy environment. Despite these advancements, Alloy users, similar to developers in imperative languages, encounter challenges in debugging and correcting subtle bugs that may arise during specification writing, particularly for complex systems.

While the Alloy Analyzer aids in automatic property verification and counterexample generation, debugging and rectifying issues in Alloy specifications remain laborious and manual tasks. Unlike the rich literature and techniques available for automatic program repair (APR) in imperative languages, the landscape for APR in declarative languages is relatively sparse. Early approaches like ARepair~\cite{arepair_icse} rely mainly on failing test cases to identify and rectify bugs, assuming the presence of tests for verification. However, this paradigm does not align well with Alloy's assertion-based specification approach, where users articulate expectations using assertions rather than tests. 
Existing APR techniques like ARepair~\cite{arepair_icse} may succumb to overfitting issues when the test coverage is insufficient, compromising the correctness of generated repairs. Despite efforts to mitigate these challenges with techniques like BeAFix~\cite{BeAFix}, which leverages assertions as correctness oracles, there is room for improvement in terms of efficiency and effectiveness.

The emergence of Large Language Models (LLMs) has revolutionized various domains, including natural language processing and code generation. These pre-trained models, such as GPT-3.5-Turbo and GPT-4, exhibit remarkable accuracy in predicting natural language and generating code. Inspired by these advancements and the persistent challenges in repairing formal specifications, we embark on a systematic empirical study to explore the potential of using LLMs to repair faulty alloy specifications. Our research aims to investigate key questions regarding the effectiveness, performance, adaptive prompting, failure characteristics, and repair costs associated with the use of LLM in \moh{Alloy} specification repair. \ma{This investigation leverages recent advancements, including LLM-agents, tools, and feedback mechanisms.}

This paper presents a systematic exploration and results of our comprehensive evaluation of the efficacy of LLMs in repairing faulty Alloy specifications, compared to state-of-the-art Alloy APR techniques. We design a repair pipeline working iteratively and integrating a dual-agent LLM framework comprising a Repair Agent and an Instructor Agent~\footnote{Our implementation and artifacts are available at: \url{https://github.com/Mohannadcse/AlloySpecRepair}}. This iterative strategy centralizes around repairing bugs in defective specifications and generating specialized prompts to guide the repair process. 

We conduct an extensive evaluation of LLM effectiveness in repairing defective Alloy specifications, comparing it against several state-of-the-art Alloy APR techniques, including ARepair\cite{arepair_icse}, ICEBAR~\cite{ICEBAR}, BeAFix~\cite{BeAFix}, and ATR~\cite{ATR}, shedding light on their capabilities, benefits, and potential limitations. Our evaluation encompassed two comprehensive sets of benchmarks, comprising 1,974 defective specifications sourced from ARepair~\cite{arepair_icse} and Alloy4Fun~\cite{Alloy4Fun}, developed by external research groups. 
For each defective specification, we assess the repair pipeline across six distinct settings, conducting up to six iterations per setting. 
Our findings contribute to advancing the field of automatic repair for declarative specifications and provide insights into leveraging LLMs for effective Alloy specification repair.

%% file: background.tex
\section{Background and Motivation} \label{sec:background}


This section provides an overview of multi-agent LLMs followed by an illustrative example of a faulty Alloy specification to motivate the research and help the reader follow the discussions that ensue.



\subsection{Multi-Agent Large Language Models}

\subsubsection{Large Language Models}
Emerging pre-trained LLMs have demonstrated impressive performance on natural language tasks, such as text generation~\cite{brown2020language,chowdhery2022palm,vaswani2017attention} and conversations~\cite{thoppilan2022lamda,openai2023gpt4}. LLMs have also been proven effective in translating natural language specifications and instructions into executable code~\cite{fan2023automated,Jigsaw,Chen2021EvaluatingLL}. These models have been trained on extensive corpora and possess the ability to execute specific tasks without the need for additional training or hard coding~\cite{bubeck2023sparks}. They are invoked and controlled simply by providing a natural language prompt. The degree to which LLMs understand tasks depends largely on the prompting techniques used to convey user-provided instructions. These prompts are categorized into two frameworks: ``zero-shot'' learning and ``few-shot'' learning~\cite{brown2020language}. Within the ``zero-shot'' learning framework, the prompt includes a description of the problem and instructions for its resolution, aiming that LLMs can tackle previously unseen tasks. Conversely, in the context of ``few-shot'' learning, LLMs are provided with examples, supplementing the guidance offered in the ``zero-shot'' prompt.

\subsubsection{Agentic LLMs}
With recent advancements in LLMs, developers are embracing the idea of creating autonomous agents that can solve various tasks and interact with environments, humans, and other agents using natural language interfaces~\cite{zhou2023agents}. These agents provide several features including planning, memory, tool usage, and multi-agent communication. Therefore, LLM-based autonomous agents have gained tremendous interest in industry and academia~\cite{wang2023survey}. Several frameworks have been developed to support harnessing multi-agent LLM applications. 
AGENTS~\cite{agents} is a unified framework and open-source library for language agents. AGENTS aims to facilitate developers to build applications with language agents. AutoGPT~\cite{AutoGPT} a multi-agent framework for LLMs, is designed to support autonomous applications of LLMs.
LangChain~\cite{langchain} supports end-to-end language agents that can automatically solve tasks specified in natural language. It also facilitates the connection between LLM agents and external tools. Langroid~\cite{langroid} is another framework that supports the development of multi-agent LLMs. It offers the capability to manage the history of messages, thus controlling the context window of LLMs. This is a crucial functionality for LLM apps operating iteratively (i.e., Self-Refine~\cite{selfrefine} and Reflexion~\cite{reflexion}). Moreover, Langroid enables seamless integration with a variety of LLMs. 

\subsubsection{Tools and function calling}
Multi-agent frameworks and recent versions of OpenAI's LLM have introduced a feature known as function calling\footnote{https://platform.openai.com/docs/guides/text-generation/function-calling}. This feature enables users to provide function descriptions to the LLM. In turn, the LLM responds with a structured response (i.e., JSON data containing the requisite arguments for invoking any available functions). 
These functions serve as action executors and provide the option to supply APIs that the LLM can query to obtain essential information for responding to users. For example, when a user inquires about the current weather in a specific city and provides the LLM with a weather API call description, the LLM can augment its response with information retrieved from the API. This facilitates a seamless and efficient interaction between the user and the LLM and provides rich responses to avoid hallucinations. 

\subsection{Alloy - Illustrative Example}
Alloy, a formal modeling language, provides a comprehensible syntax inspired by object-oriented notations and is grounded on first-order relational logic~\cite{Jackson2006}. Within an Alloy specification, three primary components shape its structure: data types, constraints expressed through formulas, and commands to initiate the analyzer.


\hamid{The language uses signature (\textsf{sig}) definitions to define sets of elements, with fields specifying relationships between these sets.}
Additionally, Alloy employs \textsf{fact} to introduce constraints that \moh{hold} in every instance of the specification. These constraints restrict the model space, ensuring its consistency. Further structuring of formulas is achieved through \textsf{pred} and \textsf{fun}, which are named parameterized Alloy \moh{expression}, and \textsf{assert} encapsulates the properties intended for \moh{verification}.

Commands such as \textsf{check} and \textsf{run} activate the automated analyzer. \textsf{Check} verifies assertions, while \textsf{run} executes predicates, aiming to identify model instances that satisfy specified conditions.
Alloy's expressiveness stems from its use of relational logic, a first-order logic extended with relational operators. These include \textsf{all}, \textsf{some}, \textsf{one}, and \textsf{lone} quantifiers, along with operators like relational join and transitive closure.

\moh{To illustrate the Alloy language and provide motivation for this research, we examine the specification shown in Listing~\ref{lst:grade-motivating-ex} from the ARepair benchmark~\cite{arepair_icse}. This model represents a university context with students, professors, classes, and assignments. 
The specification defines relationships such as teaching assistants (students assigned as assistants), instructors (professors), and the association of assignments with both classes and students. 
The predicate \texttt{PolicyAllowsGrading} determines who can grade assignments, allowing only instructors or teaching assistants (TAs) of a class. However, a bug in the policy allows a student to grade their own assignment if they are TA for the same class. The fix is explicitly preventing students from grading their own assignments by adding the condition (at line 15). This condition ensures a person ``s" who qualifies as a grader (either as a TA or an instructor) must not be among the students assigned that specific assignment.
The assertion \texttt{repair\_assert\_1} and predicate \texttt{repair\_pred\_1} at the end formalize and check the requirement that no individual is allowed to grade an assignment assigned to themselves.
}

\moh{Reasoning about and addressing this defect through direct prompting of LLMs can be challenging, as demonstrated in~\cite{hasan2023automated}. Additionally, conventional state-of-the-art Alloy repair techniques, such as ARepair~\cite{arepair_icse} and BeAFix~\cite{BeAFix}, also struggled to effectively address this defect. Given recent advances in LLM agents and prompt engineering techniques, this study aims to evaluate the capabilities of LLMs in light of these developments} to repair such cases and evaluate their effectiveness in contrast to state-of-the-art Alloy repair techniques. In the following sections, we explore the ability of LLMs to address such challenges and the specific conditions under which they can do so.

\input{grade}

\hamid{
\subsection{Underconstrained and Overconstrained Specifications}

\subsubsection{Underconstrained Specifications}

Underconstrained specifications arise when essential constraints are omitted, permitting behaviors that deviate from the intended semantics. This class of defects often results in models that allow for unintended or invalid configurations.
As illustrated in Listing~\ref{lst:underconstrained}, the specification enforces that each \texttt{Tree} must have at least one \texttt{root}, but does not restrict the number of root nodes. Consequently, multiple nodes could simultaneously serve as roots for a single tree, contradicting the common assumption of a unique root in tree structures. Repairs addressing this issue typically involve replacing \texttt{some} with \texttt{one}, or introducing explicit uniqueness constraints to ensure the intended behavior.
}

\begin{lstlisting}[style=alloystyle,
caption={Example of an underconstrained specification that permits multiple roots.},
label={lst:underconstrained}]
sig Tree {
root: set Node
}
sig Node {
children: set Node
}
fact WellFormed {
all t: Tree | some t.root
}
\end{lstlisting}

\hamid{
\subsubsection{Overconstrained Specifications}

In contrast, overconstrained specifications impose overly restrictive conditions, unintentionally excluding valid model instances. This type of defect typically results in models that are too limited, preventing the representation of legitimate scenarios.
As shown in Listing~\ref{lst:overconstrained}, while the first constraint correctly enforces a unique root, the second constraint forbids any \texttt{Node} from having children—effectively eliminating all hierarchical structures. As a result, the specification precludes valid tree configurations. Repairs in such cases often involve relaxing the restrictive constraint, such as substituting \texttt{no} with \texttt{some}, to reintroduce permissible structure while preserving correctness.
}

\begin{lstlisting}[style=alloystyle,
caption={Example of an overconstrained specification that disallows any node hierarchy.},
label={lst:overconstrained}]
fact WellFormed {
all t: Tree | one t.root
all n: Node | no n.children
}
\end{lstlisting}

%% file: grade.tex
\begin{lstlisting}[style=alloystyle, caption={\moh{A snippet of the flawed \texttt{grade} Alloy specification sourced from the ARepair benchmark\protect\footnotemark}.}, label=lst:grade-motivating-ex]
abstract sig Person {}
sig Student extends Person {}
sig Professor extends Person {}
sig Class {
	assistant_for: set Student,  // as in : "is TA for"
	instructor_of: one Professor
}
sig Assignment {
	associated_with: one Class,
	assigned_to: some Student
}

pred PolicyAllowsGrading(s: Person, a: Assignment) {
	s in a.associated_with.assistant_for or s in a.associated_with.instructor_of
	// Fix: add "s !in a.assigned_to".
}

assert repair_assert_1 {
	all s : Person | all a: Assignment | PolicyAllowsGrading[s, a] implies not s in a.assigned_to
}

check repair_assert_1

pred repair_pred_1 {
	all s : Person | all a: Assignment | PolicyAllowsGrading[s, a] implies not s in a.assigned_to
}

run repair_pred_1
\end{lstlisting}\footnotetext{https://github.com/guolong-zheng/atmprep/blob/master/benchmark/arepair/grade1.als}

%% file: approachV2.tex
\section{Methodology} \label{sec:approach}
This section outlines the repair pipeline, detailing its architecture and workflow. The pipeline is derived from the APR phases described in~\cite{apr_phases_survey}, including localization, repair, and verification. \ma{The LLM is responsible for the localization and repair steps, while the verification step is facilitated by granting the LLM access to the Alloy analyzer. Consequently, the LLM can autonomously execute all APR phases to repair the defective model}. Our implementation of this pipeline utilizes the Langroid framework~\cite{langroid}.

\subsection{Overall Workflow}
This section presents the design of the repair pipeline specifically designed for iterative self-refinement, 
wherein the repair process for a defective Alloy model operates within a predetermined budget, defined by the number of iterations allocated. Previous research has indicated that this iterative prompting method produces superior results compared to traditional single-step approaches, improving task efficiency by an average of 20\%~\cite{selfrefine}. Should the repair process exceed this limit without success, the pipeline halts further attempts. After each unsuccessful repair iteration, feedback is collected and used to update the prompt, refining the initial draft generated by the LLM.

\begin{figure}[tb!]
    \raggedright
    \centering
    \includegraphics[width=0.9\linewidth,keepaspectratio]{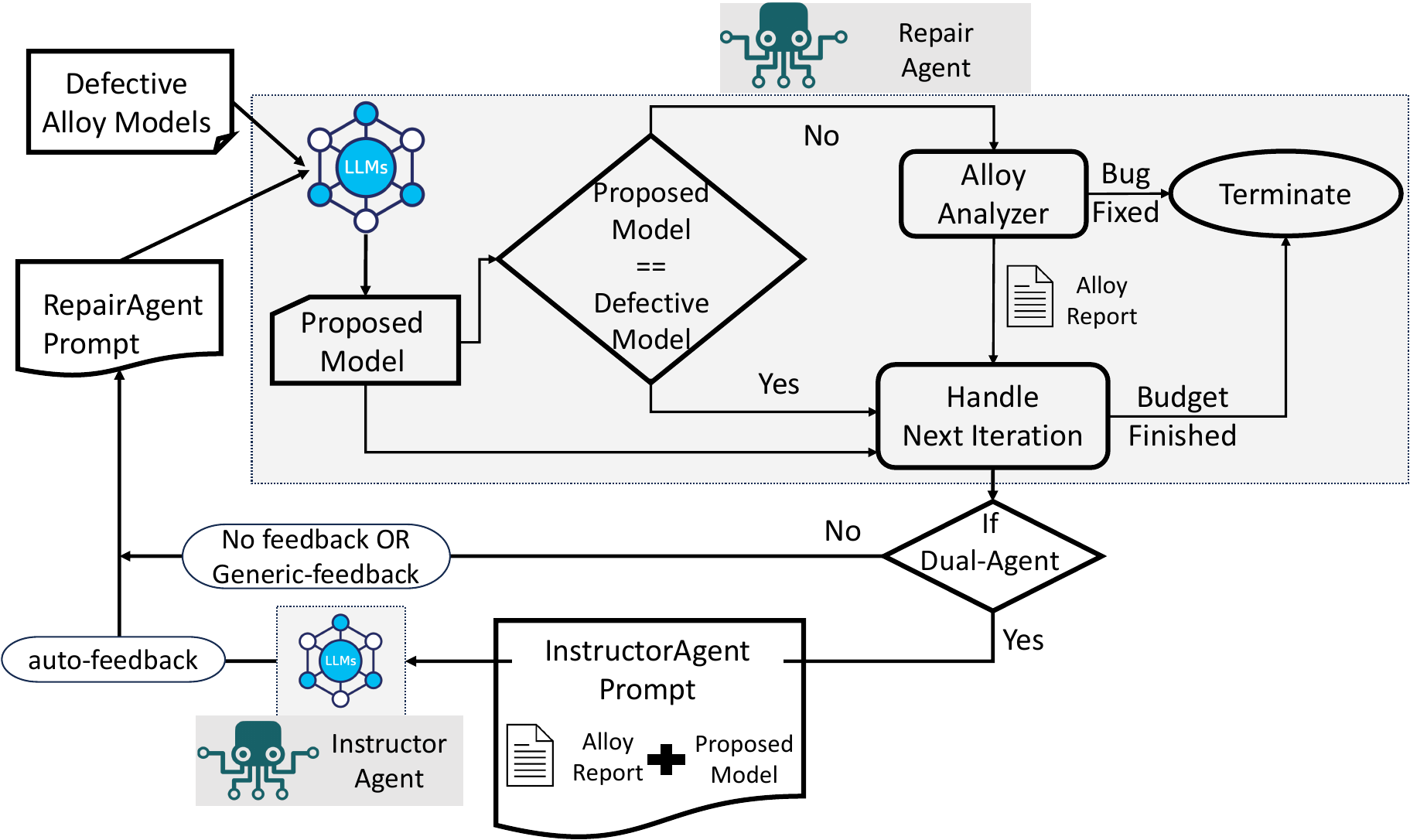}
    \vspace{-0.3cm}
    \caption{APR Pipeline for Alloy specifications. \ma{It supports Single-Agent (\textsf{Repair Agent} Only) and Dual-Agent (\textsf{Repair Agent} and \textsf{Instructor Agent}) paradigms.}}
    \label{fig:workflow}
\end{figure}

The workflow of our APR pipeline, as depicted in Figure~\ref{fig:workflow}, comprises two primary components: the \textsf{Repair Agent} and the \textsf{Instructor Agent}. Each agent maintains its context, tools, and prompts, but they collaborate to adaptively refine the \textsf{Repair Agent} prompt for enhanced effectiveness, as elaborated below. Our pipeline supports various feedback levels, employing a zero-shot prompting approach~\cite{paul2023enhancing}, where tasks are performed without explicit examples, as detailed in Table~\ref{tab:prompts}.

Upon receiving defective Alloy specifications and the prompt, our APR pipeline dispatches these elements to the LLM, which generates a patched version aimed at rectifying the defect. Subsequently, the \textsf{Repair Agent} compares the proposed model with the buggy ones to evaluate the ability of LLM to locate faults and adhere to instructions. The accuracy of the proposed model is then assessed by triggering the Alloy Analyzer tool, which runs the proposed model for validation. If the Alloy analyzer confirms the bug's resolution, the repair process concludes. Conversely, if errors persist, the pipeline initiates another iteration, depending on the remaining repair budget. If the budget is depleted, the pipeline terminates operations.
The adaptive prompt is designed to evolve and refine based on different settings (detailed in Section~\ref{sec:feedback}), ensuring continuous improvement after each unsuccessful attempt. 
The following sections provide a comprehensive overview of each component of the repair pipeline.

\input{prompts}

\subsection{Pipeline Agents}
As depicted in Figure~\ref{fig:workflow}, the pipeline comprises two agents. This design facilitates evaluating the recent deployment of LLM apps as multi-agent applications.

\subsubsection{\textbf{Repair Agent}}
This agent is the core component of the APR pipeline. 
It has access to tools/functions (i.e., ``run\_alloy\_analyzer''), maintains its context (i.e., history of messages) over the iterations, and decides the process for handling subsequent iterations. 

The agent's prompt encompasses a set of messages combined as a system message. These messages are listed in Table~\ref{tab:prompts}. 
\begin{itemize}
    \item \textbf{Agent Instruction.} This message provides a general guideline to the \textit{Repair Agent}, commencing with the agent's persona as an ``expert in repairing Alloy specifications''. It proceeds to notify LLM of defects within the provided Alloy model, without disclosing the defects' locations. The message finally describes the tools (i.e., ``run\_alloy\_analyzer'') to which this agent has access. In conclusion, it enumerates a set of procedural steps to be executed by the LLM.
    \item \textbf{Tool Instruction.} Offers instructions delineate the conditions and procedures for utilizing the tool ``run\_alloy\_analyzer''. It further specifies the data required from the LLM to ensure the proper activation of this tool. 
    This instruction guides the LLM to transmit a JSON formatted response, encapsulating two fields: (1) \texttt{request}, denoting the tool designation intended for use by the LLM, and 
    (2) \texttt{proposed\_specification}, representing the LLM's suggested version for rectifying the identified bug.
\end{itemize}

\subsubsection{\textbf{Instructor Agent}}
This agent has its LLM settings and context and generates feedback based on the report generated by Alloy Analyzer and the proposed specification. \ma{This agent leverages the AutoPrompt concept~\cite{autoprompt} by automatically constructing a prompt using LLMs.}
The produced feedback (i.e., the automatically constructed prompt) guides the \textit{Repair Agent} on the methods for rectifying the bug.
The system message that provides the instructions to this agent is described in Table~\ref{tab:prompts}. Unlike the system message of the \textit{Repair Agent}, the agent's persona is described as ``Expert in Analyzing Alloy Analyzer reports''.
This agent does not share the same context or prompts with the \textit{Repair Agent}. But the response of the \textit{Instructor Agent} is appended to the prompt that will be used in the next iteration by the \textit{Repair Agent}, as illustrated in Figure~\ref{fig:workflow}.

\subsection{Feedback Messages} \label{sec:feedback}
The initial prompt contains only \textit{Agent Instruction} and \textit{Tool Instruction} defined by the \texttt{Repair Agent}, as described in Table~\ref{tab:prompts}. 
The pipeline refines the prompts in response to the behavior of the proposed specification.
This refinement process is contingent upon repetitive buggy specifications, errors that emerge after executing these specifications through the Alloy analyzer, or LLM failure to send a response based on the required JSON format. 

To this end, Table~\ref{tab:feedback} presents the feedback messages that are supported by the pipeline. These feedback messages will be appended to the initial prompt. Following is a description of these:
\begin{itemize}
    \item \textbf{Tool fallback.} This message informs LLM that the response received does not comply with the tool's JSON format, impeding the proposed specification's extraction. To address this issue, we developed a parser atop the Langroid JSON parser to retrieve Alloy specifications from the response. Nonetheless, the parser cannot handle all scenarios, resulting in the dispatch of this feedback to the LLM, soliciting adherence to the mandated JSON format. 
    \item \textbf{Repeated Spec.} This message is triggered when the proposed spec is the same as the buggy specification. In this situation, this message tells the LLM not to repeat the buggy specification.
    \item \textbf{Alloy Analyzer Report.} This message relays the feedback of the verification process using Alloy Analyzer. Table~\ref{tab:feedback} presents three feedback levels.  
\end{itemize}

\noindent
In the following, we discuss the details of the various feedback levels supported by the pipeline. 

\input{feedback}

\textbf{Alloy Analyzer Feedback Levels:}
\label{sec:settings}
The verification stage involves running the proposed specifications with the Alloy Analyzer. Subsequently, feedback conveying the Alloy Analyzer's report is generated, available at different levels of detail. This process aids in assessing the effectiveness of various refinements made to the prompts.

In this study, we consider three feedback levels to refine the prompt as outlined in Table~\ref{tab:feedback}.
These levels reflect errors, counterexamples, and instances identified after running the Alloy Analyzer in each iteration.
They also mimic different repair scenarios, as described below:

\begin{enumerate}
        \item \textit{No-feedback}. In this setting, the Alloy Analyzer agent returns only a single response (i.e., \texttt{The proposed specification DID NOT fix the bug.}) to LLM without describing the details of the report generated by the Alloy Analyzer. This setting is designed to demonstrate the ability of LLMs to identify and fix errors in a faulty model solely based on the content of the buggy model itself, mirroring the function of a programming language compiler.
        \item \textit{Generic-feedback}. We use a template to send the Alloy Analyzer report to LLM. This report summarizes counterexamples, instances, and errors. This scenario represents a common situation where the developer shares a summary of the issue on a question-and-answer platform (such as Stack Overflow) to get help. 
        \item \textit{Auto-feedback}. When this setting is enabled, the feedback response to LLM will be generated by another LLM agent (i.e.,\texttt{ Instructor agent}), \ma{which represents the dual-agent repair pipeline}. We forward the report generated by the Alloy Analyzer and the proposed Alloy specifications to the prompt agent, which generates a prompt to instruct the \texttt{Repair Agent}. 
    \end{enumerate}

\noindent 
\ma{Noteworthy, the \textit{No-feedback} and \textit{Generic-feedback} mechanisms are used under the single-agent paradigm, whereas the \textit{Auto-feedback} approach enables the dual-agent paradigm.}
The different reports reflect the ability of LLMs to (1) locate bugs and (2) correctly perform the repair process. For example, \textit{Generic-feedback} and \textit{Auto-feedback} show the ability of LLM to perform its reasoning based on a human-created prompt versus an LLM-generated prompt, which has no access to the repair context. Also, the latter reflects a real scenario wherein the user cannot repair the buggy specification, then they consult experts or other forums to get assistance.

%% file: prompts.tex
\begin{table*}[htbp]
\centering
\caption{Zero-shot prompts used by the repair pipeline}
\scalebox{0.76}{
\begin{tabular}{|p{1.2cm}|p{2.3cm}|p{15cm}|}
\hline
\textbf{Agent} & \textbf{Message Type} & \textbf{Message content} \\ \hline
\multirow{6}{*}{
\parbox[c]{1cm}{\centering Repair Agent}} 
& Agent-Instruction & You are an expert in repairing Alloy declarative specifications. You will be presented with Alloy $<$Faulty\_SPECIFICATIONS$>$. Your task is to FIX/REPAIR the $<$Faulty\_SPECIFICATIONS$>$. Use the tool `run\_alloy\_analyzer` to demonstrate and validate the $<$FIXED\_SPECIFICATIONS$>$. Wait for my feedback, which may include error messages or Alloy solver results. You will have 5 trials to fix the $<$Faulty\_SPECIFICATIONS$>$. **Adhere to the Following Rules**: - The $<$FIXED\_SPECIFICATIONS$>$ should be consistent (having instances) and all the assertions should be valid (no counterexample). - DO NOT REPEAT the $<$FIXED\_SPECIFICATIONS$>$ that I sent you. - DO NOT provide any commentary and always send me anything ONLY using the tool `run\_alloy\_analyzer`. - The $<$FIXED\_SPECIFICATIONS$>$ MUST be different than the $<$Faulty\_SPECIFICATIONS$>$. \\ \cline{2-3} 
& Tool-instruction & ALL AVAILABLE TOOLS and THEIR JSON FORMAT INSTRUCTIONS: You have access to the following TOOLS to accomplish your task: TOOL: run\_alloy\_analyzer, PURPOSE: To show a $<$FIXED\_SPECIFICATIONS$>$ to the user. Use this tool whenever you want to SHOW or VALIDATE the $<$FIXED\_SPECIFICATIONS$>$. NEVER list out a $<$FIXED\_SPECIFICATIONS$>$ without using this tool. JSON FORMAT: {"type": "object",    "properties": {"request": {           "default": "run\_alloy\_analyzer",            "type": "string"},"specification": {            "type": "string"}}, "required": [        "specification", "request"]}. When one of the above TOOLs is applicable, you must express your request as "TOOL:" followed by the request in the above JSON format. \\ \cline{2-3} 
\hline 
\parbox[c]{1cm}{\centering Instructor Agent}
& Agent-Instruction & You are Expert in Analyzing Alloy Analyzer reports. Can you describe concisely and precisely the modifications needed to fix the error in at most 2 sentences? Based on this report from Alloy Analyzer: \{Alloy\_report\_msg\} After running this Alloy Model is: \{proposed\_spec\} \\ 
\hline
\end{tabular}
}
\label{tab:prompts}
\end{table*}

%% file: feedback.tex
\begin{table*}[htbp]
\centering
\caption{Feedback Messages used by the repair pipeline}
\scalebox{0.7}{
\begin{tabular}{|l|p{2cm}|p{15cm}|}
\hline
Feedback Type & Generated By & Message Content \\ \hline
Tool-fallback & \multirow{4}{*}{Repair Agent} & You must use the CORRECT format described in the tool `run\_alloy\_analyzer` to send me the fixed specifications. You either forgot to use it, or you used it with the WRONG format. Make sure all fields are filled out. \\ \cline{1-1} \cline{3-3} 
Repeated spec &  & The proposed $<$FIXED\_SPECIFICATIONS$>$ is IDENTICAL to the Alloy $<$Faulty\_SPECIFICATIONS$>$ that I sent you. **DO NOT** send Alloy specifications that I sent you again. ALWAYS USE the tool `run\_alloy\_analyzer` to send me a new $<$FIXED\_SPECIFICATIONS$>$. \\ \cline{1-1} \cline{3-3} 
No-feedback &  & The proposed specification DID NOT fix the bug. \\ \cline{1-1} \cline{3-3} 
Generic-feedback &  & Below are the results from the Alloy Analyzer. Fix all Errors and Counterexamples before sending me the next $<$FIXED\_SPECIFICATIONS$>$. \\ \hline
Auto-feedback & Instructor Agent & The message content will be dynamically constructed by the agent based on the supplied information. \\ \hline
\end{tabular}
}
\label{tab:feedback}
\end{table*}

%% file: experimentDesign.tex
\section{Experiment Design}

In this study, we address the following research questions (RQs):

\begin{itemize}
    \item \textbf{RQ1 (Effectiveness):} How effective is the APR pipeline in repairing compared to the state-of-the-art \ma{and how various repair settings contribute to the effectiveness}?
    \item \textbf{RQ2 (Performance):} What is the repair performance when employing pre-trained LLMs?
    \item \textbf{RQ3 (Failure Characteristics):} What are the characteristics of failures encountered during the repair process?
    \item \textbf{RQ4 (Repair Costs):} What is the cost associated with using the APR pipeline with various LLMs?
\end{itemize}

\subsection{Dataset} In our evaluation, we utilize two distinct benchmark suites: ARepair~\cite{arepair_icse} and Alloy4Fun~\cite{Alloy4Fun}. These benchmark suites have undergone extensive study and developed by independent research groups, enabling a fair comparison of various techniques.
The benchmark datasets consist of specifications varying from tens to hundreds of lines, featuring real bugs authored by humans. The defects within the benchmarks span a diverse range, from straightforward issues that can be addressed by modifying a single operator to complex challenges necessitating the synthesis of novel expressions and the substitution of entire predicate bodies.

The ARepair benchmark~\cite{arepair_icse} encompasses 38 faulty specifications extracted from a total of 12 Alloy problems. Within this benchmark, both single and multi-line errors are present, distributed across 28 faulty models with single-line errors and 10 faulty models with multi-line errors.
The Alloy4Fun benchmark~\cite{Alloy4Fun} comprises a collection of 1,938 handwritten defective models sourced from student submissions across six different Alloy problems. All bugs within this benchmark are single-line bugs. Both benchmarks are accompanied by correct versions of the specifications, serving as ground truths for verifying the accuracy of the obtained results.

\subsection{Pre-processing Benchmark Dataset} \label{sec:preprocessing}
As mentioned earlier, the benchmark datasets include comments indicating the locations of the bugs and their corresponding fixes. To ensure a fair evaluation, \moh{we perform two steps: (1) determining the uniqueness of Alloy specifications and (2) removing fix }comments from the defective models.

\rashed{\textbf{Uniqueness of Alloy Specifications.}} 
\hamid{To ensure that all specifications in the ARepair and Alloy4Fun benchmarks are unique, we applied a systematic normalization and hashing process. Each specification was processed in its entirety, preserving the original content. We removed block comments, and collapsed all whitespace (spaces, tabs, line breaks) to a single space. This normalization step eliminates superficial differences such as formatting and comments, allowing us to focus on the core logic of each Alloy specification. After normalization, we computed an MD5 hash for each specification, treating the hash as a unique fingerprint. Specifications with identical hashes were considered duplicates. We grouped specifications by hash and maintained counts for each benchmark and for the combined dataset.
Our analysis found that all 38 ARepair specifications and all 1936 Alloy4Fun files were unique—no duplicates were detected in the combined set of 1974 specifications. This confirms the diversity and uniqueness of the Alloy specifications in both datasets.
}

\textbf{Removal of Fix comments.} We have removed these comments from the defective models. Specifically, we have eliminated all lines starting with the phrase ``\textit{Fix:}" in the defective models. This line typically outlines the necessary changes required to resolve the bug. Models with multiple ``\textit{Fix:}" comments are categorized as having multi-line bugs, whereas those with only one ``\textit{Fix:}" comment are considered to have single-line bugs. Removing comments is essential to prevent LLMs from receiving explicit clues about bug locations and fixes. This ensures an accurate assessment of their repair capabilities and effectiveness in localizing bugs within defective models. \ma{Importantly, this reflects a realistic scenario in which the user employs LLMs without prior knowledge of the bug location and the fix.}

\subsection{Implementation} 
Our implementation of the APR pipeline was realized using Python 3.11, with the assistance of Langroid to facilitate the integration of multi-agent LLMs. This implementation encompasses several crucial functionalities. Firstly, it offers support for various LLMs, including local LLMs, ensuring flexibility in model selection. Secondly, it incorporates a message history control mechanism, essential for preventing context-length limitations, particularly during iterative repair processes. Lastly, our implementation facilitates the creation and customization of tools, allowing for the definition of response formats and fields. Additionally, we developed a specialized parser to address issues caused by special situations that could impede the extraction of proposed specifications.

The experiments were carried out on a system equipped with a 2.3 GHz Quad-Core Intel Core i7 processor, 32 GB of RAM, and running macOS Sonoma. Furthermore, the system was configured with Oracle Java SE Development Kit 8u202 (64-bit version), ensuring compatibility and optimal performance throughout the execution of the APR pipeline.

\subsection{Subject LLMs} \label{sec:llms}
To assess the performance of existing pre-trained LLMs, we have carefully chosen three representative models. \ma{We explored local LLMs such as Llama-2 and Mistral, but they lacked support for tools and JSON responses, making it challenging to process their responses, which often contained incomplete Alloy specifications. Following is a description of the selected models, which are summarized in Table~\ref{tab:llms}}:

\begin{itemize}
\item \textbf{GPT-3.5-Turbo.} This standard LLM, provided by OpenAI and utilized in ChatGPT, boasts a sophisticated architecture with 175 billion parameters, endowing it with extensive capabilities. Engineered to tackle a wide array of natural language processing (NLP) tasks, including text generation and completion, GPT-3.5-Turbo epitomizes advanced computational linguistics. 
\item \textbf{GPT-4-32k.} This model contains over 1.76 trillion parameters, demonstrating significantly enhanced capabilities compared to GPT-3.5. For our study, we leverage the GPT-4-32k-0613 model version.
\item \textbf{GPT-4-Turbo.} Another iteration of the GPT-4 model, GPT-4-Turbo features an updated knowledge cutoff as of April 2023 and introduces a 128k context window. Moreover, it offers cost-effectiveness compared to GPT-4, alongside notable enhancements such as improved instruction following, JSON mode, and reproducible outputs~\cite{gpt4turbo}.
\item \moh{\textbf{GPT-4o.} This model shares features with \textbf{GPT-4-Turbo} but is more cost-effective and has an updated knowledge base as of October 2023.}
\end{itemize}



\begin{table}[h]
\centering
\caption{Characteristics of Pre-trained LLMs}
\scalebox{0.8}{
\begin{tabular}{|l|l|l|l|l|l|}
\hline
\textbf{\textbf{Model}} & \textbf{\textbf{Version}} & \textbf{\textbf{Cut-off}} & \textbf{\textbf{\begin{tabular}[c]{@{}l@{}}Context Window\\ (Tokens)\end{tabular}}} & \begin{tabular}[c]{@{}l@{}}Input Cost\\ per 1M tokens\end{tabular} & \begin{tabular}[c]{@{}l@{}}Output Cost\\ per 1M tokens\end{tabular} \\ \hline
GPT-3.5 Turbo & 1106 & Sep 2021 & 16,385 & \$1 & \$2 \\ \hline
GPT-4-32k & 0613 & Sep 2021 & 32,768 & \$60 & \$120 \\ \hline
GPT-4 Turbo & 1106-preview & Apr 2023 & 128k & \$10 & \$30 \\ \hline
GPT-4o & 2024-05-01-preview & Oct 2023 & 128k & \$5 & \$15 \\ \hline
\end{tabular}
}
\label{tab:llms}
\end{table}

\subsection{Subject SOTA Systems}\label{sec:sota}
We conducted a comparative analysis of repair performance against several state-of-the-art Alloy repair tools: ARepair~\cite{arepair_icse}, ICEBAR~\cite{ICEBAR}, BeAFix~\cite{BeAFix}, ATR~\cite{ATR}, and Hasan et al.~\cite{hasan2023automated}. Each tool employs a distinct development approach and tackles specification attributes differently.

ARepair~\cite{arepair_icse} generates fixes for Alloy specifications that violate test cases. ICEBAR~\cite{ICEBAR} utilizes ARepair as a backend tool to repair faulty Alloy specifications based on a predefined set of Alloy tests. ATR~\cite{ATR} adopts a template-based methodology to enhance the repair process, leveraging fault localization strategies to identify potentially erroneous Alloy expressions that lead to assertion failures. BeAFix~\cite{BeAFix} relies on user input to identify faulty statements. It exhaustively explores all possible repair candidates up to a certain threshold through mutation and employs Alloy counterexamples to evaluate the feasibility of generalization. Finally, Hasan et al.~\cite{hasan2023automated} employed ChatGPT (GPT-3.5-Turbo) for repairing faulty specifications across five scenarios. \ma{This approach employs a purely LLM-based method, wherein a single repair iteration is conducted without the implementation of agent setup or feedback mechanisms}. For our comparison, we exclude scenarios where ChatGPT is provided with ``\textit{Fix:}'' comments, as this represents an unrealistic setting and does not align with our ``Benchmark Pre-processing" step (cf.~Section~\ref{sec:preprocessing}).


\hamid{
Since the state-of-the-art tools—ARepair, ICEBAR, BeAFix, and ATR—were  evaluated on static Alloy models prior to the release of Alloy 6, our experimental analysis focused exclusively on static Alloy models to ensure a fair comparison. These tools do not support mutable specifications or recently introduced features such as linear temporal logic (LTL), making them incompatible with models that leverage Alloy 6’s new capabilities. Additionally, due to structural dependencies and the absence of updated implementations, these tools could not be reliably executed on datasets incorporating Alloy 6-specific features. Nevertheless, our repair technique is not inherently limited to static specifications and, in principle, can be extended to support the dynamic capabilities introduced in Alloy 6.}

\subsection{Experimental Configuration and Settings}

\textbf{LLM Temperature.} To balance between deterministic progression and iterative refinement, we set the temperature parameter of the Large Language Models (LLMs) to 0.2, allowing for a moderate level of randomness while maintaining some level of determinism throughout the repair process~\cite{contrastrepair,autocoderover}.

\noindent
\textbf{Number of Iterations.} Initially, we conducted a preliminary experiment using the ARepair benchmark framework with GPT-4-32k, allocating a budget of ten iterations. However, we observed diminishing returns after six iterations in most cases. Therefore, we opted for a six-iteration budget in the APR pipeline.

\noindent
\textbf{Metric for Comparing Repair Performance of LLMs.} We employ the \textit{Correct@k} metric~\cite{llmcompdroid} to evaluate the success rate of the techniques in repairing defective Alloy specifications. This metric quantifies the number of defects successfully repaired within a maximum of $k$ iterations, where $k$ is set to 6 in our study (denoted as \textit{Correct@6}). The formula for this metric is described in Equation~\ref{eq:correct_at_k}.

\begin{equation}
\footnotesize
\text{Correct@}k = \left( \frac{\text{\# of bugs successfully repaired within } k \text{ iterations}}{\text{Total number of bugs evaluated}} \right) \times 100
\label{eq:correct_at_k}
\end{equation}

\noindent
\textbf{Experiment Settings and Baseline.} Our evaluation encompasses a total of 1,974 defective models. We assess three distinct LLMs, as outlined in Table~\ref{tab:llms}. The APR pipeline supports three levels of report granularity and permits multiple iterations, resulting in a total of 106,596 repair attempts (calculated as the product of 1,974 benchmarks, three LLMs, three feedback levels, and up to six iterations). \ma{The Single-agent paradigm utilizes the No-feedback and Generic-
feedback mechanisms, whereas the dual-agent paradigm employs the Auto-feedback approach within the repair pipeline}.

\hamid{
The repair experiment on the Alloy4Fun benchmark leverages GPT-4o—the most cost-effective option among GPT-4-32k and GPT-4-Turbo—for a randomly sampled subset of 357 models, maximizing both efficiency and performance. For comprehensive coverage, GPT-3.5-Turbo is applied to the entire Alloy4Fun dataset, taking advantage of its substantially lower cost compared to the GPT-4 family. This experimental design enables a broad and cost-efficient evaluation across all models, while also allowing for targeted, high-performance analysis using GPT-4o on a representative subset. 
Table~\ref{table:combinations} presents the settings corresponding to different combinations of LLMs and feedback levels.
}

\begin{table}[!ht]
\centering
\caption{Settings corresponding to various combinations of LLMs and feedback levels}
\begin{tabular}{cccl}
\toprule
\textbf{Setting No.} & \textbf{\# Agents} & \textbf{LLM} & \textbf{Feedback Level} \\ \midrule
Setting-1 & Single & GPT-4-32k & No-Feedback \\ 
Setting-2 & Single &  GPT-4-32k & Generic-Feedback \\ 
Setting-3 & Dual &  GPT-4-32k & Auto-Feedback \\ 
Setting-4 & Single &  GPT-4-Turbo & No-Feedback \\ 
Setting-5 & Single &  GPT-4-Turbo& Generic-Feedback \\ 
Setting-6 & Dual &  GPT-4-Turbo& Auto-Feedback \\ 
Setting-7 & Single &  GPT-4o & No-Feedback \\ 
Setting-8 & Single &  GPT-4o & Generic-Feedback \\ 
Setting-9 & Dual &  GPT-4o & Auto-Feedback \\ 
Setting-10 & Single &  GPT-3.5-Turbo & No-Feedback \\ 
Setting-11 & Single &  GPT-3.5-Turbo & Generic-Feedback \\ 
Setting-12 & Dual &  GPT-3.5-Turbo & Auto-Feedback \\ 
\bottomrule
\end{tabular}
\label{table:combinations}
\end{table}

%% file: experimentResults.tex
\section{Experimental Results} \label{sec:evaluation}

 
This section summarizes 
the data we collected, its interpretation, and our results.

\subsection{Results for RQ1: Effectiveness}
\ma{This research question investigates the effectiveness of LLMs in repairing Alloy specifications under various settings within the repair pipeline. To address this, we (1) compare the repair pipeline with state-of-the-art Alloy repair tools and (2) assess the impact of different feedback levels, LLMs, and agent paradigms on repair performance, 
}\moh{and study the influence of LLMs on the repaired models.}

\textbf{Comparing with SoTA.}
\hamid{
Tables~\ref{tab:ARepair-Compare-SoTA} and \ref{tab:A4F-Sample-Compare-SoTA} present a comprehensive comparison of our approach against state-of-the-art Alloy repair techniques, including 
ARepair~\cite{arepair_icse}, ICEBAR~\cite{ICEBAR}, BeAFix~\cite{BeAFix}, ATR~\cite{ATR}, and Hasan et al.~\cite{hasan2023automated}, on the ARepair and Alloy4Fun benchmarks, respectively. Each table reports the number of defects addressed and repairs achieved by each tool, as well as by various LLM settings described in Table~\ref{table:combinations}. For the ARepair benchmark (Table~\ref{tab:ARepair-Compare-SoTA}), results are shown across all settings, enabling a detailed evaluation of repair effectiveness. For the Alloy4Fun benchmark (Table~\ref{tab:A4F-Sample-Compare-SoTA}), GPT-4o is applied to a representative subset of models to balance performance and computational cost, while GPT-3.5-Turbo is evaluated on the full dataset.
}


\hamid{The ``Defects Count" column reports the total number of bugs identified within each model category, providing insight into the distribution of defects across specifications. When the value in ``Defects Count" is equal to ``Total \#specs," it indicates that each specification contains exactly one defect, suggesting the presence of only single-line bugs. In contrast, if ``Defects Count" exceeds ``Total \#specs," this reveals that some specifications contain multiple defects, indicative of multi-line bugs. For example, in Table~\ref{tab:ARepair-Compare-SoTA}, the \texttt{balancedBST} model exhibits 8 defects across 3 specifications, clearly demonstrating the occurrence of multi-line bugs.}

\hamid{
The subsequent columns present the number of correctly repaired specifications for each state-of-the-art Alloy repair tool, with the remaining columns reporting the repair outcomes under the various settings described in Table~\ref{table:combinations}.}

\hamid{
The results highlight the superior performance of the GPT-4 family—particularly GPT-4o and GPT-4-Turbo—over both GPT-4-32k and traditional state-of-the-art Alloy repair tools. In contrast, GPT-3.5-Turbo (Settings 10–12) demonstrates significantly lower repair effectiveness than traditional tools, which aligns with previous findings by Hasan et al.~\cite{hasan2023automated}.}

\hamid{
Furthermore, across all evaluated LLMs and benchmarks, the Auto-feedback configuration (Settings 3, 6, 9, and 12 in Tables~\ref{tab:A4F-Sample-Compare-SoTA} and \ref{tab:ARepair-Compare-SoTA}) consistently delivers the strongest repair performance. Notably, the Generic-feedback configuration (Setting 8) achieves results comparable to Setting 9 on the Alloy4Fun sampled benchmark with GPT-4o, a phenomenon attributed to the inherent stochasticity of LLMs. To further investigate, we re-executed repair experiments on five models that were initially only repaired under Generic-feedback; two of these were successfully repaired upon rerunning with Auto-feedback. This outcome underscores the effectiveness of LLM-driven prompt construction over generic, human-crafted prompts and supports the adoption of multi-agent LLM configurations for automated specification repair.}

\input{ARepair-Compare-SoTA}

\input{A4F-Sample-Compare-SoTA}

\hamid{
The repair results for the ARepair benchmark, as presented in Table~\ref{tab:ARepair-Compare-SoTA}, show that Settings 6 and 9 deliver substantially superior performance compared to all SoTA tools, including traditional program repair approaches. Notably, the APR pipeline is able to repair specifications that remain unresolved by most, or even all, existing SoTA techniques. For instance, the \texttt{grade} specification (see Listing~\ref{lst:grade-motivating-ex}), which could not be repaired by the majority of SoTA tools, was successfully addressed under both Settings 6 and 9. Similarly, the \texttt{farmer} specification, which contains a multi-line bug and was not repaired by any SoTA tool, was successfully fixed by all APR configurations. The \texttt{ctree} specification, which was addressed by only a single SoTA tool, was also repaired by all APR configurations leveraging GPT-4 models.
}

\hamid{
These results demonstrate that models from the GPT-4 family are particularly effective at handling complex specifications, especially those involving multi-line bugs. In particular, the findings underscore the promising potential of agentic LLMs enhanced with feedback mechanisms. As shown in Table~\ref{tab:correct@k}, which provides a comparative analysis of repair performance across all evaluated LLMs using the \textit{Correct@6} metric, the dual-agent paradigm (Auto-Feedback) consistently outperforms the single-agent approach across all models. Within this paradigm, GPT-4o achieved the highest repair performance.}

\begin{table}[!h]
\centering
\caption{
\hamid{Comparison of repair performance for different LLMs across various feedback levels—No-Feedback (NF), Generic-Feedback (GF), and Auto-Feedback (AF)—as measured by the Correct@6 metric.}
}
\scalebox{0.8}{
\begin{tabular}{llc|llc|llc|llc}
\toprule
\multicolumn{3}{c}{GPT-3.5-Turbo} & \multicolumn{3}{c}{GPT-4-32k} & \multicolumn{3}{c}{GPT-4-Turbo} & \multicolumn{3}{c}{GPT-4o}\\ \midrule
\multicolumn{2}{c}{Single-agent} & \multicolumn{1}{c}{Dual-agent} & \multicolumn{2}{c}{Single-agent} & \multicolumn{1}{c}{Dual-agent} & \multicolumn{2}{c}{Single-agent} & \multicolumn{1}{c}{Dual-agent} & \multicolumn{2}{c}{Single-agent} & \multicolumn{1}{c}{Dual-agent} \\ \cmidrule(lr){1-3} \cmidrule(lr){4-6} \cmidrule(lr){7-9} \cmidrule(lr){10-12}
\multicolumn{1}{l}{NF} & \multicolumn{1}{l}{GF} & AF & \multicolumn{1}{l}{NF} & \multicolumn{1}{l}{GF} & AF & \multicolumn{1}{l}{NF} & \multicolumn{1}{l}{GF} & AF & \multicolumn{1}{l}{NF} & \multicolumn{1}{l}{GF} & AF\\ 
\multicolumn{1}{l}{10.5} & \multicolumn{1}{l}{15.8} & 47.4 & \multicolumn{1}{l}{39.5} & \multicolumn{1}{l}{42.1} & 57.9 & \multicolumn{1}{l}{44.7} & \multicolumn{1}{l}{50.0} & 73.4 & \multicolumn{1}{l}{63.2} & \multicolumn{1}{l}{71.1} & 73.4\\ \bottomrule
\end{tabular}
}
\label{tab:correct@k}
\end{table}

\input{Size_of_variables_LLM}


To gain deeper insights into the complementarity of these methods, the Venn diagrams in Figure~\ref{fig:venn_diagrams_arepair} illustrate cases fixed exclusively by one tool but not by others, as well as the overlap between techniques (i.e., cases correctly repaired by multiple tools). In particular, the repair pipeline successfully addresses faulty models that other SoTA repair tools cannot resolve. 

\input{venn_diagrams_arepair}

\noindent

\hamid{
\textbf{Influence of Defective Model Complexity/Size on Repair Performance.}
To investigate the relationship between model complexity and repair performance, we assessed complexity based on the number of variables clauses in the propositional formulas generated by the Alloy Analyzer from LLM-produced repairs (see Tables~\ref{tab:variables-clauses-arepair} and \ref{tab:variable-clauses-a4f}) and conducted a detailed analysis thereof.}

\hamid{
GPT-4o consistently outperforms GPT-3.5-Turbo in repairing faulty specifications, despite the latter generating more complex formulas with significantly more clauses. The clause-to-variable ratio further highlights this difference (2.87 for GPT-3.5-Turbo vs. 1.36 for GPT-4o), indicating that GPT-4o constructs more concise and efficient logical expressions. This suggests that GPT-4o achieves more efficient and accurate repairs by leveraging simpler logical structures while maintaining correctness.} 

\hamid{
The tendency of GPT-3.5-Turbo to generate more verbose and constraint-heavy formulas, even with comparable variable counts, underscores the limitations imposed by its smaller context window and less sophisticated reasoning capabilities. The consistently higher clause-to-variable ratios across benchmarks for GPT-3.5-Turbo further support this conclusion. These findings confirm that model capability—especially in context comprehension and logical reasoning—is a critical factor for successful Alloy repair. GPT-4o’s superior performance, as highlighted in the \%repair row of Table~\ref{tab:A4F-Sample-Compare-SoTA}, demonstrates how advances in LLM architecture directly enhance formal reasoning and model repair effectiveness.}


\hamid{
\textbf{Discussing LLMs repair performance. }
The evaluation reveals a clear performance gap between GPT-3.5-Turbo and the GPT-4 family across all experimental configurations. GPT-3.5-Turbo consistently exhibits the lowest repair rates, largely due to its architectural limitations such as a smaller context window and less advanced reasoning capabilities. These constraints hinder its ability to effectively incorporate iterative feedback and maintain coherence over successive repair attempts, often causing context overflows and semantic inconsistencies. Additionally, GPT-3.5-Turbo frequently generates Alloy-specific errors, including incorrect operator usage, type mismatches, and hallucinated constructs, that are less common in specialized repair tools designed with native Alloy semantics. These shortcomings persist even in simpler Alloy4Fun models, limiting its practical utility in automated specification repair. In contrast, GPT-4 models, including GPT-4-32k, GPT-4-Turbo, and GPT-4o, demonstrate significantly improved repair performance across all settings and benchmarks. Benefiting from larger context windows and enhanced reasoning abilities, these models handle complex specifications and multi-line bugs more effectively. Among them, GPT-4o and GPT-4-Turbo achieve comparable results, with GPT-4o often leading in agentic (Auto-Feedback) configurations that leverage multi-agent prompt construction and iterative feedback. This dual-agent paradigm consistently boosts repair success, highlighting the advantages of advanced LLM architectures combined with adaptive prompting strategies.} 

\begin{tcolorbox}[colback=gray!20, colframe=gray!50!black, rounded corners]
\moh{The results for RQ1 indicate that an agentic LLM incorporating an iterative feedback mechanism achieves improved repair performance compared to both traditional state-of-the-art Alloy repair tools and non-agentic LLM configurations. Notably, GPT-4o consistently outperformed other LLMs, yielding the most effective repair outcomes across both agentic and non-agentic settings.
}
\end{tcolorbox}

\subsection{Results for RQ2: Performance}

This research question delves into the repair performance of various pre-trained LLMs concerning factors such as repair iteration budget, bug type, localization capabilities, and adherence to instructions. 
\hamid{
This analysis leverages a comprehensive benchmark that includes all evaluated LLMs and a diverse range of defect types, encompassing both single-line and multi-line bugs. As shown in Table 5, each defective model was tested across up to 12 distinct settings and up to 6 repair iterations, resulting in a total of 2,736 repair trials—effectively corresponding to the repair of 2,736 defective models. This extensive experimental setup provides a robust and representative evaluation of the repair capabilities of the LLMs across a wide variety of scenarios.}

\begin{figure*}[htbp!]
    \centering
    \begin{subfigure}[b]{0.32\textwidth}
        \centering
        (a)
         \includegraphics[width=\textwidth]{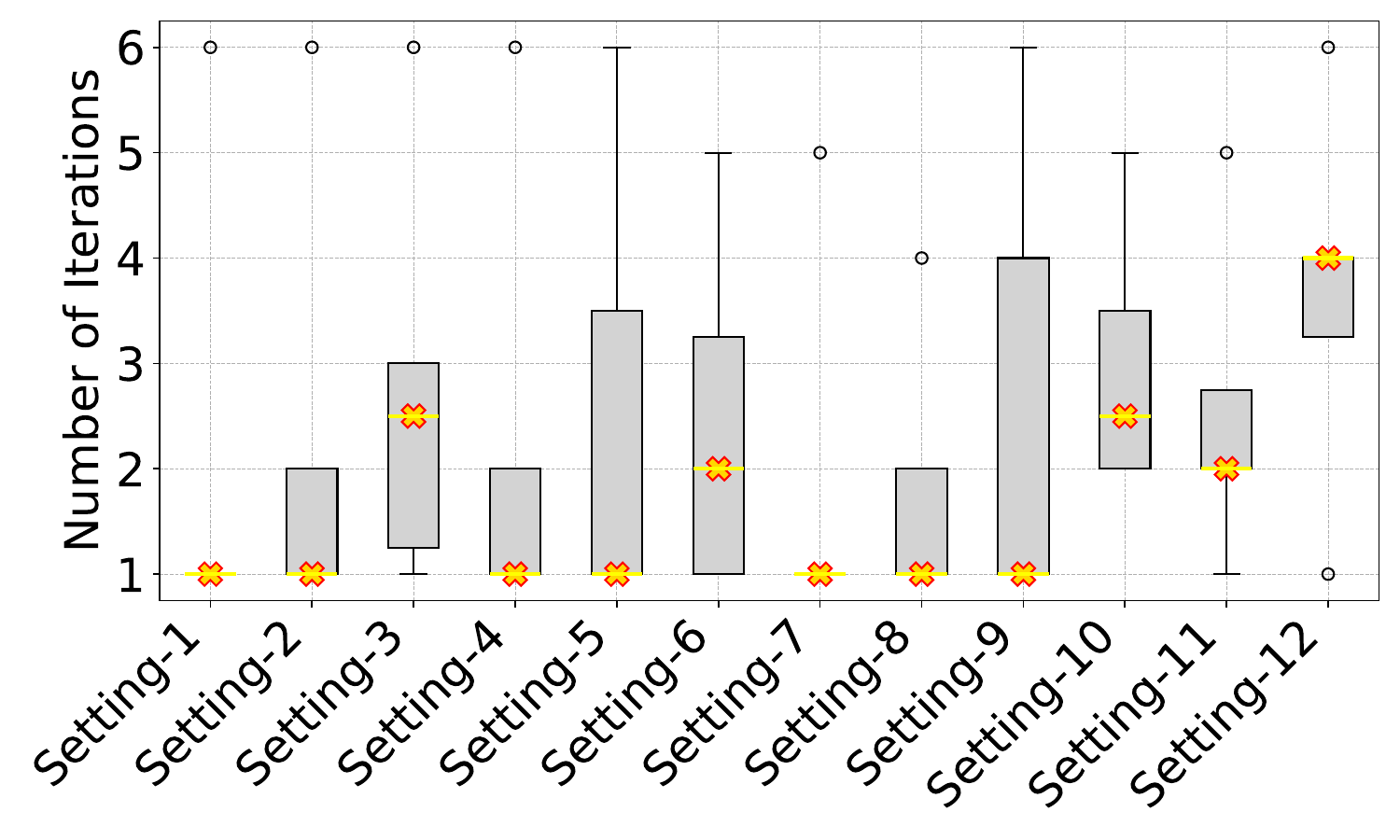}
        \label{fig:iterationAnalysis}
    \end{subfigure}
    \hfill
    \begin{subfigure}[b]{0.33\textwidth}
        \centering
        (b)        
        \includegraphics[width=\textwidth]{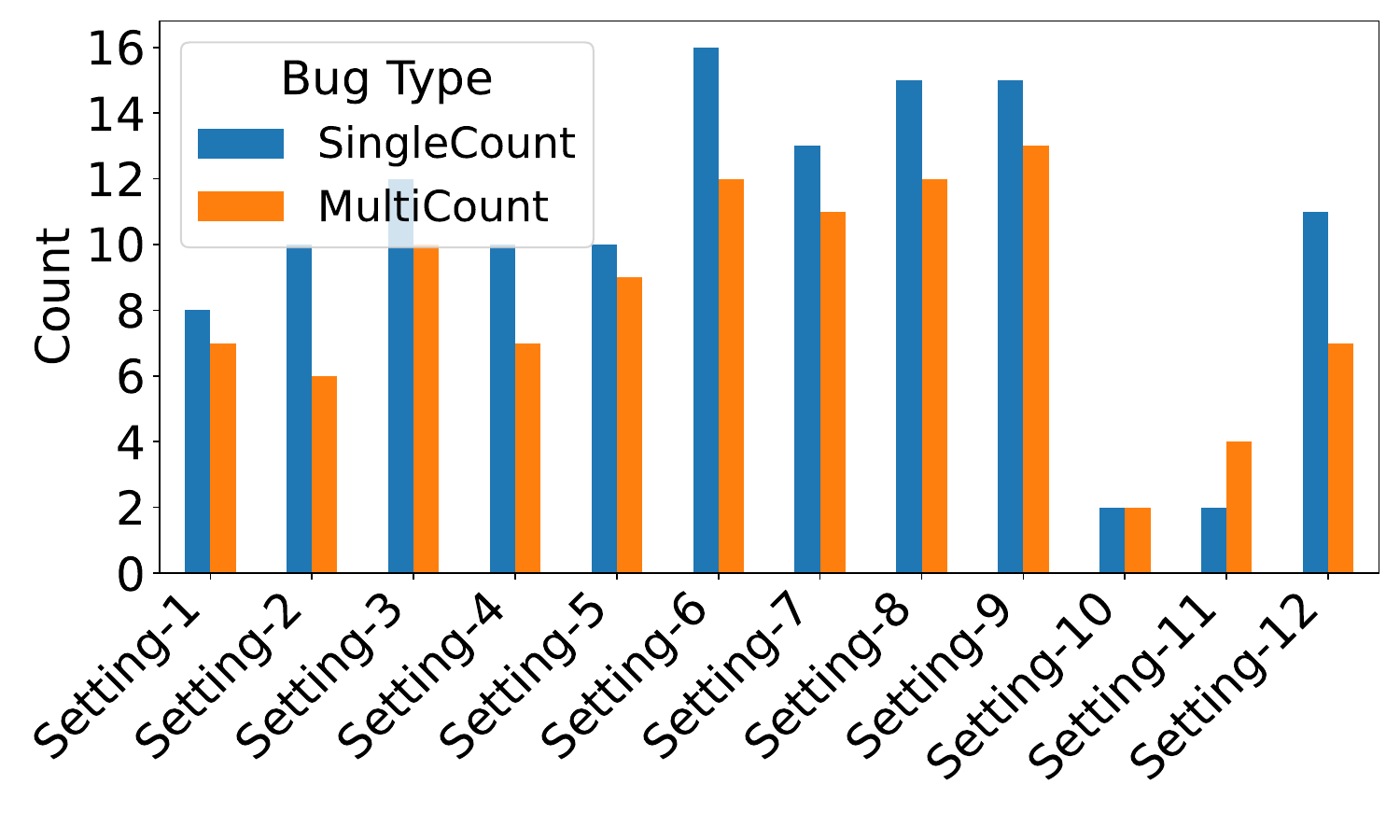}
        \label{fig:bugtype}
    \end{subfigure}
    \hfill
    \begin{subfigure}[b]{0.26\textwidth}
        \centering
        (c)        
        \includegraphics[width=\textwidth]{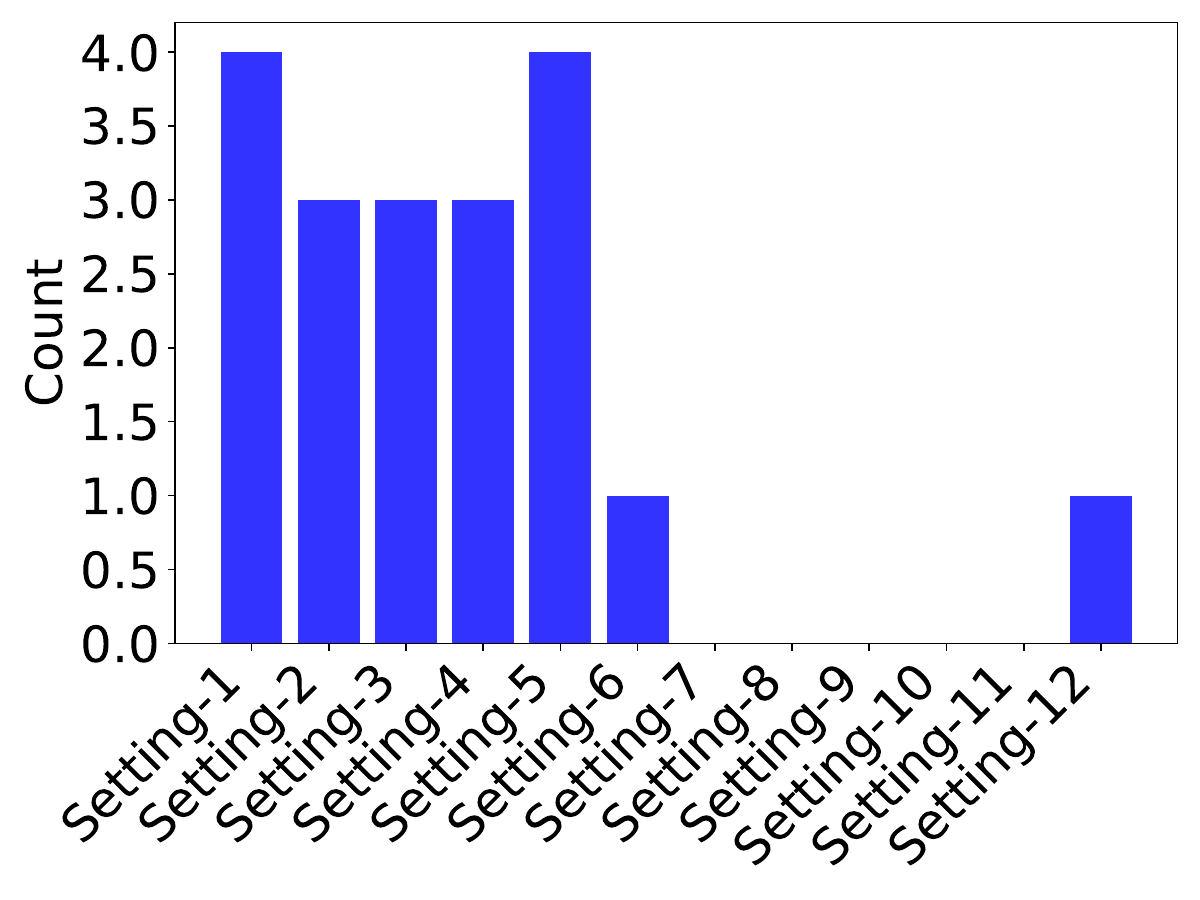}
        \label{fig:repeated1stIter}
    \end{subfigure}
    \vspace{-0.5cm}
        \caption{(a) Iteration count distribution for repairing specifications across various settings; (b) Repair distribution categorized by bug type: single-line or multi-line (higher values preferred); (c) The incident count for initial repair attempts mirroring the buggy specification (lower values preferred).}
\label{fig:RQ2}
\end{figure*}

\noindent
\textbf{Repair Iteration Budget.}
The box-whisker plot in 
Figure~\ref{fig:RQ2}(a)
illustrates the iterations needed by different settings to rectify faulty specifications. Across all settings, the median number of iterations predominantly centers around 1.0, except for Settings 3, 6\moh{, and 10-12}. Although \moh{Settings 3 and 6} exhibit superior repair capabilities, they may require a marginally higher number of iterations to achieve optimal results.
However, median values for Settings 1, 2, 4, and 5 remain at 1.0, suggesting that at least half of the issues are resolved within the first iteration. \moh{In contrast, the median values for Settings 7–9 are consistently 1.0, indicating that such settings require the fewest repair iterations, independent of the employed feedback mechanism.}

\noindent
\textbf{Repairing Based on Bug Type.}
\moh{Figure~\ref{fig:RQ2}(b) illustrates the repair effectiveness of various configurations in the repair pipeline with respect to bug types—specifically, single-line and multi-line defects. The dual-agent configurations in Setting-6 (auto-feedback, GPT-4-Turbo) and Setting-9 (auto-feedback, GPT-4o) demonstrated the highest overall repair rates, successfully fixing 16 single-line and 12 multi-line bugs, and 15 single-line and 13 multi-line bugs, respectively. In comparison, the best-performing configuration for GPT-4-32k (Setting-3, auto-feedback) showed relatively stronger performance on single-line bugs, repairing 12 Alloy models, versus 10 for multi-line bugs.} 

\noindent
\textbf{Following Instructions.}
\moh{Adherence to instructions is crucial, in tasks  such as automated repair, to ensure repair quality and control costs. 
Figure~\ref{fig:RQ2}(c)
reveals that both GPT-4 models sometimes repeat buggy models in their initial repair iterations, despite system instructions to avoid this. GPT-4o generally shows better compliance, under all settings, with zero number of repeated buggy specifications. Interestingly, GPT-3.5-Turbo exhibited similar behavior to GPT-4o.}

\begin{tcolorbox}[colback=gray!20, colframe=gray!50!black, rounded corners]
\moh{The evaluated LLMs exhibited varying behaviors across key factors, including instruction adherence, the number of repair iterations, and effectiveness in handling both single-line and multi-line bugs. Among all models and across all settings, GPT-4o consistently outperformed its counterparts across all measured dimensions. Notably, the dual-agent configuration generally resulted in an increased number of iterations for all models—except GPT-4o.}
\end{tcolorbox}

\subsection{Results for RQ3: Characteristics of Failed Repairs}

\begin{wrapfigure}{r}{0.45\textwidth}
    \centering
    \includegraphics[width=\linewidth]{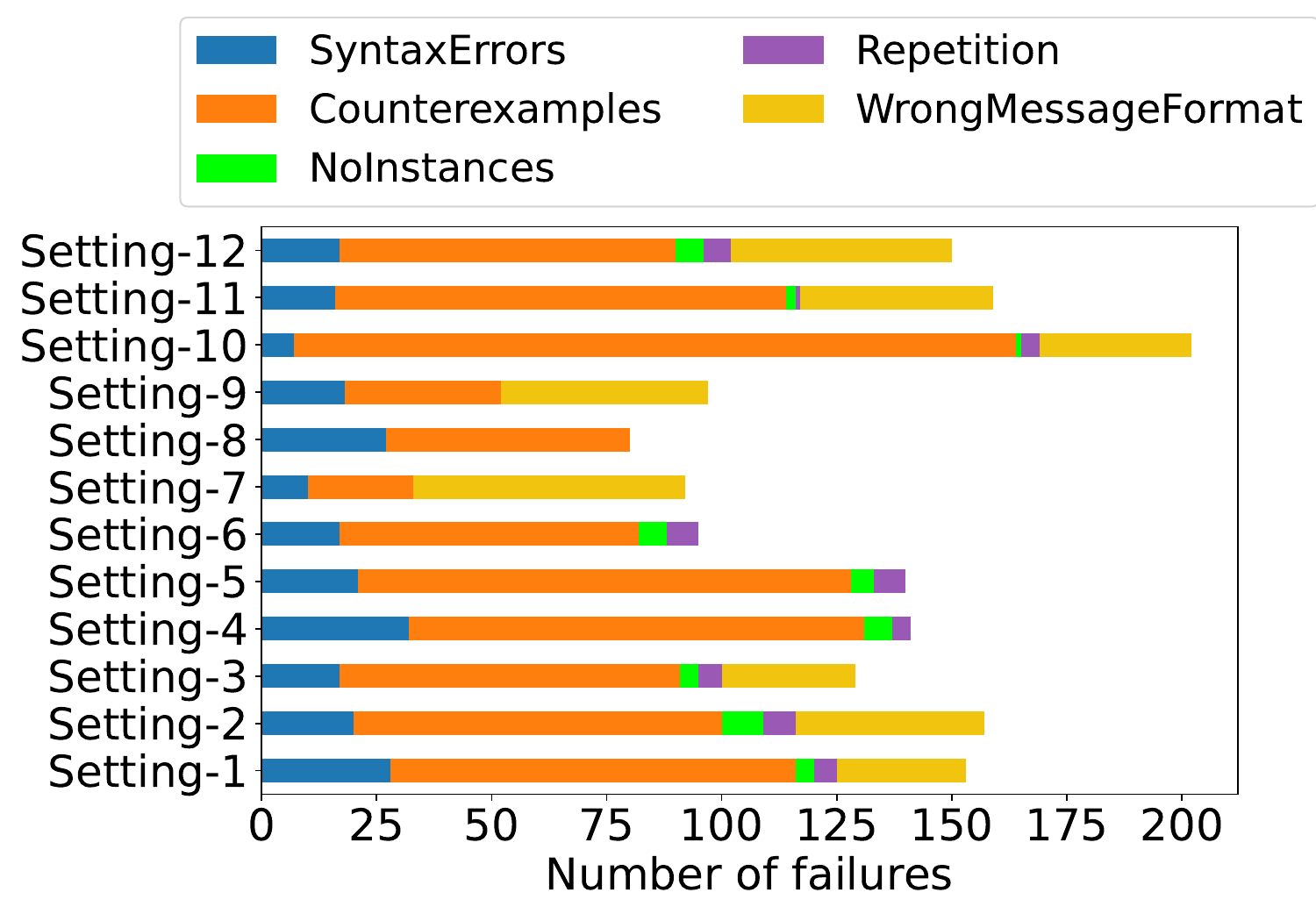}
    \vspace{-0.5cm}
    \caption{Error types observed in failed repair iterations across all settings}
    \vspace{-0.7cm}
    \label{fig:failures}
\end{wrapfigure}

\ma{This research question explores the underlying causes of unsuccessful repair iterations.} Figure~\ref{fig:failures} illustrates the distribution of failed iterations in different settings and provides a comprehensive breakdown of the reasons for the failure. Predominantly, failures arise when the proposed repairs fail to fulfill the assertions, resulting in the generation of counterexamples. Additionally, syntax errors in proposed repairs impede the compilation process by Alloy Analyzer.
Settings 1-3, employing GPT-4-32k, manifest instances of failures attributed to incorrect message formats, a phenomenon absent in Settings 4-6 utilizing GPT-4-Turbo. Such disparities suggest discrepancies in the compliance of LLM responses with the JSON format mandated by the alloy\_analyzer\_tool. Categories such as ``Repetition" and ``No instances" exhibit lower prevalence. ``Repetition" denotes instances where proposed repairs same as the supplied defective models, while ``No instances" signify scenarios where proposed models fail to generate instances, indicative of an incapacity to meet model constraints.


\begin{figure}[htbp]
    \centering
    \includegraphics[width=0.9\textwidth]{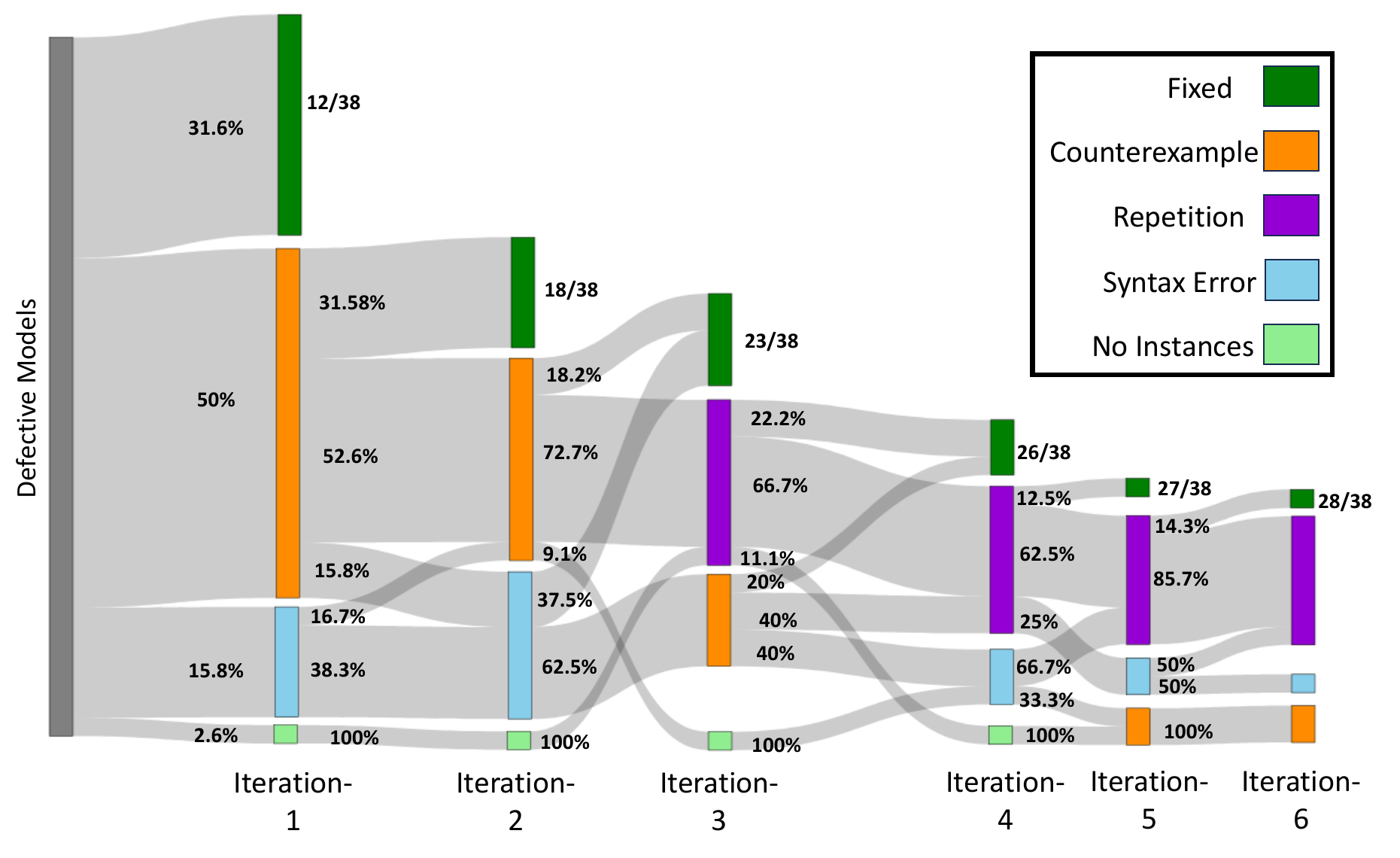}
    \caption{Repair progress in dual-agent Setting-6 (GPT-4-Turbo and Auto-Feedback) across iterations. Percentage numbers show transition volumes between statuses in subsequent iterations. Cumulative counts of fixed models follow each ``Fixed'' block.}
    \label{fig:iter_status}
\end{figure}

Figure~\ref{fig:iter_status} tracks the results across all iterations in Setting-6, revealing the evolving nature of failure categories during the iterative prompting process. Although instances of ``Counterexample'' failures are effectively addressed, challenges persist to rectify issues classified under ``Repetition''. 

\ma{We investigated the repair behavior of the ten defective models that remained unresolved after six iterations. Among these models, five were categorized as ``Counterexample,'' four exhibited syntax errors, and one was identified as a ``No instance''. Notably, the statuses of these ten specifications have remained largely unchanged since iterations 3 and 4. Particularly, the defective model \texttt{arr1} consistently exhibited the same status from the first iteration onwards. This persistent behavior corroborates our preliminary analysis regarding the number of iterations, wherein no further progress was observed beyond the sixth iteration.}

\noindent
\rashed{\textbf{Constraint-Related Issues.}}  
\rashed{Our analysis of LLM-generated results reveals an important nuance in constraint-related issues: underconstrained and overconstrained specifications often coexist with various error manifestations. These fundamental constraint problems can simultaneously present as syntax errors, type errors, absence of valid instances, and counter examples. Rather than being distinct categories, constraint issues and error types frequently intersect. Addressing these interrelated challenges requires sophisticated specification engineering techniques that can balance constraint expressiveness with feasibility, ultimately leading to more reliable and robust LLM-based specification repairs.}

\input{ConstraintsTable}

\rashed{Our analysis proceeds in two principal steps to identify specification constraint defects across LLM-generated models. First, we detect \textit{overconstrained} specifications by inspecting counterexamples and the presence of instances. If a counterexample is present and no instance is found—or if only a counterexample is detected, indicating overly restrictive constraints—we classify the model as \textit{overconstrained}. This suggests that the model’s constraints are too strict to allow any valid instance. Second, for specifications not flagged as \textit{overconstrained}, we identify \textit{underconstrained} issues by comparing each model against our ground‐truth benchmarks. If there is a discrepancy in the constraints—such as a missing constraint or an implementation that fails to reflect the intended semantics—we label the model as \textit{underconstrained}. This two‐stage procedure ensures that each model is categorized into exactly one defect class or deemed correct if neither condition applies.}

\rashed{Table \ref{tab:A4F-Constraint-Type} summarizes the distribution of overconstrained and underconstrained specifications among all unfixed issues across various LLM configurations on the ARepair and Alloy4Fun benchmarks, respectively. Notably, the results reveal that the constraint tendencies of each LLM can vary substantially depending on the benchmark context. On ARepair, GPT-4-32k and GPT-4-Turbo most frequently produce overconstrained specifications, with over 56\% of unfixed issues falling into this category across most settings. GPT-4o shows mixed behavior, with some settings demonstrating a greater tendency toward underconstraining, particularly settings 7 and 9 (57.1\% and 50.0\% respectively).}

\rashed{ In contrast, on the Alloy4Fun benchmark, GPT-4o consistently produces overconstrained specifications (60-62\% of unfixed issues), while GPT-3.5-Turbo maintains a similar ratio of overconstrained issues (approximately 60\%). This reversal in behavior between benchmarks is particularly notable, suggesting that its constraint tendencies are heavily influenced by the specific characteristics and complexities of the benchmark rather than being an intrinsic property of the model. These findings underscore a consistent trade-off between generating overly strict and overly permissive constraints, with distinct patterns emerging for each model and benchmark. Our systematic categorization method robustly captures these nuanced behaviors, providing insight into both model capabilities and the influence of benchmark design on LLM-driven specification repair.}

\begin{tcolorbox}[colback=gray!20, colframe=gray!50!black, rounded corners]
Analysis of failure characteristics during the repair process highlights common issues such as proposed repairs failing assertions and syntax errors, with observed transitions between failure categories over iterative prompting. Therefore, increasing the number of iterations does not necessarily result in a successful repair of the model. 
\end{tcolorbox}

\subsection{Results for RQ4: Repair Costs}
We consider two primary cost factors to assess the financial implications of utilizing the APR pipeline with different LLMs: (i) time taken for bug resolution and (ii) monetary expenses associated with token consumption, based on OpenAI's pricing as of March 2024.

\begin{figure}[ht]
    \centering
    \begin{minipage}[b]{0.49\columnwidth}
        \includegraphics[width=\linewidth]{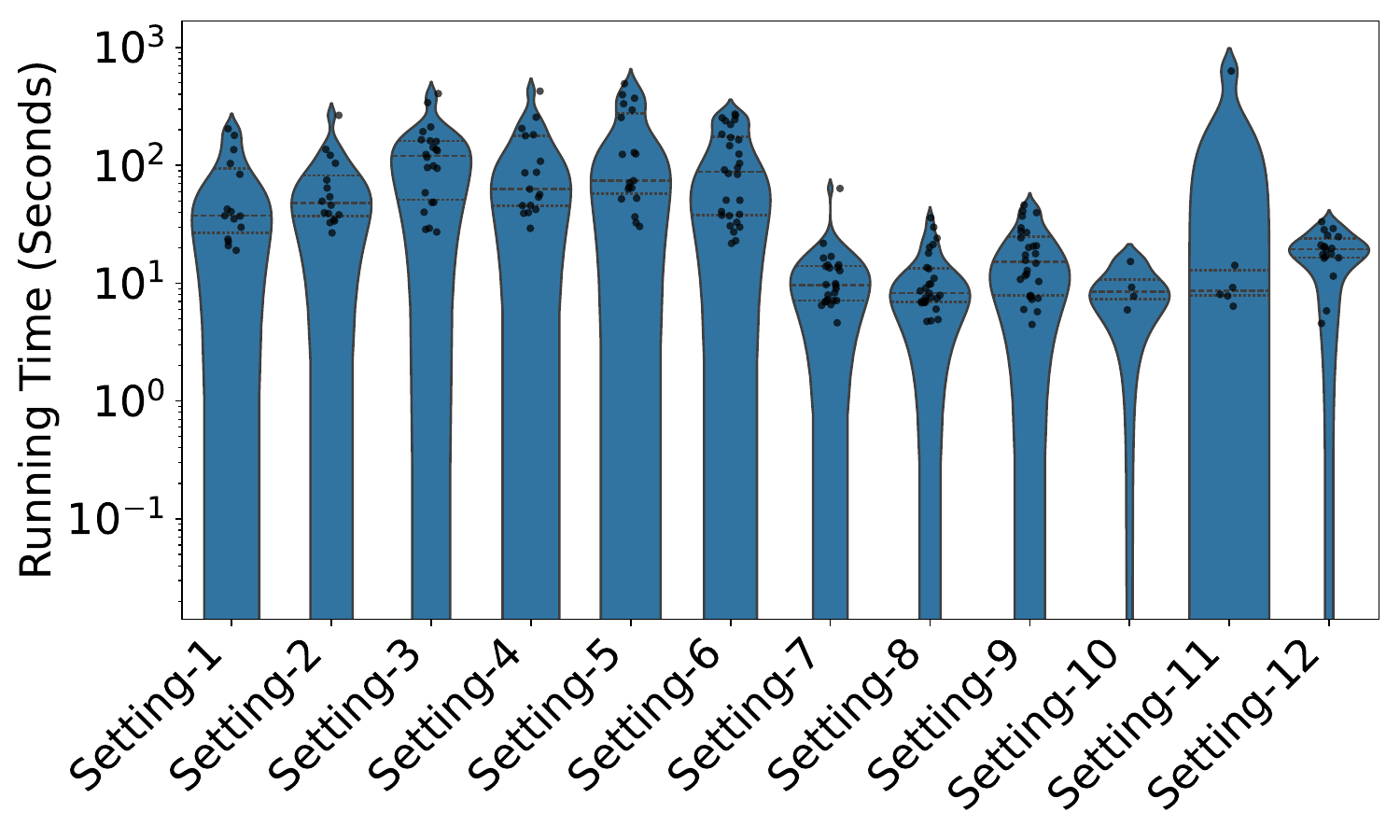}
    \end{minipage}
    \begin{minipage}[b]{0.49\columnwidth}
        \includegraphics[width=\linewidth]{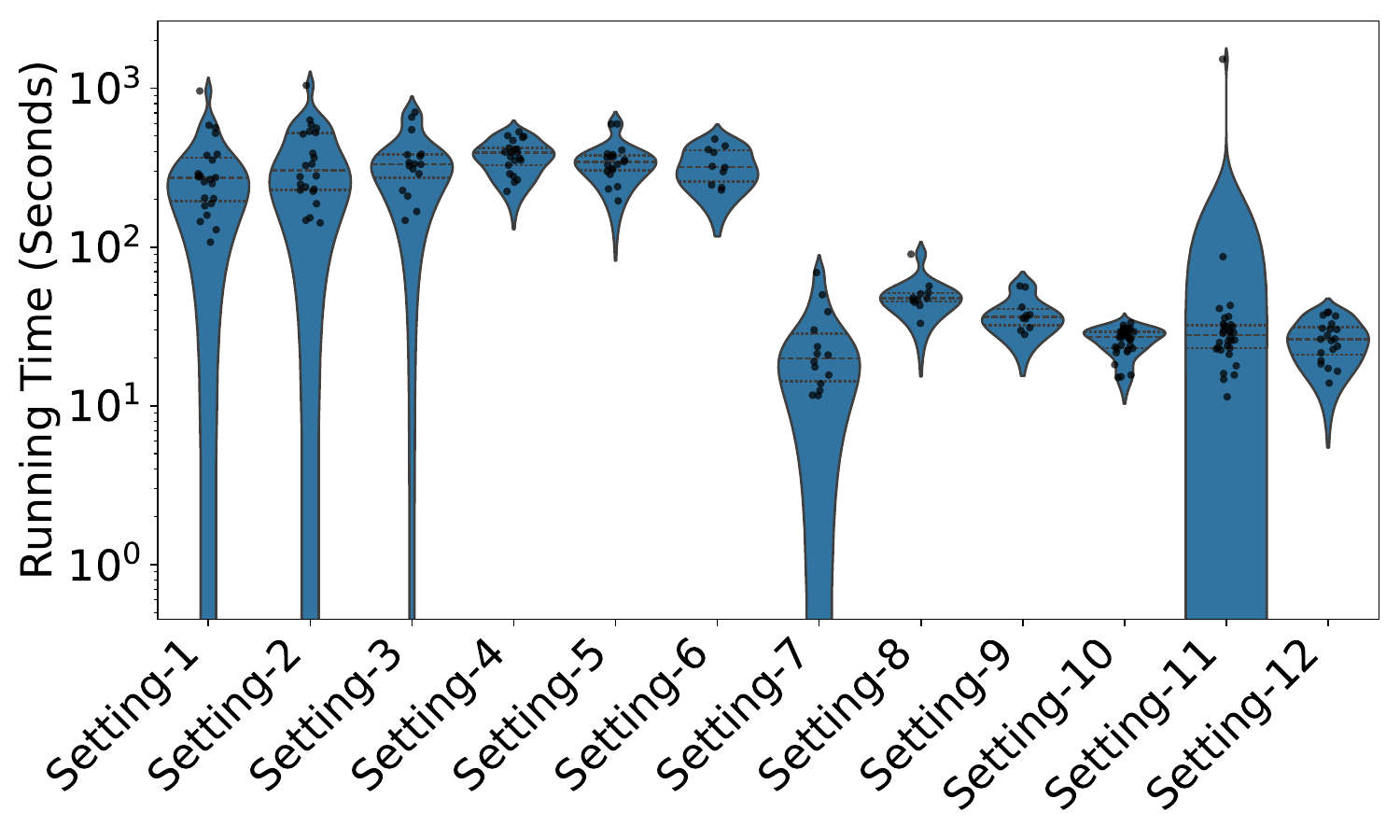}
    \end{minipage}
    \caption{Distribution of running time across all settings for fixed bugs (left) and unfixed bugs (right).}
    \label{fig:run_time_no_fixed}
\end{figure}

\noindent
\textbf{Time Analysis:}
We measure the time taken for bug resolution using the results presented in Table~\ref{tab:ARepair-Compare-SoTA}. Figure~\ref{fig:run_time_no_fixed} illustrates the runtime analysis, contrasting fixed and unfixed scenarios. Across different settings, the median duration required for bug repair varies from 37.3 seconds in Setting-1 to 120.3 seconds in Setting-3. Notably, the maximum time spent on repairs reached 493.28 seconds in Setting-5. Conversely, the median duration associated with unfixed bugs amounted to 273.8 seconds and 392.2 seconds in Settings-1 and 4, respectively. \moh{Interestingly, all settings supporting the \textit{dual-agent} setup (i.e., 3, 6, 9, and 12) consistently exhibit increased running time compared to the \textit{single-agent} configuration, across all LLMs and under both fixed and unfixed bug scenarios.}

\begin{figure}[ht]
    \centering
    \begin{minipage}[b]{0.49\columnwidth}
        \includegraphics[width=\linewidth]{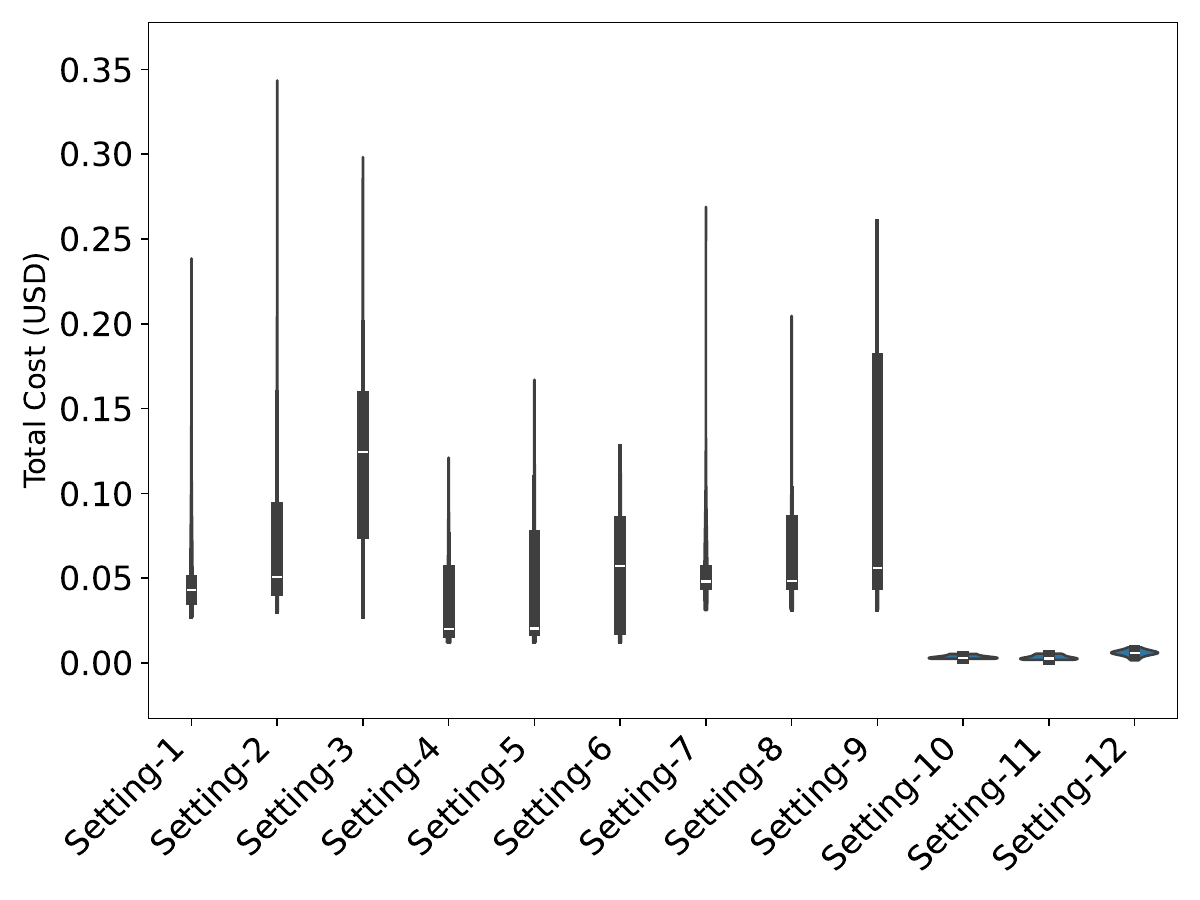}
    \end{minipage}
    \begin{minipage}[b]{0.49\columnwidth}
        \includegraphics[width=\linewidth]{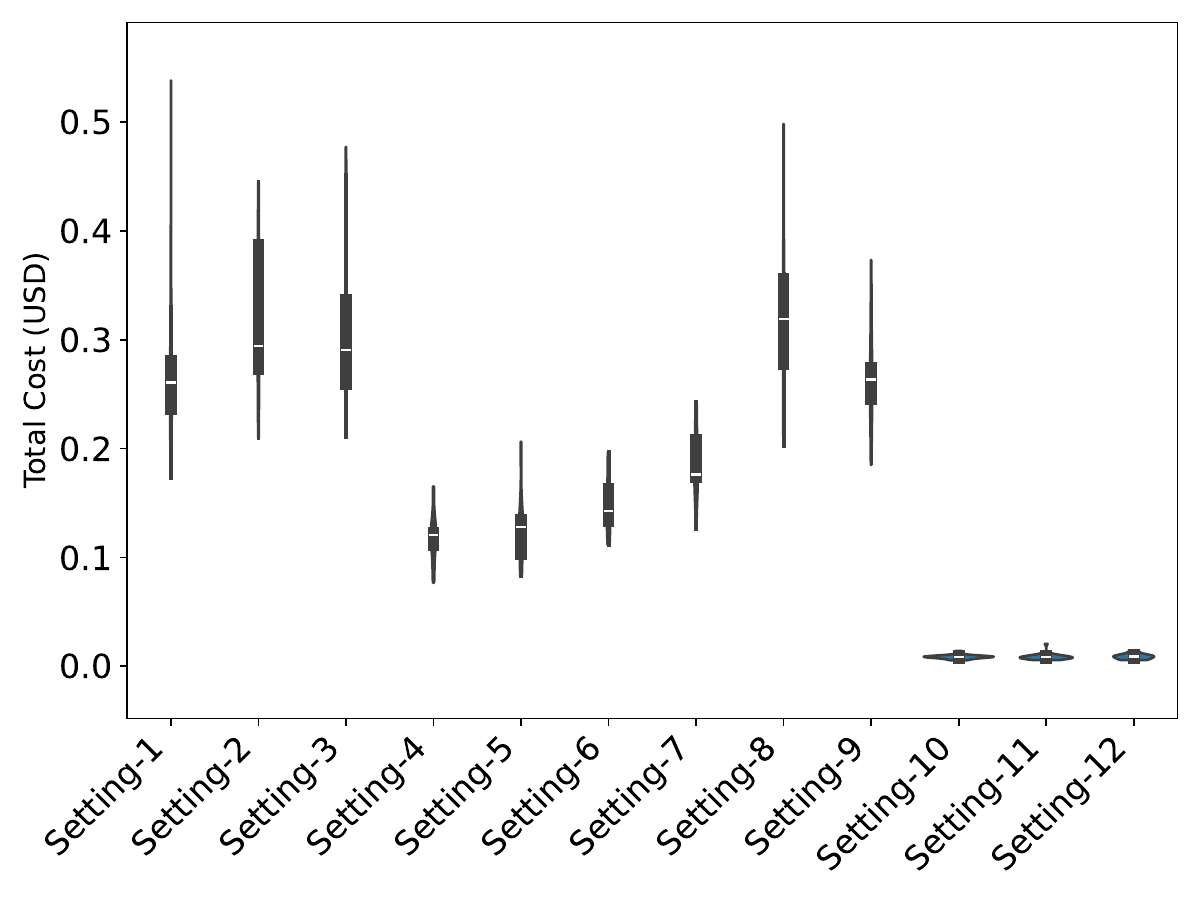}
    \end{minipage}
    \caption{Monetary distribution across all settings for fixed bugs (left) and unfixed bugs (right).}
    \label{fig:cost}
\end{figure}

\noindent
\textbf{Monetary Costs:}
Figure~\ref{fig:cost} presents the monetary costs incurred under each setting for both fixed and unfixed models. Settings utilizing GPT-3.5-Turbo show the lowest costs for both fixed and unfixed models, attributed to the cheaper cost per token for GPT-3.5-Turbo compared to GPT-4 family models (see Table~\ref{tab:llms}). Additionally, settings employing the \textit{dual-agent} (3, 6, 9, and 12) exhibit marginally higher costs than their counterparts utilizing the same LLM under the \textit{single-agent} setup.

\begin{tcolorbox}[colback=gray!20, colframe=gray!50!black, rounded corners]
\moh{The analysis of repair costs in RQ4 reveals notable variations in both execution time and monetary expenditure across different configurations. Among the models evaluated, GPT-3.5-Turbo consistently incurs the lowest cost for bug resolution. Within the GPT-4 family, GPT-4o demonstrates superior efficiency, achieving the lowest repair cost and runtime. Additionally, settings employing the \textit{dual-agent} paradigm tend to exhibit slightly higher expenses compared to their single-agent counterparts.}
\end{tcolorbox}

%% file: ARepair-Compare-SoTA.tex
\begin{table*}[ht]
  \caption{
  \hamid{Comparison of state-of-the-art Alloy repair techniques—ARepair~\cite{arepair_icse}, ICEBAR~\cite{ICEBAR}, BeAFix~\cite{BeAFix}, ATR~\cite{ATR}, and Hasan et al.~\cite{hasan2023automated}—on the ARepair benchmark, using the settings detailed in Table~\ref{table:combinations}.}
  }
  \scalebox{0.47}{
  \begin{tabular}{lc@{}c@{}cccccccccccccccccc}
    \toprule
    \textbf{Model} & \textbf{Total} & \multicolumn{1}{c}{\textbf{Defects}} & \textbf{} & \textbf{ARepair} & \textbf{ICEBAR} & \textbf{BeAFix} & \textbf{ATR} & \textbf{Hasan et al.} & \multicolumn{12}{c}{\textbf{\#repairs Under Setting}} \\
    & \textbf{\#specs} & \multicolumn{1}{c}{\textbf{Count}} &  & \textbf{\#repairs} & \textbf{\#repairs} & \textbf{\#repairs} & \textbf{\#repairs} & \textbf{\#repairs} & \multicolumn{3}{c}{\textbf{GPT-4-32k}} & \multicolumn{3}{c}{\textbf{GPT-4-Turbo}} & \multicolumn{3}{c}{\textbf{GPT-4o}} & \multicolumn{3}{c}{\textbf{GPT-3.5-Turbo}} \\
    \cmidrule(lr){10-12} \cmidrule(lr){13-15} \cmidrule(lr){16-18} \cmidrule(lr){19-21}
    & & & & & & & & & \textbf{1} & \textbf{2} & \textbf{3} & \textbf{4} & \textbf{5} & \textbf{6} & \textbf{7} & \textbf{8} & \textbf{9} & \textbf{10} & \textbf{11} & \textbf{12} \\
    
    \midrule

    addr        & 1  & 1 & & 1  & 1  & 1  & 1  & 0  & 0 & 0 & 1 & 0 & 1 & 1 & 0 & 0 & 0 & 0 & 0 & 0 \\ 
    arr         & 2  & 2 & & 2  & 2  & 1  & 0  & 0  & 0 & 0 & 0 & 1 & 0 & 1 & 1 & 2 & 1 & 0 & 1 & 1 \\ 
    balancedBST & 3  & 8 & & 1  & 2  & 1  & 1  & 0  & 1 & 1 & 2 & 1 & 1 & 1 & 1 & 1 & 1 & 0 & 0 & 1 \\ 
    bempl       & 1  & 1 & & 0  & 1  & 0  & 1  & 1  & 0 & 0 & 0 & 1 & 1 & 1 & 0 & 0 & 0 & 0 & 0 & 0 \\ 
    cd          & 2  & 3 & & 0  & 2  & 2  & 2  & 0  & 1 & 2 & 2 & 2 & 1 & 2 & 2 & 2 & 2 & 1 & 0 & 0 \\ 
    ctree       & 1  & 1 & & 1  & 0  & 0  & 0  & 0  & 1 & 1 & 1 & 1 & 1 & 1 & 1 & 1 & 1 & 0 & 0 & 0 \\ 
    dll         & 4  & 8 & & 0  & 3  & 3  & 2  & 0  & 4 & 4 & 4 & 2 & 2 & 4 & 4 & 4 & 4 & 1 & 3 & 3 \\ 
    farmer      & 1  & 2 & & 0  & 0  & 0  & 0  & 0  & 1 & 1 & 1 & 1 & 1 & 1 & 1 & 1 & 1 & 1 & 1 & 1 \\ 
    fsm         & 2  & 2 & & 2  & 2  & 1  & 2  & 0  & 1 & 1 & 2 & 0 & 0 & 1 & 1 & 1 & 2 & 0 & 0 & 0 \\ 
    grade       & 1  & 1 & & 0  & 1  & 0  & 1  & 0  & 0 & 0 & 0 & 0 & 0 & 1 & 0 & 0 & 1 & 0 & 0 & 0 \\ 
    other       & 1  & 1 & & 0  & 0  & 1  & 1  & 1  & 0 & 0 & 0 & 0 & 0 & 1 & 1 & 1 & 1 & 1 & 0 & 1 \\ 
    student     & 19 & 34 & & 2  & 7  & 13 & 10 & 2  & 6 & 6 & 9 & 8 & 11 & 13 & 12 & 14 & 14 & 0 & 1 & 11 \\ 
    
    \midrule
    \textbf{Summary} & \textbf{38} & \textbf{64} & & \textbf{9} & \textbf{21} & \textbf{24} & \textbf{22} & \textbf{4} & \textbf{15} & \textbf{16} & \textbf{22} & \textbf{17} & \textbf{19} & \textbf{28} & \textbf{24} & \textbf{27} & \textbf{28} & \textbf{4} & \textbf{6} & \textbf{18}\\
    \textbf{\%repair} & & & & \textbf{23.7} & \textbf{55.3} & \textbf{63.2} & \textbf{57.8} & \textbf{10.5} & \textbf{39.5} & \textbf{42.1} & \textbf{57.9} & \textbf{44.7} & \textbf{50} & \textbf{73.4} & \textbf{63.2} & \textbf{71.1} & \textbf{73.4} & \textbf{10.5} & \textbf{15.8} & \textbf{47.4}\\
    
    \bottomrule
  \end{tabular}
  }
  \label{tab:ARepair-Compare-SoTA}
\end{table*}

%% file: A4F-Sample-Compare-SoTA.tex
\begin{table*}[ht]
  \caption{
  \hamid{Comparison of state-of-the-art Alloy repair techniques—ARepair~\cite{arepair_icse}, ICEBAR~\cite{ICEBAR}, BeAFix~\cite{BeAFix}, ATR~\cite{ATR}, and Hasan et al.~\cite{hasan2023automated}—on the Alloy4Fun benchmark, using the settings described in Table~\ref{table:combinations}. Bracketed numbers indicate the models selected for the GPT-4o repair experiment to manage computational costs.}
  }
  \scalebox{0.55}{
  \begin{tabular}{lcccccccccccccc}
    \toprule
    \textbf{Model} & \textbf{Total} & \textbf{Defects} & \textbf{ARepair} & \textbf{ICEBAR} & \textbf{BeAFix} & \textbf{ATR} & \textbf{Hasan et al.} & \multicolumn{6}{c}{\textbf{\#repairs Under Setting}} \\
    & \textbf{\#specs} & \textbf{Count} & \textbf{\#specs} & \textbf{\#repairs} & \textbf{\#repairs} & \textbf{\#repairs} & \textbf{\#repairs} & \multicolumn{3}{c}{\textbf{GPT-4o}} & \multicolumn{3}{c}{\textbf{GPT-3.5-Turbo}} \\
    \cmidrule(lr){9-11} \cmidrule(lr){12-14}
    & & & & & & & & \textbf{7} & \textbf{8} & \textbf{9} & \textbf{10} & \textbf{11} & \textbf{12} \\
    \midrule

    classroom   & 999(60)  & 999(60)  & 88(3)  & 424(18) & 387(16) & 688(29) & 88(4)  & (0)  & (0)  & (6)  & 0   & 1   & 21 \\ 
    cv          & 138(60)  & 138(60)  & 2(0)   & 86(26)  & 82(20)  & 38(0)   & 4(3)   & (56) & (58) & (59) & 38  & 32  & 70 \\ 
    graphs      & 283(60)  & 283(60)  & 19(8)  & 237(39) & 232(38) & 260(44) & 20(39) & (22) & (40) & (22) & 23  & 51  & 35 \\ 
    lts         & 249(60)  & 249(60)  & 1(0)   & 73(20)  & 41(9)   & 70(15)  & 21(32) & (38) & (48) & (46) & 68  & 115 & 106 \\ 
    production  & 61(60)   & 61(60)   & 27(23) & 36(33)  & 56(53)  & 43(46)  & 12(12) & (59) & (60) & (58) & 25  & 32  & 34 \\ 
    trash       & 206(57)  & 206(57)  & 48(9)  & 195(46) & 183(49) & 187(34) & 2(2)   & (1)  & (0)  & (10) & 0   & 0   & 12 \\ 

    \midrule
    \textbf{Summary} 
    & \textbf{1936(357)} & \textbf{1936(357)} & \textbf{185(43)} & \textbf{1051(182)} & \textbf{981(185)} & \textbf{1286(168)} & \textbf{147(92)} 
    & \textbf{(176)} & \textbf{(206)} & \textbf{(201)} & \textbf{154} & \textbf{231} & \textbf{278} \\ 

    \textbf{\%repair} 
    &  &  & \textbf{9.5(12)} & \textbf{54.2(50.9)} & \textbf{50.6(51.8)} & \textbf{66.4(47)} & \textbf{7.6(25.7)} 
    & \textbf{(49.2)} & \textbf{(55.7)} & \textbf{(56.3)} & \textbf{7.6} & \textbf{11.9} & \textbf{14.3} \\ 

    \bottomrule
  \end{tabular}
  }
  \label{tab:A4F-Sample-Compare-SoTA}
\end{table*}

%% file: Size_of_variables_LLM.tex
\begin{table*}[ht]
  \centering
  \caption{
\hamid{
  Average number of variables and clauses in propositional formulas generated by the Alloy Analyzer from LLM-produced repairs on the ARepair benchmark. Results are reported as overall averages across all evaluated settings for each LLM model.}
  }
  \label{tab:variables-clauses-arepair}
  \scalebox{0.7}{
  \begin{tabular}{l|c|cccc}
    \toprule
    \multirow{2}{*}{\textbf{Model}} & \multirow{2}{*}{\makecell{\textbf{Total} \\ \textbf{\# spec}}} & \multicolumn{4}{c}{\textbf{LLM Models}} \\
    \cmidrule(lr){3-6}
    & & \textbf{GPT-4-32k} & \textbf{GPT-4-Turbo} & \textbf{GPT-4o} & \textbf{GPT-3.5-Turbo} \\
    \midrule
    addr & 1 & variables: 333 & variables: 567 & variables: 66 & variables: 187 \\
         &   & clauses: 368 & clauses: 627 & clauses: 74 & clauses: 215 \\
    \cline{1-6}\noalign{\vskip 2pt}
    arr & 2 & variable: 1,973 & variable: 7,578 & variables: 6,364 & variables: 1,482 \\
        &   & clauses: 4,969 & clauses: 16,284 & clauses: 6,096 & clauses: 3,314 \\
    \cline{1-6}\noalign{\vskip 2pt}
    balancedBSt & 3 & variables: 2,001 & variables: 15,312 & variables: 4,155 & variables: 1,797 \\
                &   & clauses: 3,720 & clauses: 52,602 & clauses: 8,616 & clauses: 3,306 \\
    \cline{1-6}\noalign{\vskip 2pt}
    bempl & 1 & variables: 2,249 & variables: 700 & variables: 744 & variables: 694 \\
          &   & clauses: 3,489 & clauses: 1,012 & clauses: 1,071 & clauses: 1,003 \\
    \cline{1-6}\noalign{\vskip 2pt}
    cd & 2 & variables: 294 & variables: 280 & variables: 288 & variables: 576 \\
       &   & clauses: 196 & clauses: 164 & clauses: 326 & clauses: 1,188 \\
    \cline{1-6}\noalign{\vskip 2pt}
    ctree & 1 & variables: 2,976 & variables: 1,130 & variables: 1,721 & variables: 1,679 \\
          &   & clauses: 2,857 & clauses: 1,035 & clauses: 1,591  & clauses: 1,524 \\
    \cline{1-6}\noalign{\vskip 2pt}
    dll & 4 & variables: 21,572 & variables: 20,012 & variables: 21,524 & variables: 6,324 \\
        &   & clauses: 66,088 & clauses: 65,248 & clauses: 50,128 & clauses: 4,108 \\
    \cline{1-6}\noalign{\vskip 2pt}
    farmer & 1 & variables: 1,468 & variables: 1,213 & variables: 2,252 & variables: 1,199 \\
           &   & clauses: 2,808 & clauses: 2,177 & clauses: 4,096 & clauses: 2,429 \\
    \cline{1-6}\noalign{\vskip 2pt}
    fsm & 2 & variables: 894 & variables: 2,224 & variables: 540 & variables: 1,080 \\
        &   & clauses: 1,816 & clauses: 4,556 & clauses: 1,108 & clauses: 2,278 \\
    \cline{1-6}\noalign{\vskip 2pt}
    grade & 1 & variables: 4,763 & variables: 2,830 & variables: 1,024 & variables: 491 \\
          &   & clauses: 7,930 & clauses: 4,578 & clauses: 1,585 & clauses: 604 \\
    \cline{1-6}\noalign{\vskip 2pt}
    other & 1 & variables: 477 & variables: 264 & variables: 113 & variables: 268 \\
          &   & clauses: 78 & clauses: 352 & clauses: 128 & clauses: 358 \\
    \cline{1-6}\noalign{\vskip 2pt}
    Student & 19 & variables: 100,548 & variables: 63,992 & variables: 31,236 & variables: 63,878 \\
            &    & clauses: 321,727 & clauses: 208,696 & clauses: 20,520 & clauses: 208,354 \\
    \cline{1-6}\noalign{\vskip 2pt}
    \textbf{Total} & \textbf{38} & \textbf{variables: 139,548} & \textbf{variables: 116,102} & \textbf{variables: 70,027} & \textbf{variables: 79,655} \\
                   &            & \textbf{clauses: 416,046} & \textbf{clauses: 357,331} & \textbf{clauses: 95,339} & \textbf{clauses: 228,681} \\
    \bottomrule
  \end{tabular}
  }
\end{table*}

\begin{table*}[]
  \centering
 \caption{
\hamid{
 Average number of variables and clauses in propositional formulas generated by the Alloy Analyzer from LLM-produced repairs on the A4F benchmark. Results are reported as overall averages across all evaluated settings for each LLM model. For a fair comparison, results are based on the same set of 357 specifications for both GPT-3.5-Turbo and GPT-4o.}
 }
  \scalebox{0.94}{
\begin{tabular}{l|c|cccc}
\toprule
  \multicolumn{2}{c|}{} & \multicolumn{2}{c|}{\textbf{GPT-4o}} & \multicolumn{2}{c}{\textbf{GPT-3.5-Turbo}} \\
  \cmidrule(lr){1-2} \cmidrule(lr){3-4} \cmidrule(lr){5-6}
  \textbf{Model} & \textbf{\# specs used} & \textbf{Variables} & & \textbf{\# specs used} & \textbf{Variables} \\
  \midrule
  classroom   & 60  & variables: 35,760 & & 60  & variables: 3,580  \\
              &     & clauses: 47,340   & &      & clauses: 4,691  \\
  \cmidrule{1-6}
  cv          & 60  & variables: 105,660 & & 60  & variables: 31,320 \\
              &     & clauses: 8,640    & &      & clauses: 34,620 \\
  \cmidrule{1-6}
  graphs      & 60  & variables: 17,040 & & 60  & variables: 7,902 \\
              &     & clauses: 37,200   & &      & clauses: 6,180 \\
  \cmidrule{1-6}
  lts         & 60  & variables: 26,100 & & 60  & variables: 88,560 \\
              &     & clauses: 26,820   & &      & clauses: 199,260 \\
  \cmidrule{1-6}
  production  & 60  & variables: 16,560 & & 60  & variables: 61,380 \\
              &     & clauses: 19,560   & &      & clauses: 124,560 \\
  \cmidrule{1-6}
  trash       & 57  & variables: 7,068  & & 57  & variables: 23,598 \\
              &     & clauses: 8,892    & &      & clauses: 43,605 \\
  \cmidrule{1-6}
  \textbf{Total} & \textbf{357} & \textbf{variables: 208,188} & & \textbf{357} & \textbf{variables: 216,340} \\
                &              & \textbf{clauses: 148,452}   & &             & \textbf{clauses: 412,916} \\
  \bottomrule
\end{tabular}
  }
  \label{tab:variable-clauses-a4f}
\end{table*}

%% file: venn_diagrams_arepair.tex
\begin{figure*}[ht!]
    \centering
    \begin{minipage}{0.32\textwidth}
        \includegraphics[width=\linewidth]{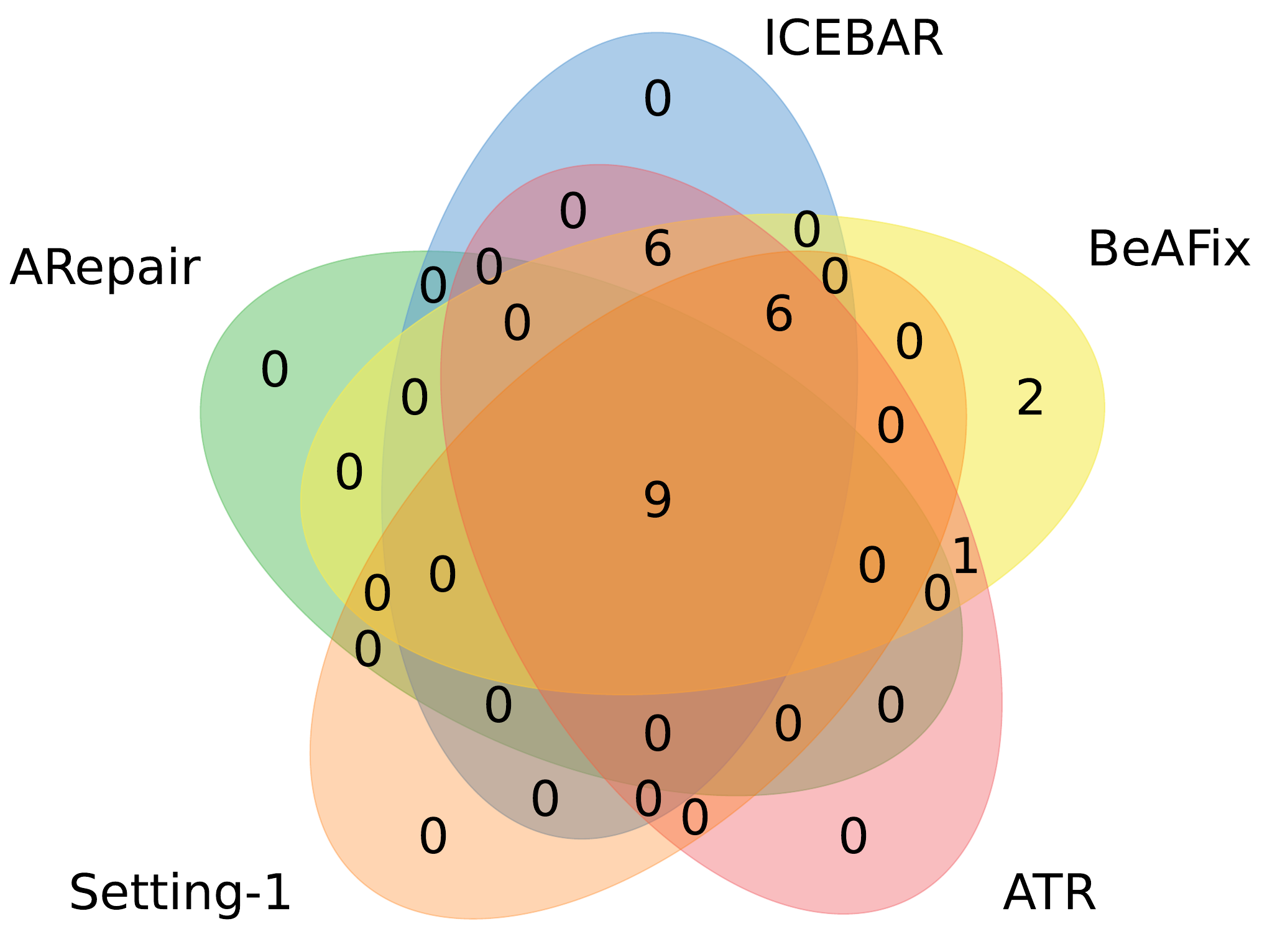}
    \end{minipage}
    \hfill
    \begin{minipage}{0.32\textwidth}
        \includegraphics[width=\linewidth]{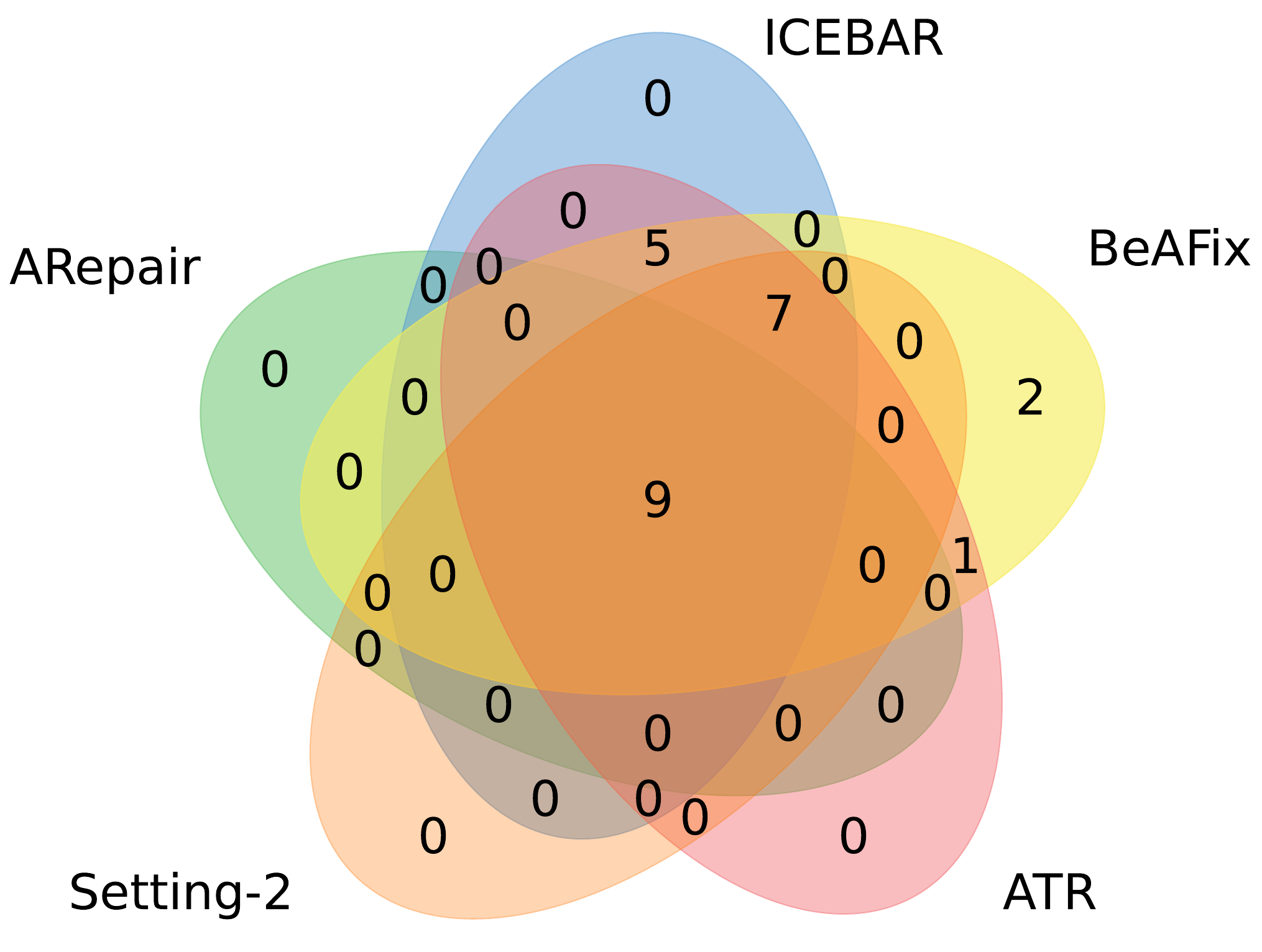}
    \end{minipage}
    \hfill
    \begin{minipage}{0.32\textwidth}
        \includegraphics[width=\linewidth]{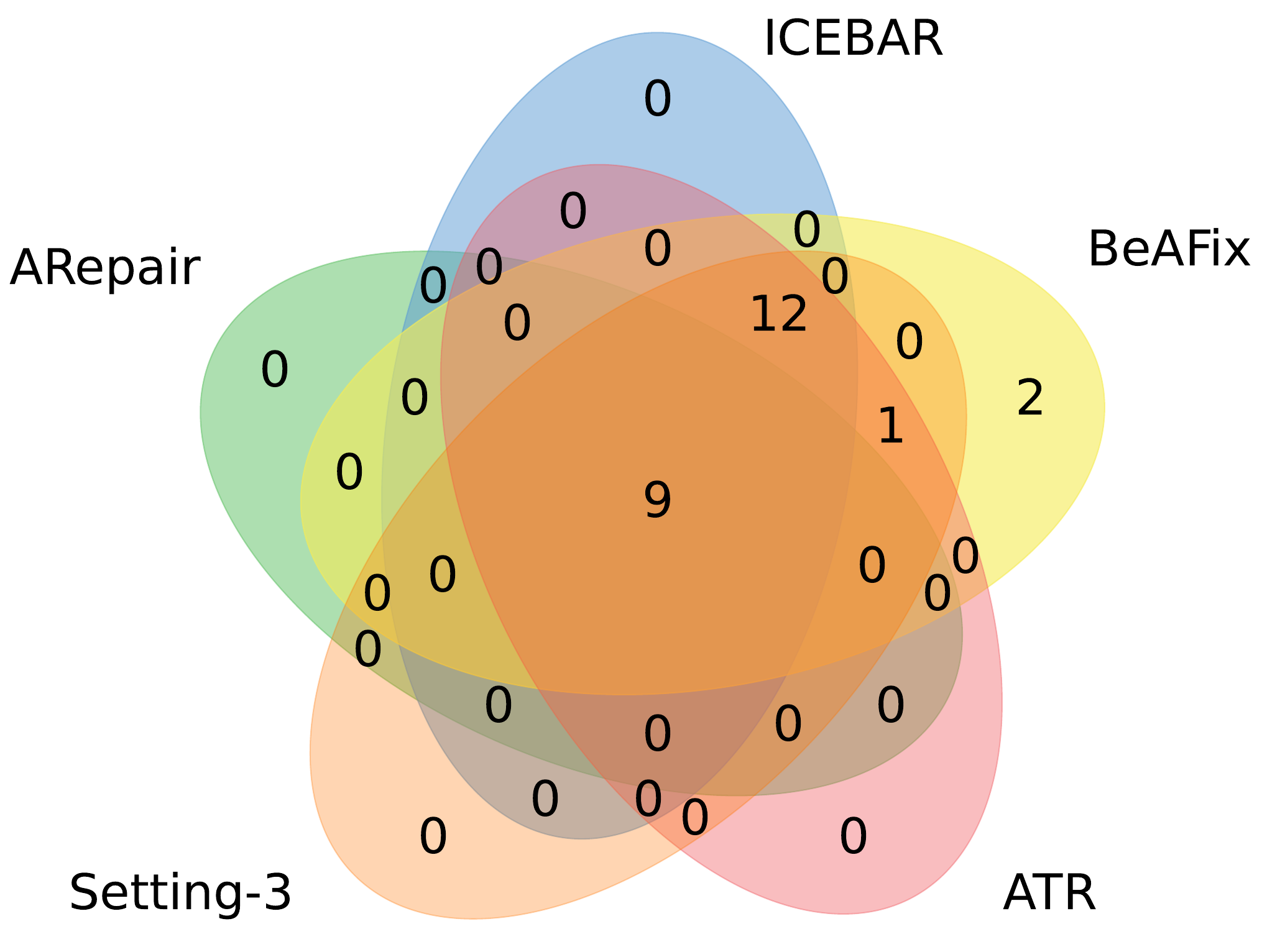}
    \end{minipage}
    
    \vspace{0.5cm} 
    
    \begin{minipage}{0.32\textwidth}
        \includegraphics[width=\linewidth]{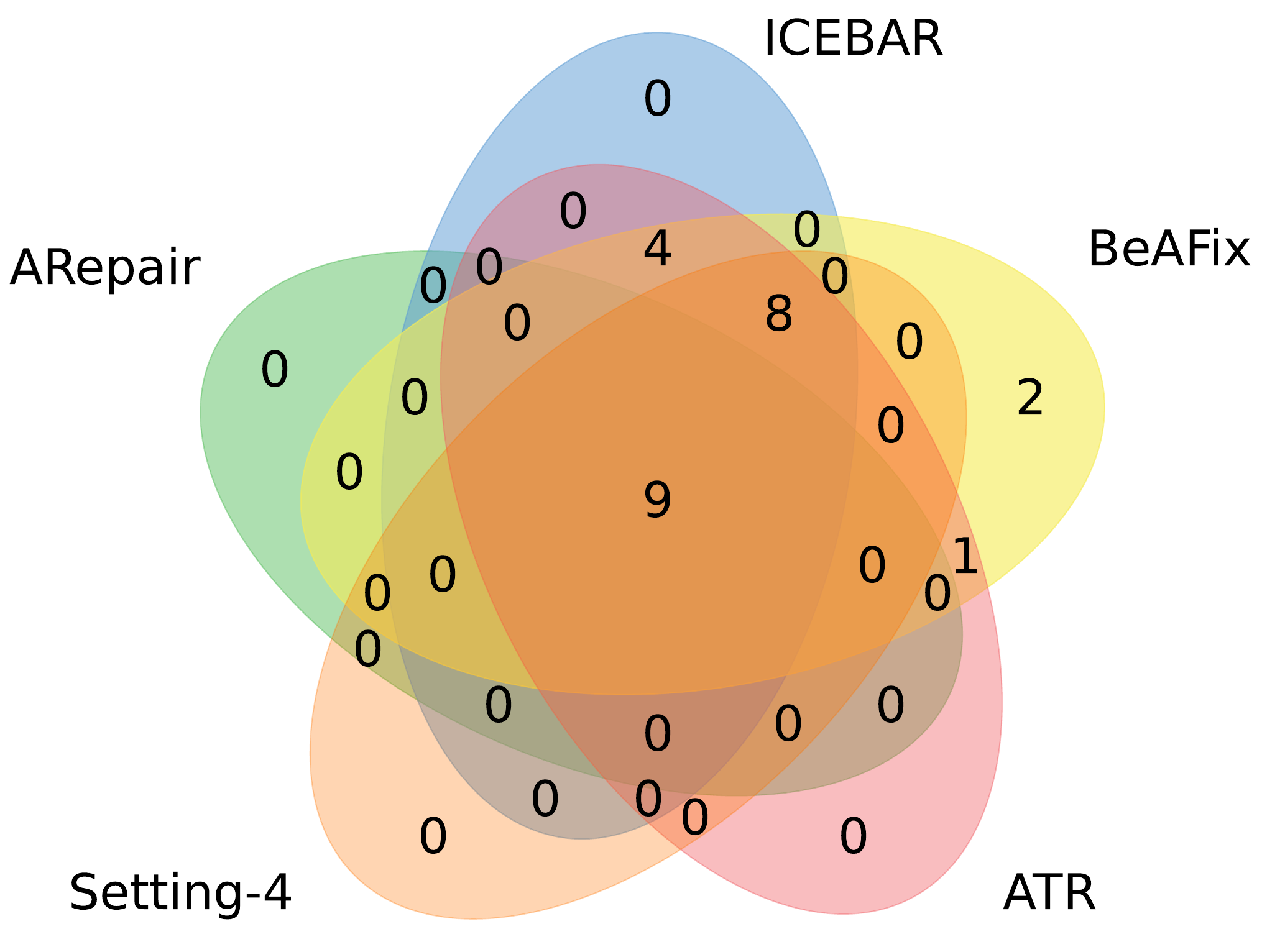}
    \end{minipage}
    \hfill
    \begin{minipage}{0.32\textwidth}
        \includegraphics[width=\linewidth]{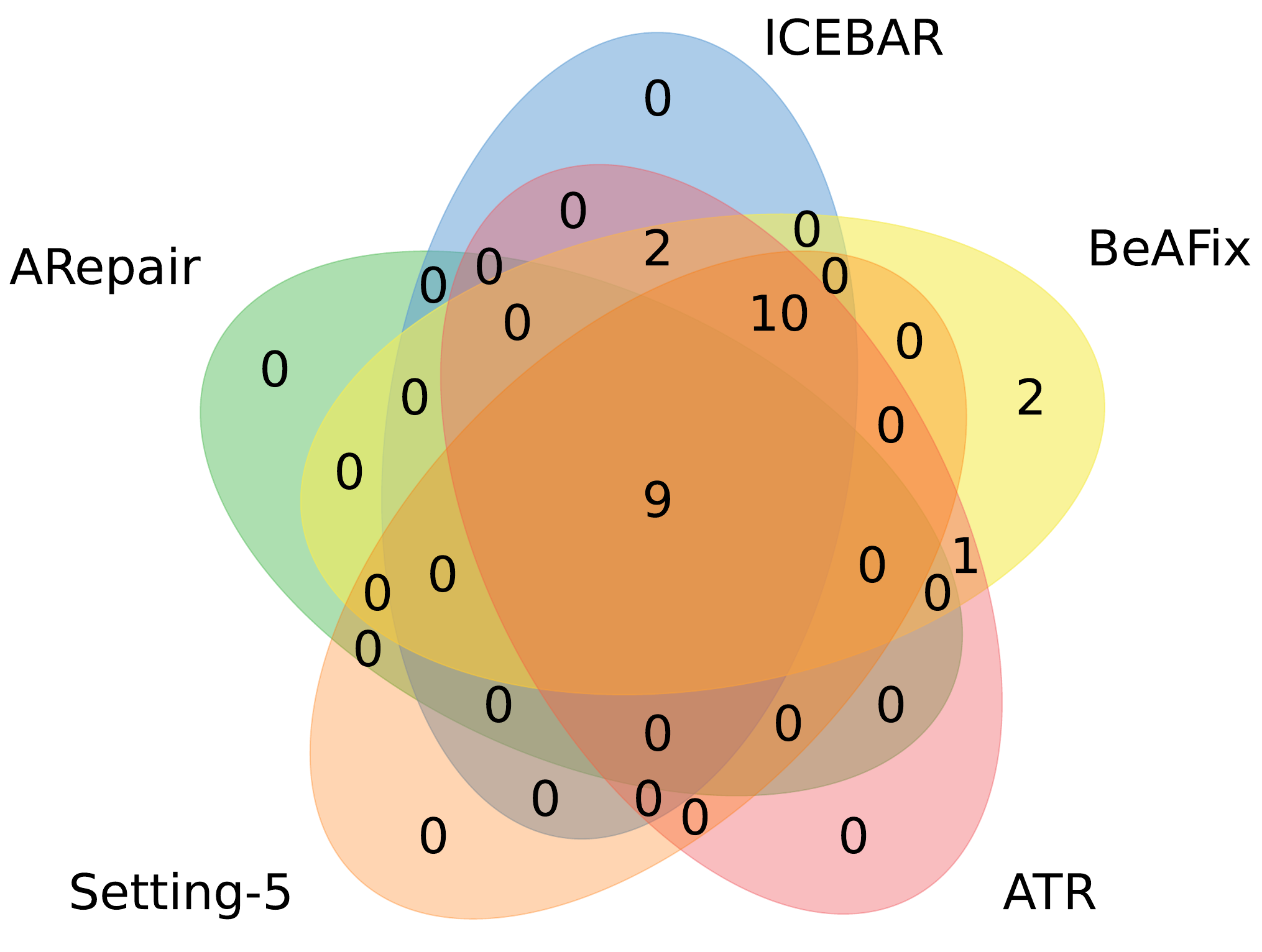}
    \end{minipage}
    \hfill
    \begin{minipage}{0.32\textwidth}
        \includegraphics[width=\linewidth]{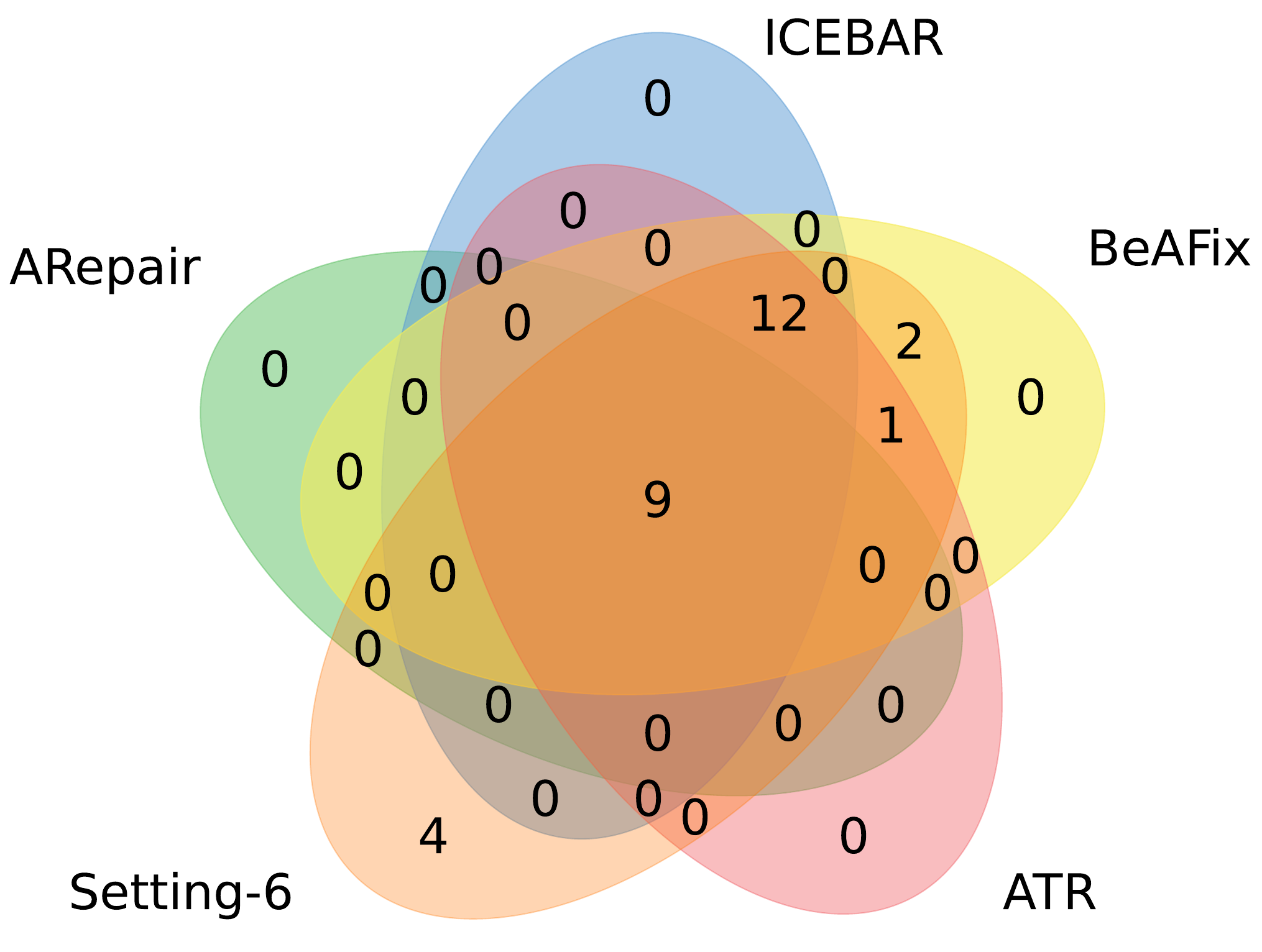}
    \end{minipage}

    \vspace{0.5cm} 
    
    \begin{minipage}{0.32\textwidth}
        \includegraphics[width=\linewidth]{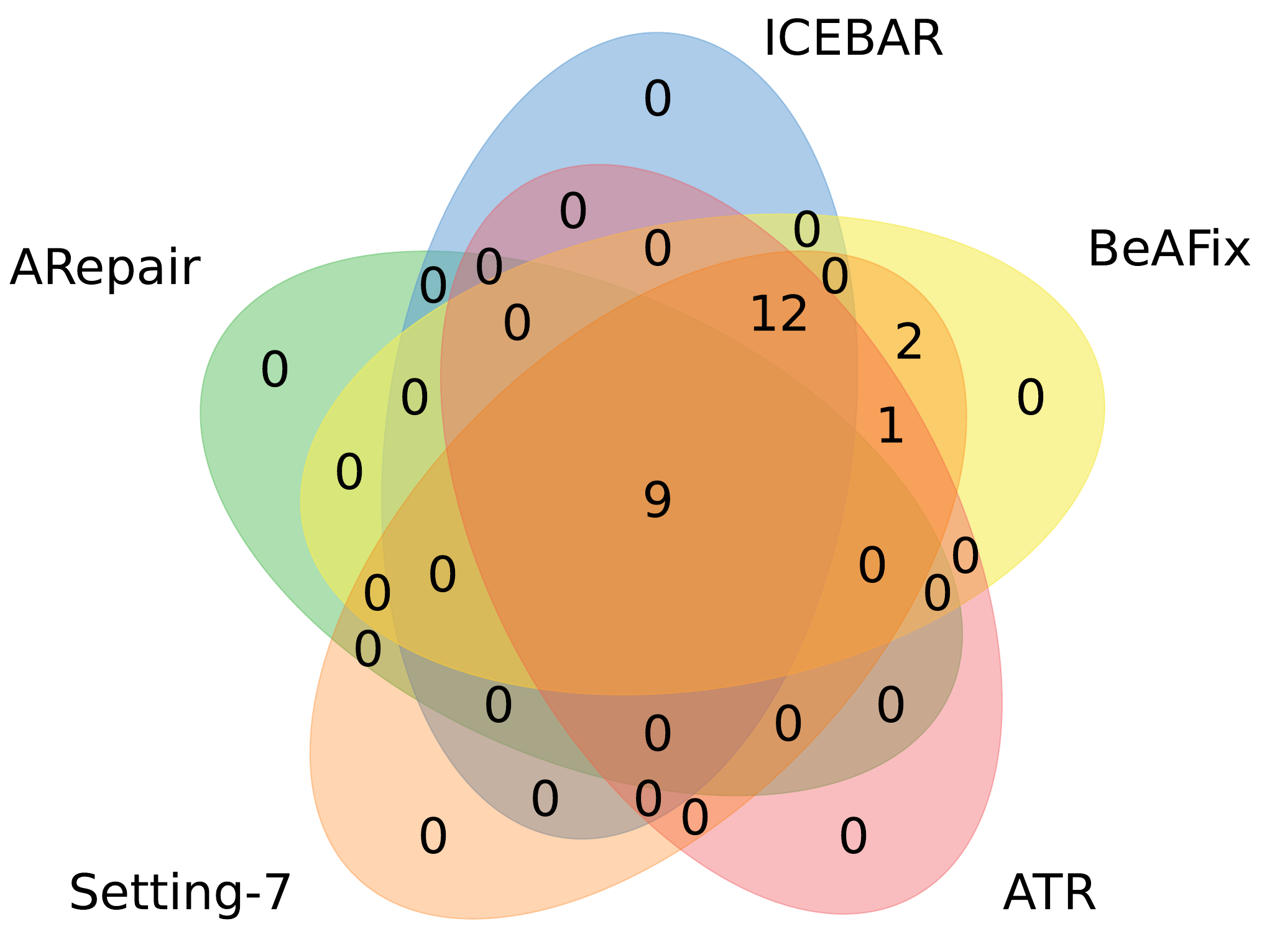}
    \end{minipage}
    \hfill
    \begin{minipage}{0.32\textwidth}
        \includegraphics[width=\linewidth]{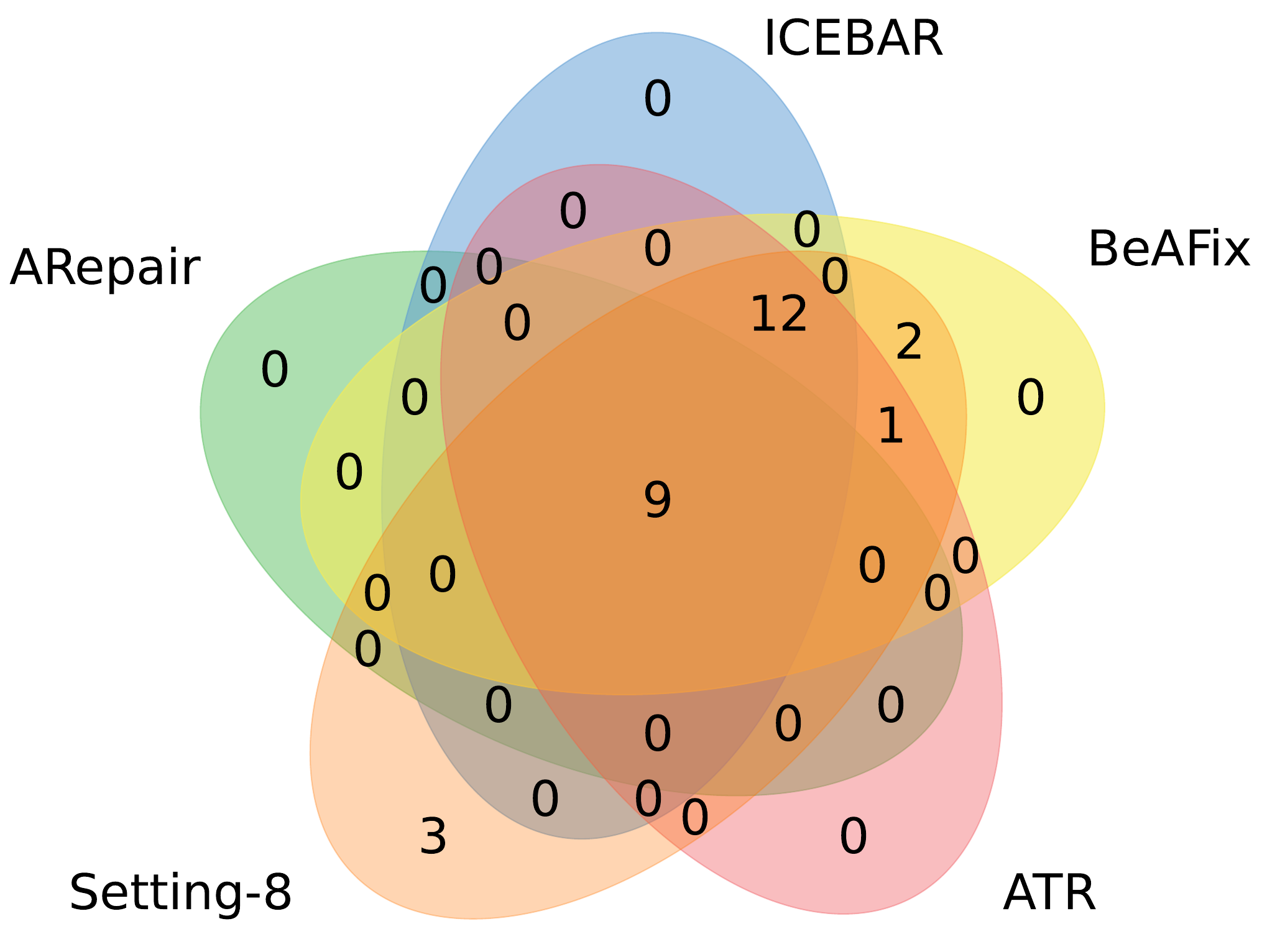}
    \end{minipage}
    \hfill
    \begin{minipage}{0.32\textwidth}
        \includegraphics[width=\linewidth]{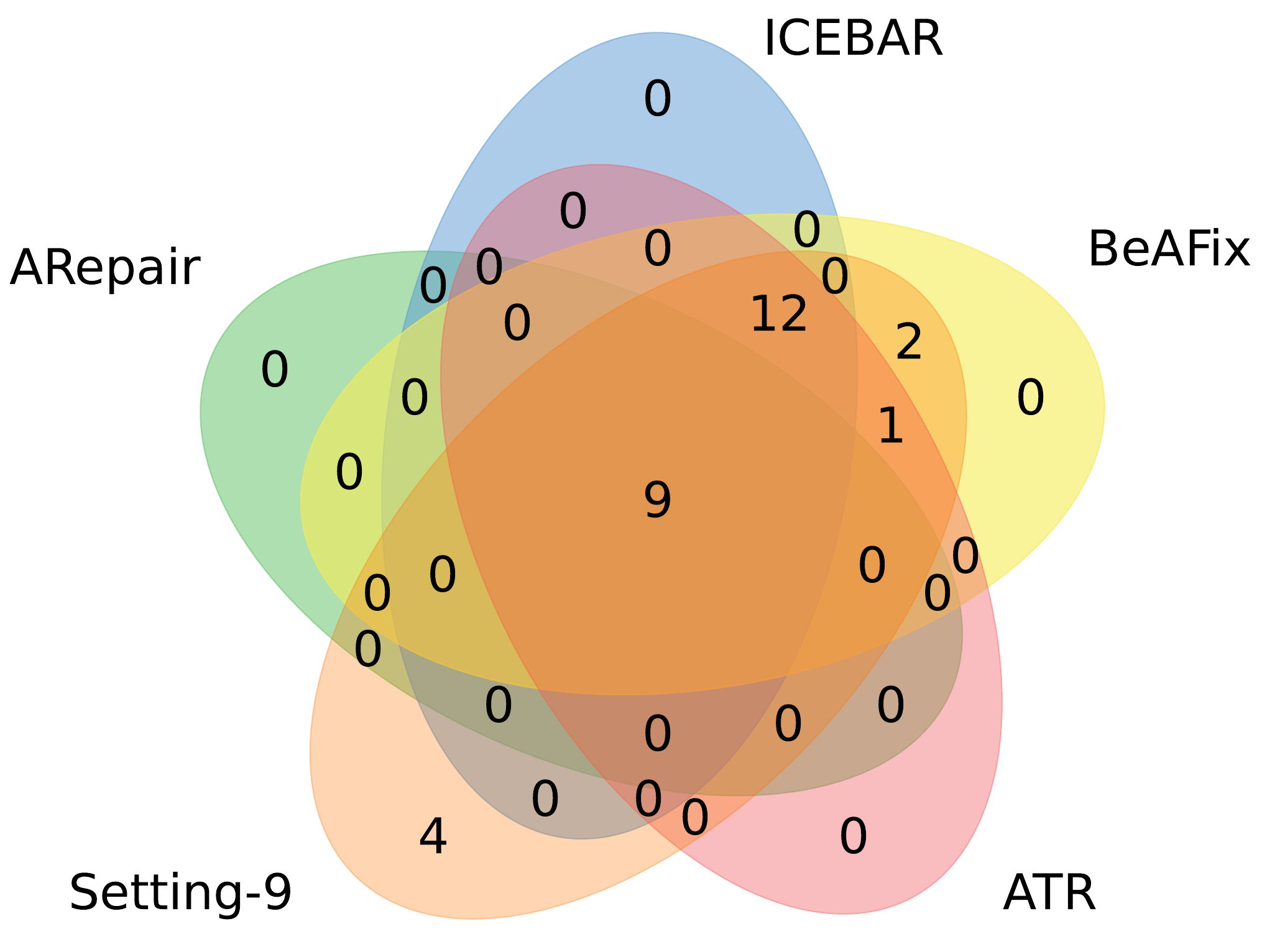}
    \end{minipage}

    \vspace{0.5cm} 
    
    \begin{minipage}{0.32\textwidth}
        \includegraphics[width=\linewidth]{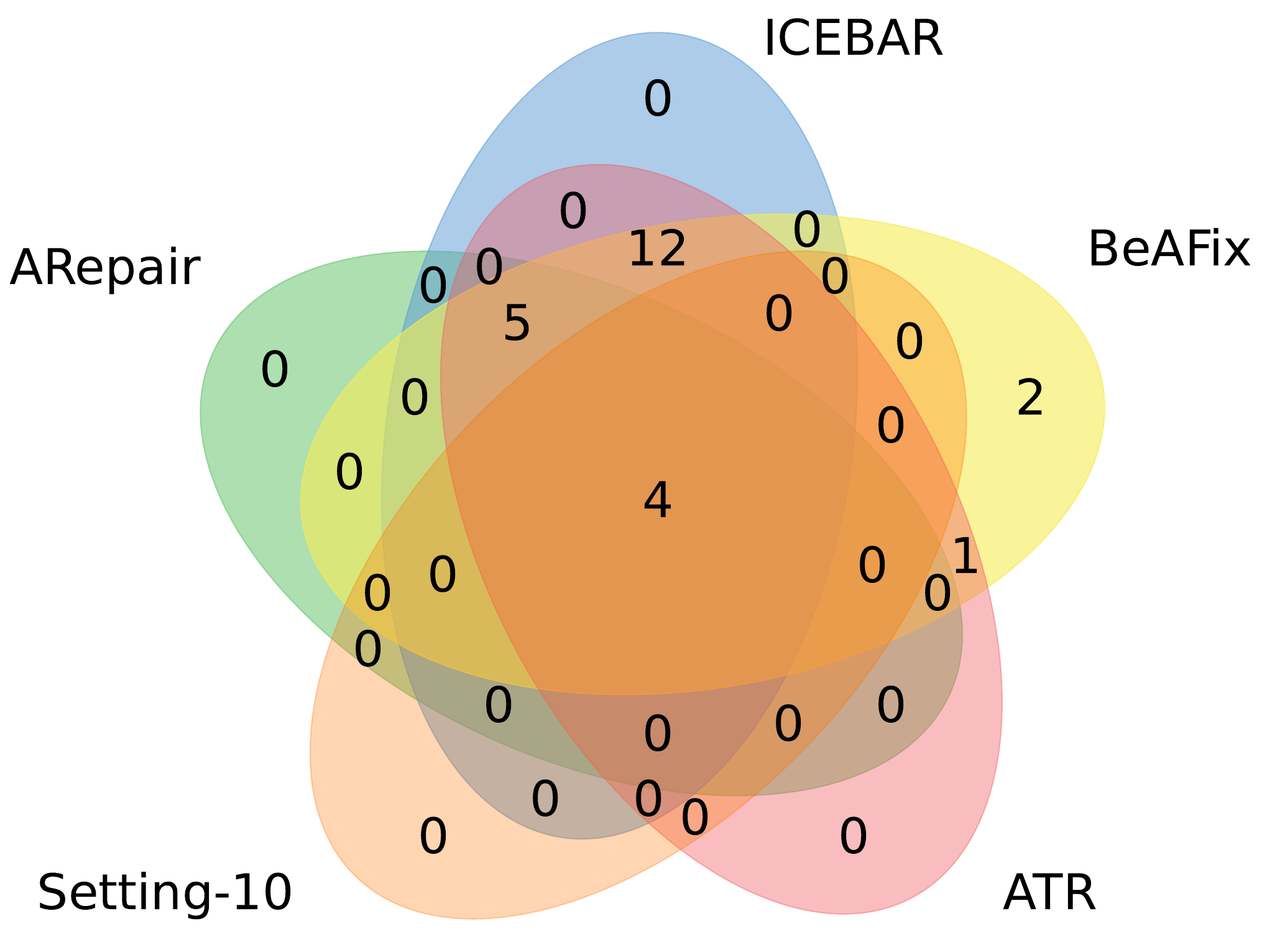}
    \end{minipage}
    \hfill
    \begin{minipage}{0.32\textwidth}
        \includegraphics[width=\linewidth]{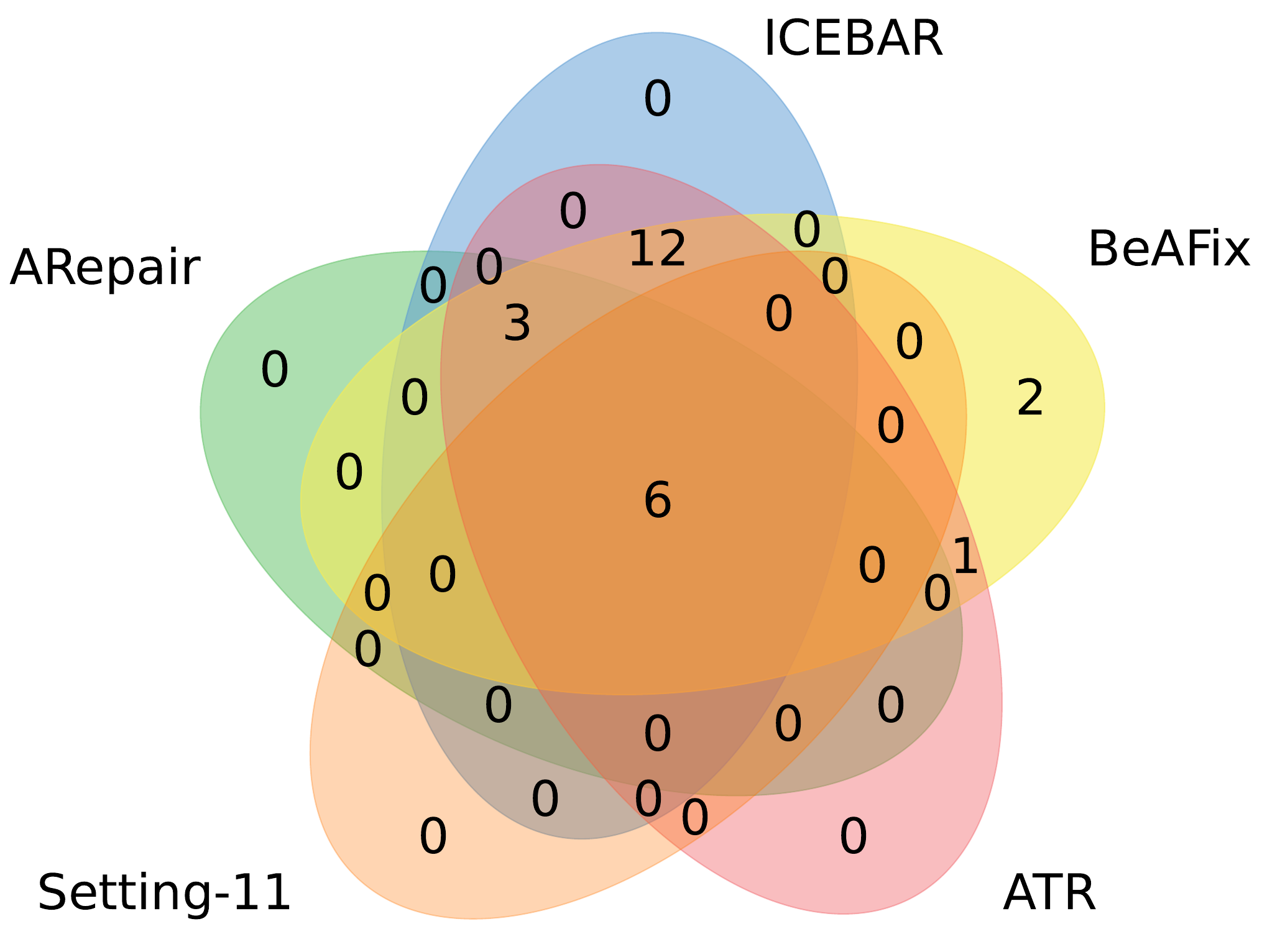}
    \end{minipage}
    \hfill
    \begin{minipage}{0.32\textwidth}
        \includegraphics[width=\linewidth]{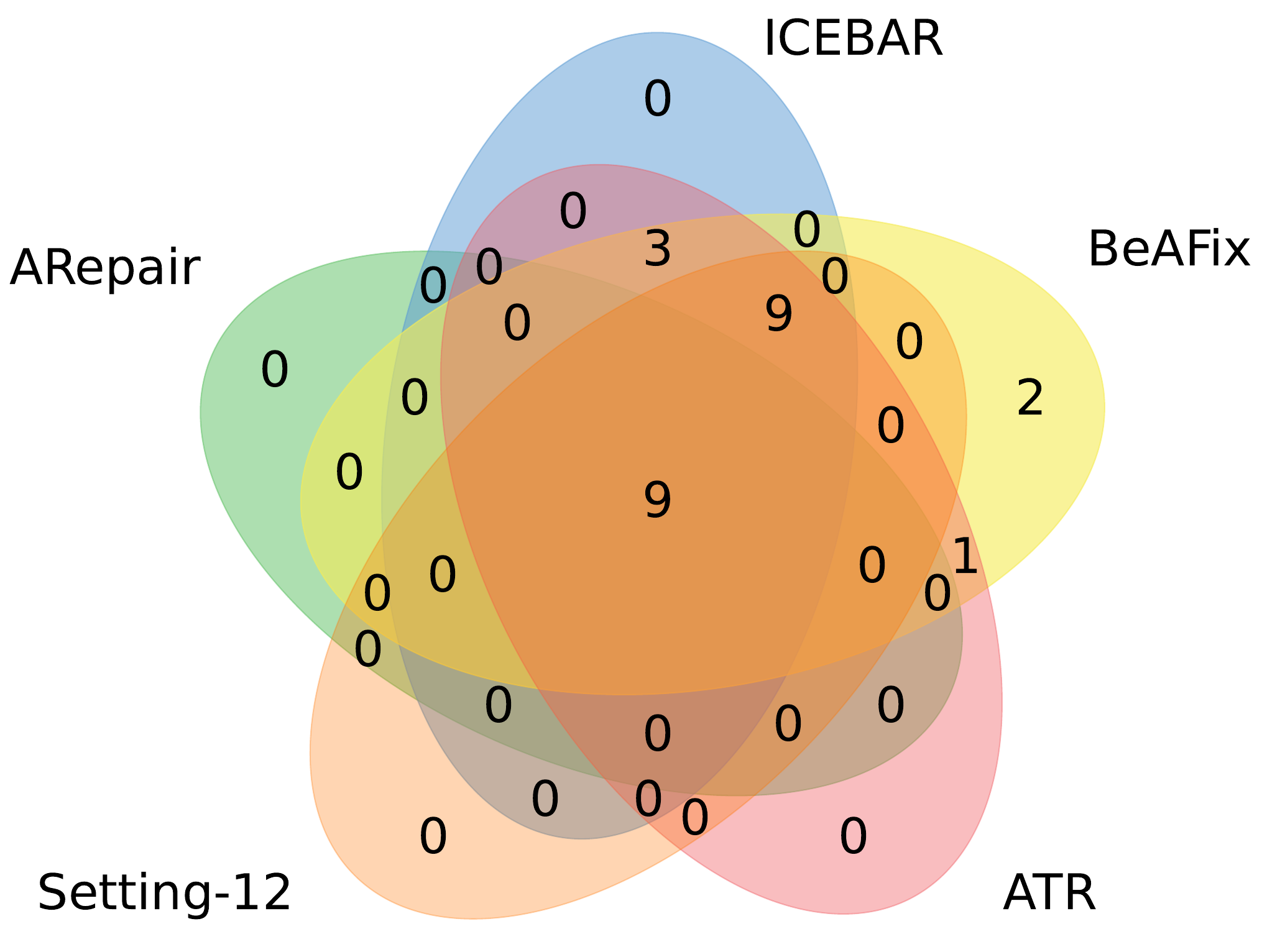}
    \end{minipage}
    
    \caption{Venn diagrams showing the exclusive and overlapping successful repairs \moh{for the ARepair benchmark} achieved by different repair methods. These illustrate the complementary and unique capabilities of the APR pipeline compared to state-of-the-art (SoTA) repair tools.}    
    \label{fig:venn_diagrams_arepair}
\end{figure*}

%% file: ConstraintsTable.tex
\begin{table}[ht]
  \caption{\rashed{Frequency of \textit{overconstrained} and \textit{underconstrained} specifications among all unfixed issues across all LLM settings on ARepair and Alloy4Fun benchmarks.}}

\begin{minipage}[t]{0.52\textwidth}
  \centering
  \scalebox{0.5}{
  \begin{tabular}{lcccccccccccc}
    \toprule
    \multicolumn{13}{c}{\textbf{ARepair Benchmark}} \\
    \midrule
    \multicolumn{13}{c}{\textbf{Part A: Over-Constrained Issues}} \\
    \midrule
    \multirow{2}{*}{\textbf{Model}} & \multicolumn{3}{c}{\textbf{GPT-4-32k}} & \multicolumn{3}{c}{\textbf{GPT-4-Turbo}} & \multicolumn{3}{c}{\textbf{GPT-4o}} & \multicolumn{3}{c}{\textbf{GPT-3.5-Turbo}} \\
    \cmidrule(lr){2-4} \cmidrule(lr){5-7} \cmidrule(lr){8-10} \cmidrule(lr){11-13}
    & \textbf{1} & \textbf{2} & \textbf{3} & \textbf{4} & \textbf{5} & \textbf{6} & \textbf{7} & \textbf{8} & \textbf{9} & \textbf{10} & \textbf{11} & \textbf{12} \\
    \midrule
    addr        & 1 & 1 & 0 & 1 & 0 & 0 & 1 & 1 & 1 & 1 & 1 & 1 \\ 
    arr         & 1 & 1 & 1 & 0 & 1 & 0 & 0 & 0 & 0 & 1 & 0 & 0 \\ 
    balancedBST & 1 & 1 & 0 & 1 & 1 & 1 & 1 & 1 & 1 & 2 & 2 & 1 \\ 
    bempl       & 0 & 0 & 0 & 0 & 0 & 0 & 0 & 1 & 1 & 0 & 0 & 0 \\ 
    cd          & 0 & 0 & 0 & 0 & 1 & 0 & 0 & 0 & 0 & 1 & 1 & 1 \\ 
    ctree       & 0 & 0 & 0 & 0 & 0 & 0 & 0 & 0 & 0 & 1 & 1 & 1 \\ 
    dll         & 0 & 0 & 0 & 1 & 1 & 0 & 0 & 0 & 0 & 2 & 0 & 0 \\ 
    farmer      & 0 & 0 & 0 & 0 & 0 & 0 & 0 & 0 & 0 & 0 & 0 & 0 \\ 
    fsm         & 0 & 0 & 0 & 1 & 1 & 0 & 0 & 0 & 0 & 1 & 1 & 1 \\ 
    grade       & 1 & 1 & 1 & 1 & 1 & 0 & 1 & 1 & 0 & 1 & 1 & 1 \\ 
    other       & 1 & 1 & 1 & 1 & 1 & 0 & 0 & 0 & 0 & 0 & 1 & 0 \\ 
    student     & 8 & 8 & 6 & 7 & 5 & 3 & 3 & 2 & 2 & 12 & 11 & 5 \\
    \midrule
    \textbf{Summary} & \textbf{13} & \textbf{13} & \textbf{9} & \textbf{13} & \textbf{12} & \textbf{4} & \textbf{6} & \textbf{6} & \textbf{5} & \textbf{22} & \textbf{19} & \textbf{11} \\
    \textbf{\%ratio} & \textbf{56.5} & \textbf{59.1} & \textbf{56.3} & \textbf{61.9} & \textbf{63.2} & \textbf{40.0} & \textbf{42.9} & \textbf{54.5} & \textbf{50.0} & \textbf{64.7} & \textbf{59.4} & \textbf{55.0} \\
    \midrule
    \multicolumn{13}{c}{\textbf{Part B: Under-Constrained Issues}} \\
    \midrule
    \multirow{2}{*}{\textbf{Model}} & \multicolumn{3}{c}{\textbf{GPT-4-32k}} & \multicolumn{3}{c}{\textbf{GPT-4-Turbo}} & \multicolumn{3}{c}{\textbf{GPT-4o}} & \multicolumn{3}{c}{\textbf{GPT-3.5-Turbo}} \\
    \cmidrule(lr){2-4} \cmidrule(lr){5-7} \cmidrule(lr){8-10} \cmidrule(lr){11-13}
    & \textbf{1} & \textbf{2} & \textbf{3} & \textbf{4} & \textbf{5} & \textbf{6} & \textbf{7} & \textbf{8} & \textbf{9} & \textbf{10} & \textbf{11} & \textbf{12} \\
    \midrule
    addr        & 0 & 0 & 0 & 0 & 0 & 0 & 0 & 0 & 0 & 0 & 0 & 0 \\ 
    arr         & 1 & 1 & 1 & 0 & 1 & 0 & 0 & 0 & 0 & 1 & 0 & 0 \\ 
    balancedBST & 1 & 1 & 0 & 1 & 1 & 1 & 1 & 1 & 1 & 1 & 1 & 1 \\ 
    bempl       & 1 & 1 & 1 & 0 & 0 & 0 & 1 & 0 & 0 & 1 & 1 & 1 \\ 
    cd          & 1 & 0 & 0 & 0 & 0 & 0 & 0 & 0 & 0 & 0 & 1 & 1 \\ 
    ctree       & 0 & 0 & 0 & 0 & 0 & 0 & 0 & 0 & 0 & 0 & 0 & 0 \\ 
    dll         & 0 & 0 & 0 & 1 & 1 & 0 & 0 & 0 & 0 & 1 & 1 & 1 \\ 
    farmer      & 0 & 0 & 0 & 0 & 0 & 0 & 0 & 0 & 0 & 0 & 0 & 0 \\ 
    fsm         & 1 & 1 & 0 & 1 & 1 & 1 & 1 & 1 & 0 & 1 & 1 & 1 \\ 
    grade       & 0 & 0 & 0 & 0 & 0 & 0 & 0 & 0 & 0 & 0 & 0 & 0 \\ 
    other       & 0 & 0 & 0 & 0 & 0 & 0 & 0 & 0 & 0 & 0 & 0 & 0 \\ 
    student     & 5 & 5 & 4 & 4 & 3 & 3 & 4 & 3 & 3 & 7 & 7 & 3 \\
    \midrule
    \textbf{Summary} & \textbf{10} & \textbf{9} & \textbf{6} & \textbf{8} & \textbf{7} & \textbf{6} & \textbf{8} & \textbf{5} & \textbf{5} & \textbf{12} & \textbf{13} & \textbf{9} \\
    \textbf{\%ratio} & \textbf{43.5} & \textbf{40.9} & \textbf{37.5} & \textbf{38.1} & \textbf{36.8} & \textbf{60.0} & \textbf{57.1} & \textbf{45.5} & \textbf{50.0} & \textbf{35.3} & \textbf{40.6} & \textbf{45.0} \\
    \bottomrule
  \end{tabular}
  }
  \label{tab:ARepair-Constraint-Type}
\end{minipage}
\hfill
\begin{minipage}[t]{0.46\textwidth}
  \centering
  \scalebox{0.5}{
  \begin{tabular}{lcccccc}
    \toprule
    \multicolumn{7}{c}{\textbf{Alloy4Fun Benchmark}} \\
    \midrule
    \multicolumn{7}{c}{\textbf{Part A: Over-Constrained Issues}} \\
    \midrule
    \textbf{Model} & \multicolumn{6}{c}{\textbf{\#issues Under Setting}} \\
    \cmidrule(lr){2-7}
    & \multicolumn{3}{c}{\textbf{GPT-4o}} & \multicolumn{3}{c}{\textbf{GPT-3.5-Turbo}} \\
    \cmidrule(lr){2-4} \cmidrule(lr){5-7}
    & \textbf{7} & \textbf{8} & \textbf{9} & \textbf{10} & \textbf{11} & \textbf{12} \\
    \midrule
    classroom   & (38) & (39) & (35) & 612 & 610 & 593 \\ 
    cv          & (3) & (1) & (0) & 60 & 65 & 41 \\ 
    graphs      & (23) & (13) & (23) & 150 & 135 & 145 \\ 
    lts         & (13) & (8) & (9) & 110 & 82 & 88 \\ 
    production  & (1) & (0) & (1) & 22 & 18 & 16 \\ 
    trash       & (32) & (33) & (27) & 120 & 120 & 114 \\ 
    \midrule
    \textbf{Summary} 
    & \textbf{(110)} & \textbf{(94)} & \textbf{(95)} & \textbf{1074} & \textbf{1030} & \textbf{997} \\ 
    \textbf{\%ratio} 
    & \textbf{(60.8)} & \textbf{(62.3)} & \textbf{(61.0)} & \textbf{60.3} & \textbf{60.4} & \textbf{60.1} \\ 
    \midrule
    \multicolumn{7}{c}{\textbf{Part B: Under-Constrained Issues}} \\
    \midrule
    \textbf{Model} & \multicolumn{6}{c}{\textbf{\#issues Under Setting}} \\
    \cmidrule(lr){2-7}
    & \multicolumn{3}{c}{\textbf{GPT-4o}} & \multicolumn{3}{c}{\textbf{GPT-3.5-Turbo}} \\
    \cmidrule(lr){2-4} \cmidrule(lr){5-7}
    & \textbf{7} & \textbf{8} & \textbf{9} & \textbf{10} & \textbf{11} & \textbf{12} \\
    \midrule
    classroom   & (22) & (21) & (19) & 387 & 388 & 385 \\ 
    cv          & (1) & (1) & (1) & 40 & 41 & 27 \\ 
    graphs      & (15) & (7) & (15) & 110 & 97 & 103 \\ 
    lts         & (9) & (4) & (5) & 71 & 52 & 55 \\ 
    production  & (0) & (0) & (1) & 14 & 11 & 11 \\ 
    trash       & (24) & (24) & (20) & 86 & 86 & 80 \\ 
    \midrule
    \textbf{Summary} 
    & \textbf{(71)} & \textbf{(57)} & \textbf{(61)} & \textbf{708} & \textbf{675} & \textbf{661} \\ 
    \textbf{\%ratio} 
    & \textbf{(39.2)} & \textbf{(37.7)} & \textbf{(39.0)} & \textbf{39.7} & \textbf{39.6} & \textbf{39.9} \\ 
    \bottomrule
  \end{tabular}
  }
  \label{tab:A4F-Constraint-Type}
\end{minipage}

\vspace{0.5cm}
\centering
\scalebox{0.8}{
\begin{tabular}{lccccccccccccccccccc}
\toprule
\textbf{Total} & \multicolumn{12}{c}{\textbf{ARepair}} & \multicolumn{6}{c}{\textbf{Alloy4Fun}} \\
\cmidrule(lr){2-13} \cmidrule(lr){14-19}
\textbf{Issues} & \textbf{1} & \textbf{2} & \textbf{3} & \textbf{4} & \textbf{5} & \textbf{6} & \textbf{7} & \textbf{8} & \textbf{9} & \textbf{10} & \textbf{11} & \textbf{12} & \textbf{7} & \textbf{8} & \textbf{9} & \textbf{10} & \textbf{11} & \textbf{12} \\
\midrule
Over-constrained & 13 & 13 & 9 & 13 & 12 & 4 & 6 & 6 & 5 & 22 & 19 & 11 & (110) & (94) & (95) & 1074 & 1030 & 997 \\
Under-constrained & 10 & 9 & 6 & 8 & 7 & 6 & 8 & 5 & 5 & 12 & 13 & 9 & (71) & (57) & (61) & 708 & 675 & 661 \\
\bottomrule
\end{tabular}
}
\end{table}

%% file: threatsToValidity.tex
\section{Threats to Validity}

\noindent
\textbf{Internal Validity.}
Variations in prompt design and format may introduce biases in the repair process, affecting the outcomes independently of the LLM's capabilities. To mitigate this threat, we carefully standardized the prompt design across all experiments, ensuring consistency in the information provided to the LLMs. Additionally, we conducted analyses to assess the impact of prompt variations on repair performance, enabling us to isolate the effects of LLM capabilities from potential biases introduced by prompt design. 

\noindent
\textbf{External Validity.}
The generalizability of our findings may be limited by the specific characteristics of the benchmarks used. To address this, we utilized diverse benchmark suites and evaluated the performance of the repair pipeline across various scenarios. Furthermore, we documented the experimental setup comprehensively to facilitate replication and external validation of our results. \moh{Finally, some repair instances in our benchmark have been fixed in prior work and may have been included in the training data of the LLMs. This raises the concern that the models might be memorizing existing repairs rather than generating novel solutions, which could limit their applicability to Alloy models they have not been trained on. As a result, this poses a threat to the generalizability of our findings, as the LLMs may perform well on known faults but struggle with previously unseen ones. Future work could further investigate this aspect, following the approach of~\cite{memorization}, and introduce a new benchmark dataset—analogous to the ConDefects dataset~\cite{condefects-fse}—designed specifically to mitigate the issue of data leakage.}

\noindent
\textbf{Construct Validity.}
Biases in the measurement of repair performance metrics could distort our assessment of LLM capabilities. To mitigate this, we employed standardized metrics and conducted sensitivity analyses to validate the robustness of our findings. Additionally, we ensured transparency and reproducibility in our methodology to enhance the validity of our measurements.

%% file: relatedwork.tex
\section{Related Work}

Several recent studies have explored the integration of Large Language Models (LLMs) into software engineering tasks, particularly in the realm of program repair. 

AlphaRepair~\cite{AlphaRepair} leverages LLMs for APR in a zero-shot setting, but it requires removing the buggy line and replacing it with masked tokens. It then queries the LLM to fill-in the masked tokens with the correct
tokens to generate patches.
Xia et al.~\cite{xia2023conversational} improve APR performance by incorporating test feedback into prompts, while Kang et al.~\cite{kang2023explainable} enabled LLMs to utilize a debugger for information gathering and patch generation. Additionally, Fuzz4All~\cite{fuzz4all} leverages LLMs as an input generation and mutation engine, employing an auto-prompting phase to generate concise input prompts. RepairAgent~\cite{repairagent} employs LLMs, agents, and dynamic prompts for APR, albeit in the context of repairing Java applications. TestPilot~\cite{TestPilot} uses LLMs to generate unit test cases for JavaScript, employing a few-shot learning approach to refine prompts with failed tests and error messages.

In contrast to these efforts, our study focuses on using LLMs for automated specification repair, a less explored area in software engineering. We adopt an approach similar to Fuzz4All's auto-prompting phase, \ma{specifically by incorporating LLM to construct the prompt~\cite{autoprompt}.} Moreover, while existing work predominantly uses LLMs for program repair in imperative languages like Java and JavaScript, our study extends the application of LLMs to the domain of Alloy specifications, addressing a gap in the literature regarding specification repair techniques for declarative languages. \moh{Consequently, a direct quantitative comparison is largely infeasible, with the exception of the tool proposed by Hasan et al.\cite{hasan2023automated}, which utilizes LLMs to repair Alloy specifications. Nevertheless, Table~\ref{tab:compare_llm_APR} presents a comparative analysis of state-of-the-art LLM-based APR tools. This comparison highlights that our APR pipeline incorporates recent advances and techniques in LLMs that are also employed by contemporary repair tools.}

\begin{table}[]
\centering
\caption{\ma{Comparing the feature that we considered in the APR pipeline with contemporary state-of-the-art LLM-APR tools, the column labeled ``Autoprompt'' denotes that the LLM generates the prompt.}}
\scalebox{0.7}{
\begin{tabular}{lccccccr}
\toprule
APR Tools & Agentic & Tools & Feedback & Autoprompt & Zero-shot & Spec Repair & \# LLMs \\ \midrule
AlphaRepair~\cite{AlphaRepair} & \no & \no & \no & \no & \yes &\no & 1 \\
ChatRepair ~\cite{ChatRepair} & \no & \yes & \yes & \no & \yes & \no & 1 \\
conversationalAPR~\cite{xia2023conversational} & \no & \yes & \yes & \no & \yes & \no & 10 \\
Xia et al.~\cite{apr_llm_eval} & \no & \no & \no & \no & \no & \no & 9 \\
Hasan et al.~\cite{hasan2023automated} & \no & \yes & \no & \no & \yes & \yes & 1 \\
RepairAgent~\cite{repairagent} & \yes & \yes & \yes & \no & \yes & \no & 1 \\
Ours & \yes & \yes & \yes & \yes & \yes & \yes & 4 \\ \bottomrule
\end{tabular}
}
\label{tab:compare_llm_APR}
\end{table}

%% file: conclusion.tex
\section{Conclusion}
In this study, we explore the potential of pre-trained LLMs to facilitate the repair of Alloy specifications, taking into account recent advancements in LLMs, including the use of agents, feedback mechanisms, zero-shot learning capabilities, and auto-prompting techniques. 
The investigation reveals that employing a dual-agent repair pipeline enhances the repair process, albeit with a marginal increase in token consumption.
The comparative evaluation highlights the superior performance of the GPT-4 model family over GPT-3.5-Turbo, underscoring their promising applicability. Overall, the findings of this research indicate a positive outlook for applying LLMs in automated program repair for declarative specifications, particularly through repair pipelines that integrate contemporary innovations in the field of LLMs.

%% file: appendix.tex

\section*{Appendix: Additional Results}
\addcontentsline{toc}{section}{Appendix: Additional Results}
This section presents the results for the Alloy4Fun benchmark based on settings 7-12.

\input{venn_diagrams_A4F}

\vspace{-1cm}

\begin{figure*}[h!]
    \centering
    \begin{subfigure}[b]{0.45\textwidth}
        \centering
        (a)
        \includegraphics[width=0.9\textwidth]{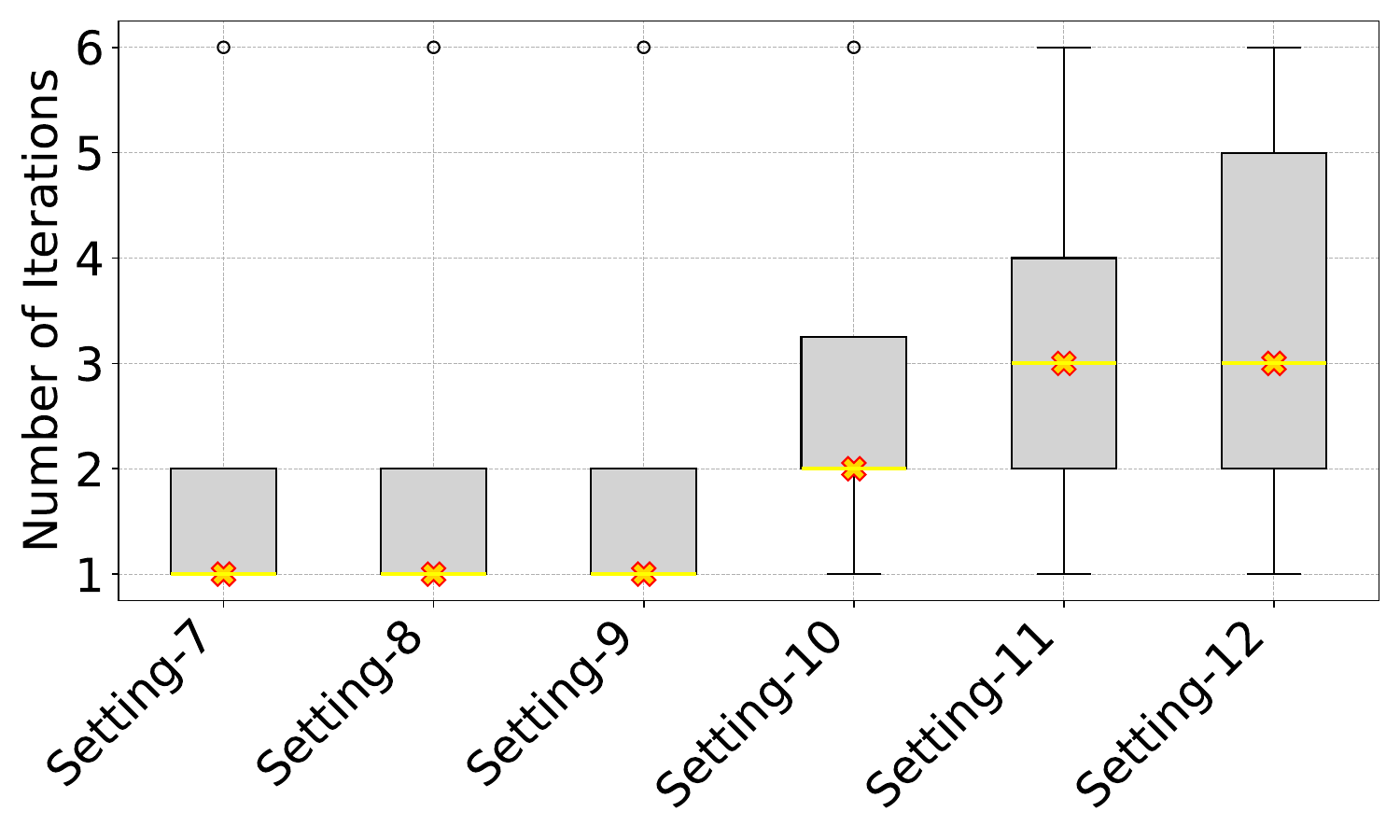}
        \label{fig:iterationAnalysis_A4F}
    \end{subfigure}
    \hfill
    \begin{subfigure}[b]{0.45\textwidth}
        \centering
        (b)        
        \includegraphics[width=0.9\textwidth]{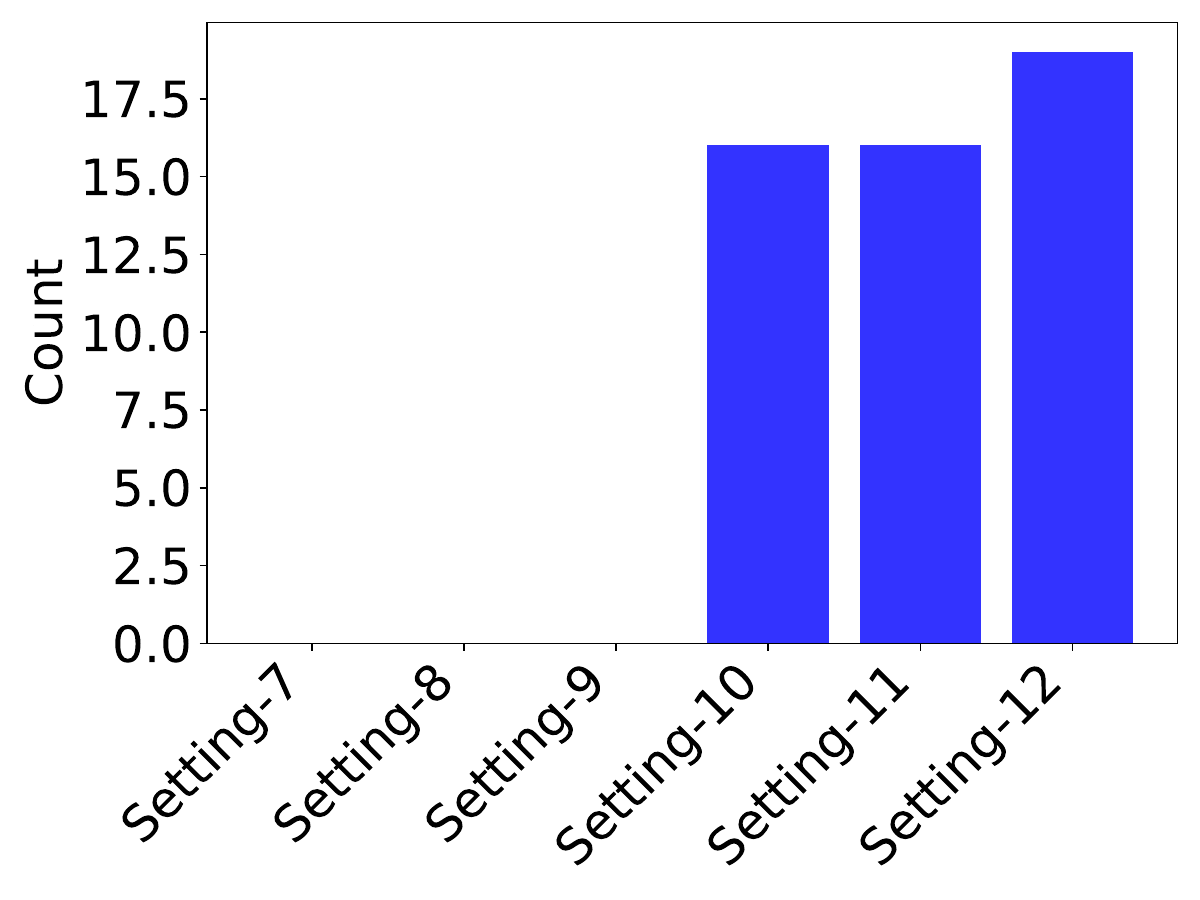}
        \label{fig:repeated1stIter_A4F}
    \end{subfigure}
    \vspace{-0.1cm}
    \caption{\moh{Alloy4Fun benchmark results: (a) Iteration count distribution for repairing specifications across various settings. (b) Incident count for initial repair attempts mirroring the buggy specification (lower values preferred)}.}
    \label{fig:RQ2-A4F}
\end{figure*}

\begin{figure}[htbp]
    \centering
    \begin{minipage}{0.45\textwidth}
        \centering
        \includegraphics[width=0.9\textwidth]{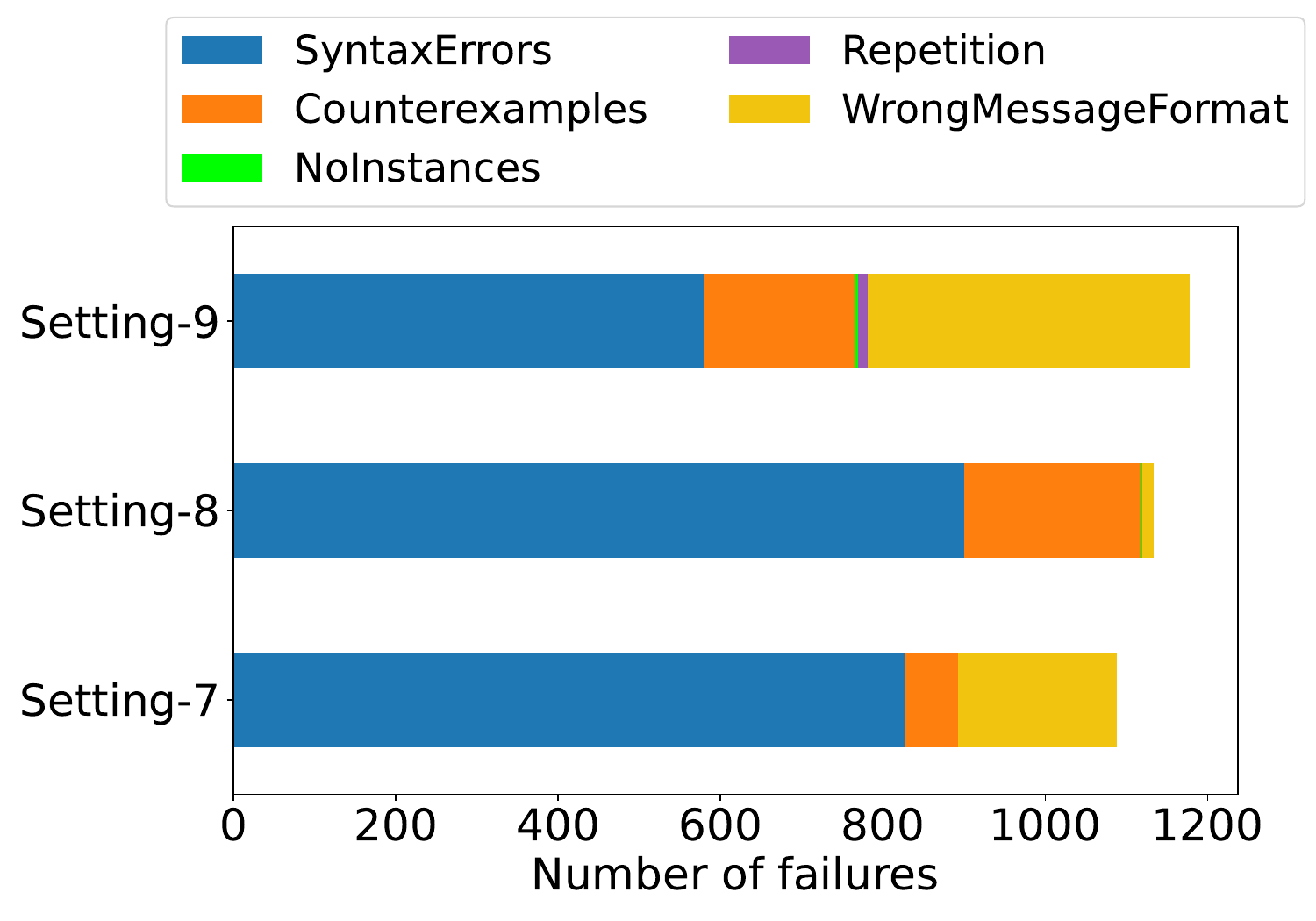}
        \caption{\moh{Error types observed in failed repair iterations for the Alloy4Fun benchmark across settings 7-9.}}
        \label{fig:failures_ARepair_S7-12}
    \end{minipage}
    \hfill
    \begin{minipage}{0.45\textwidth}
        \centering
        \includegraphics[width=0.9\textwidth]{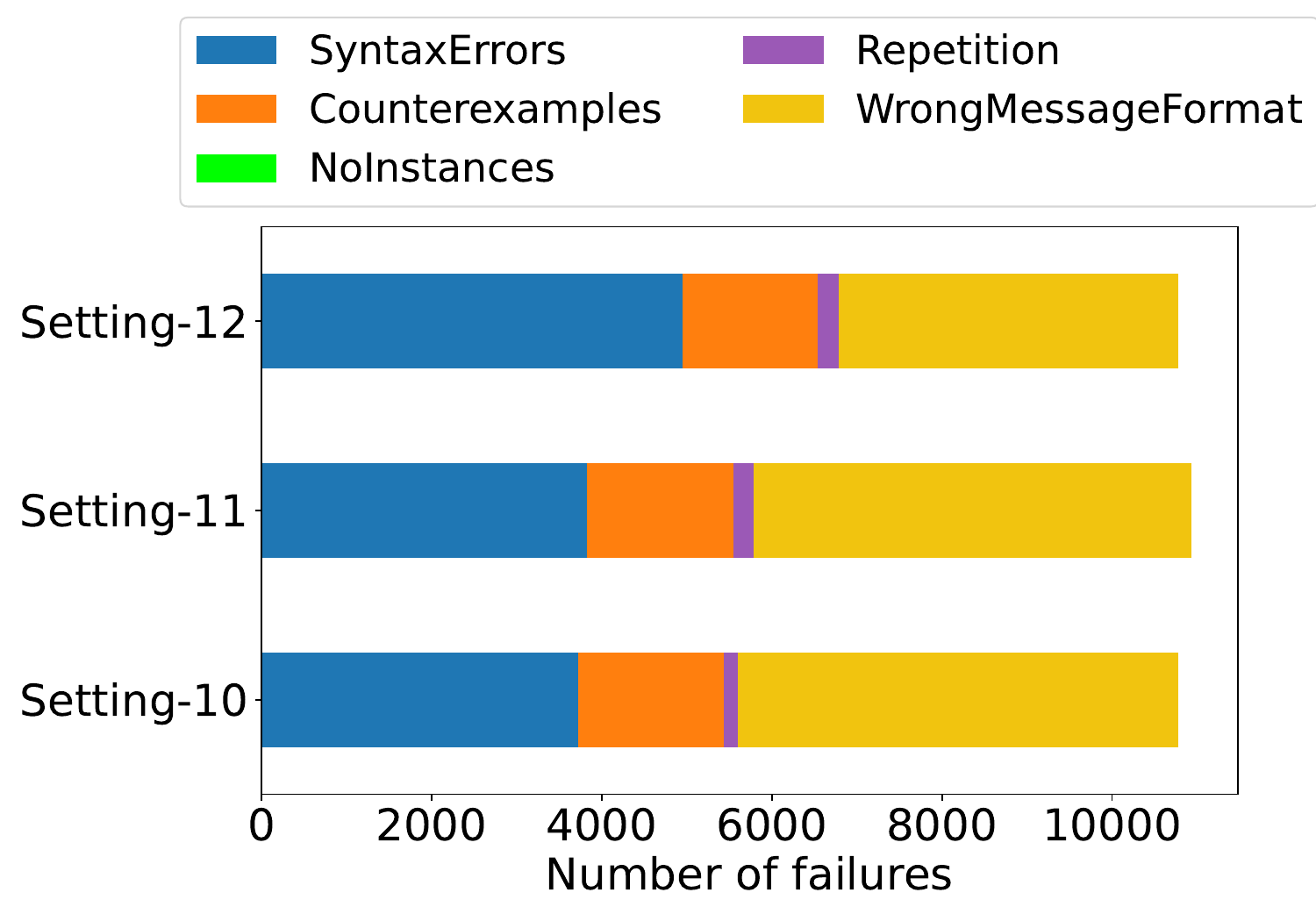}
        \caption{\moh{Error types observed in failed repair iterations for the Alloy4Fun benchmark across settings 10-12.}}
        \label{fig:failures_A4F_S7-12}
    \end{minipage}
\end{figure}
\vspace{-1cm}
\begin{figure}[ht]
    \centering
    \begin{minipage}[b]{0.49\columnwidth}
        \includegraphics[width=\linewidth]{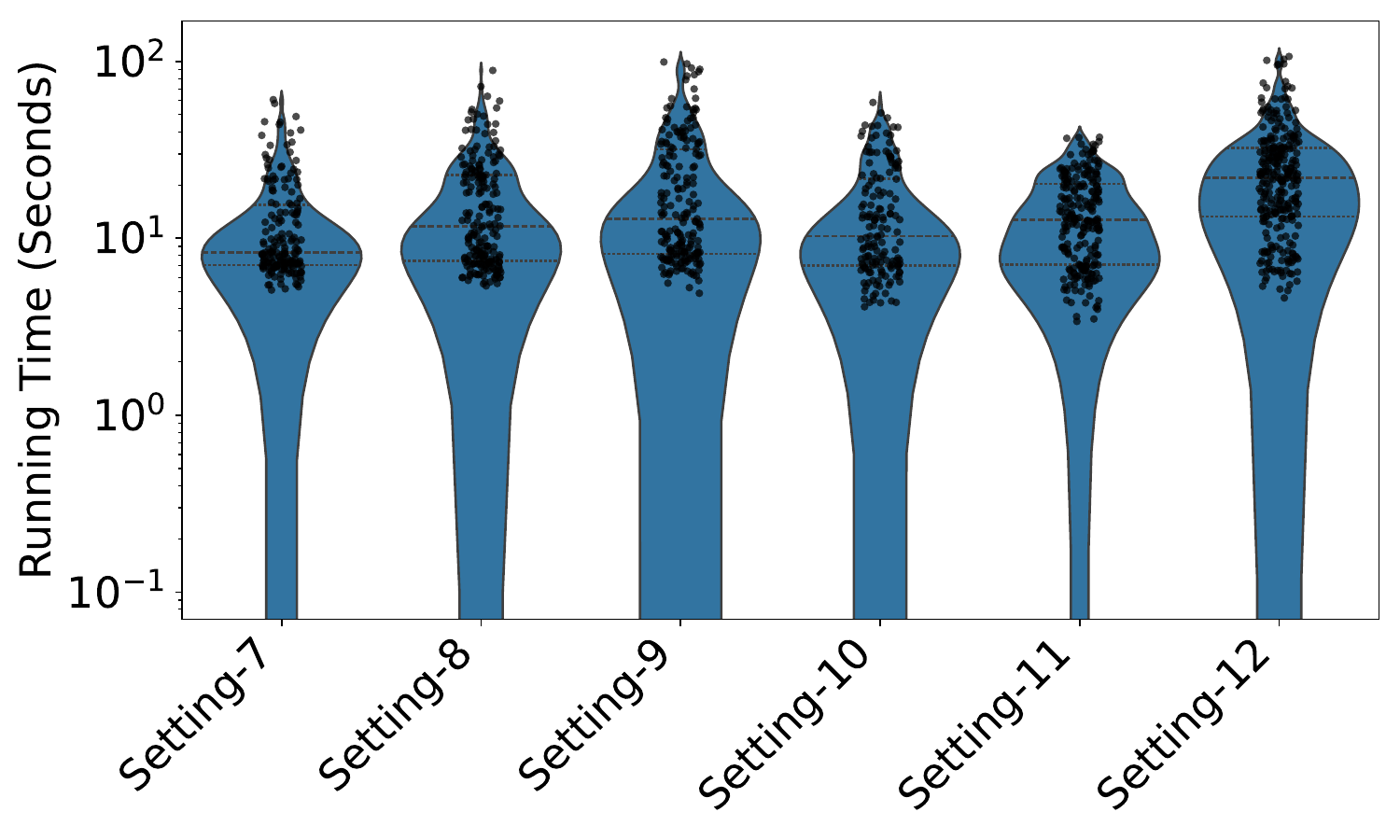}
    \end{minipage}
    \begin{minipage}[b]{0.49\columnwidth}
        \includegraphics[width=\linewidth]{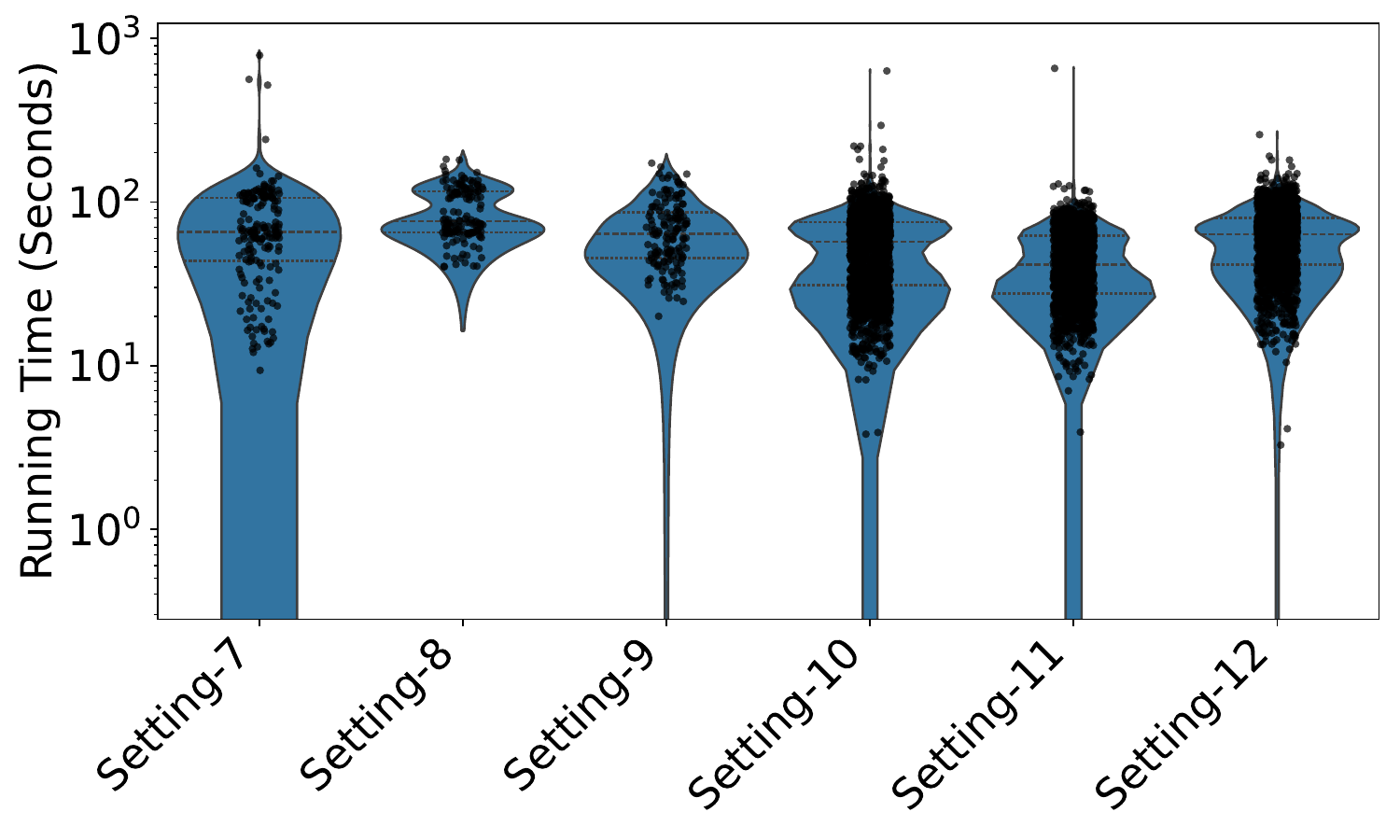}
    \end{minipage}
    \vspace{-0.3cm}
    \caption{\moh{Distribution of running time across settings 7-12 for the Alloy4Fun benchmark for fixed bugs (left) and unfixed bugs (right).}}
    \label{fig:run_time_no_fixed_A4F_S7-12}
\end{figure}

\vspace{-10cm}
\begin{figure}[h!]
    \centering
    \begin{minipage}[b]{0.49\columnwidth}
        \includegraphics[width=0.9\linewidth]{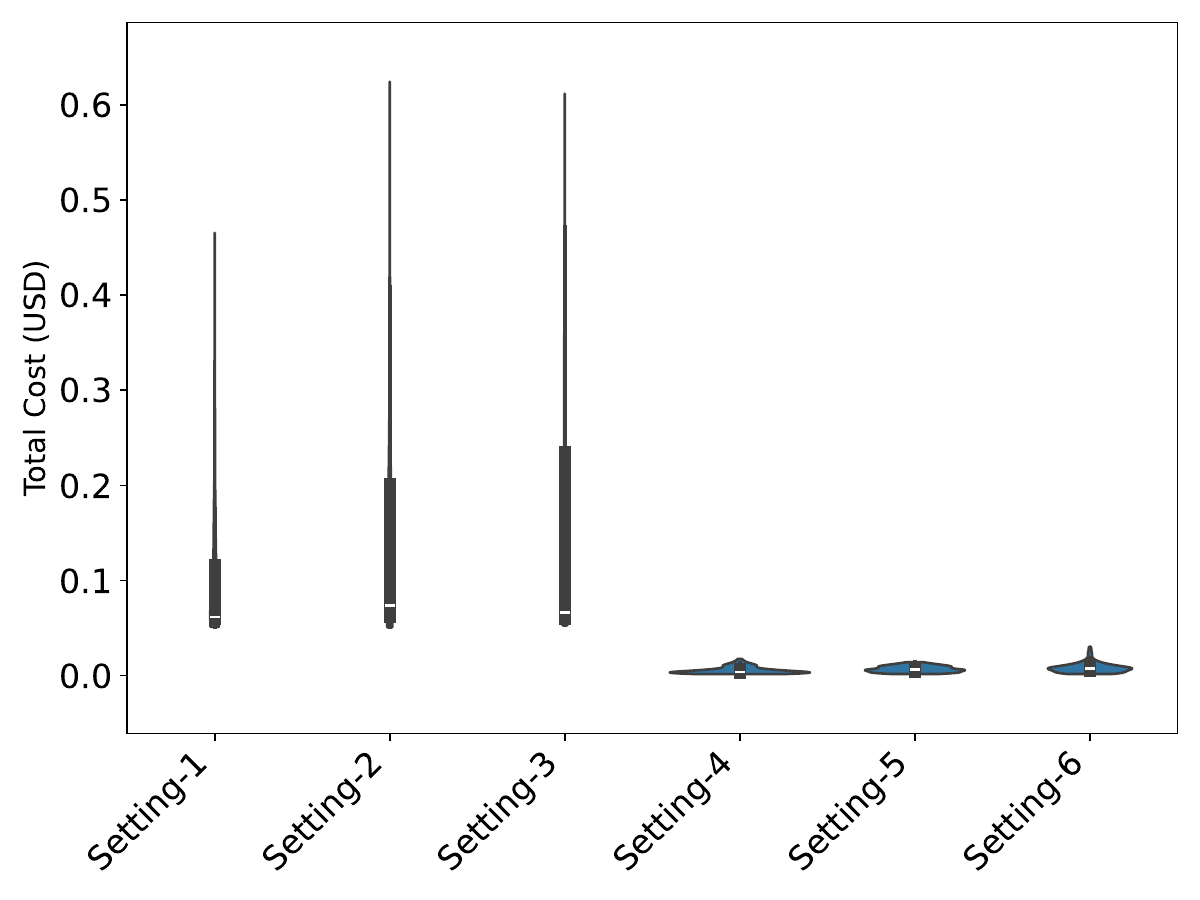}
    \end{minipage}
    \begin{minipage}[b]{0.49\columnwidth}
        \includegraphics[width=0.9\linewidth]{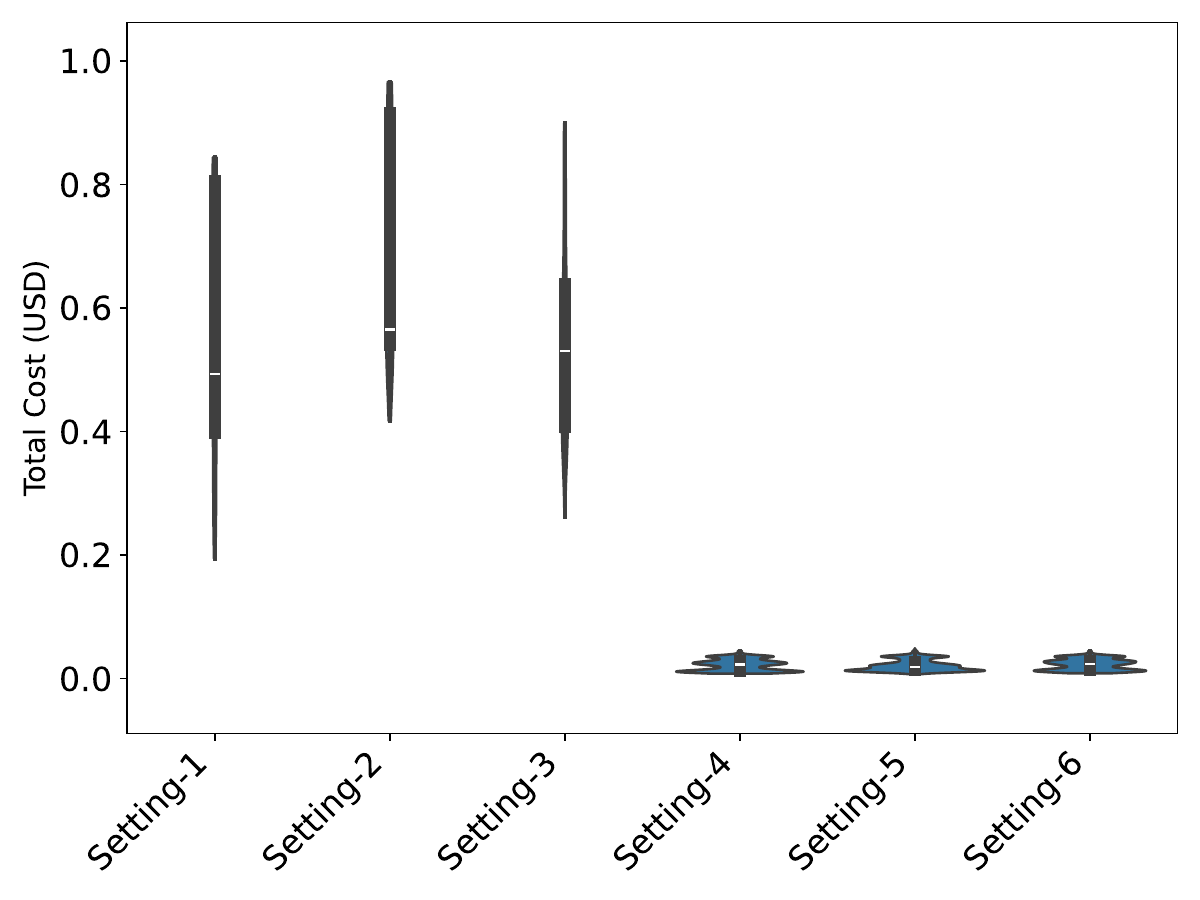}
    \end{minipage}
    \vspace{-0.3cm}
    \caption{\moh{Monetary distribution across settings 7-12 for the Alloy4Fun benchmark for fixed bugs (left) and unfixed bugs (right).}}
    \label{fig:cost_ARepair_S7-12}
\end{figure}

%% file: venn_diagrams_A4F.tex
\begin{figure*}[ht!]
    \centering
    \begin{minipage}{0.32\textwidth}
        \includegraphics[width=\linewidth]{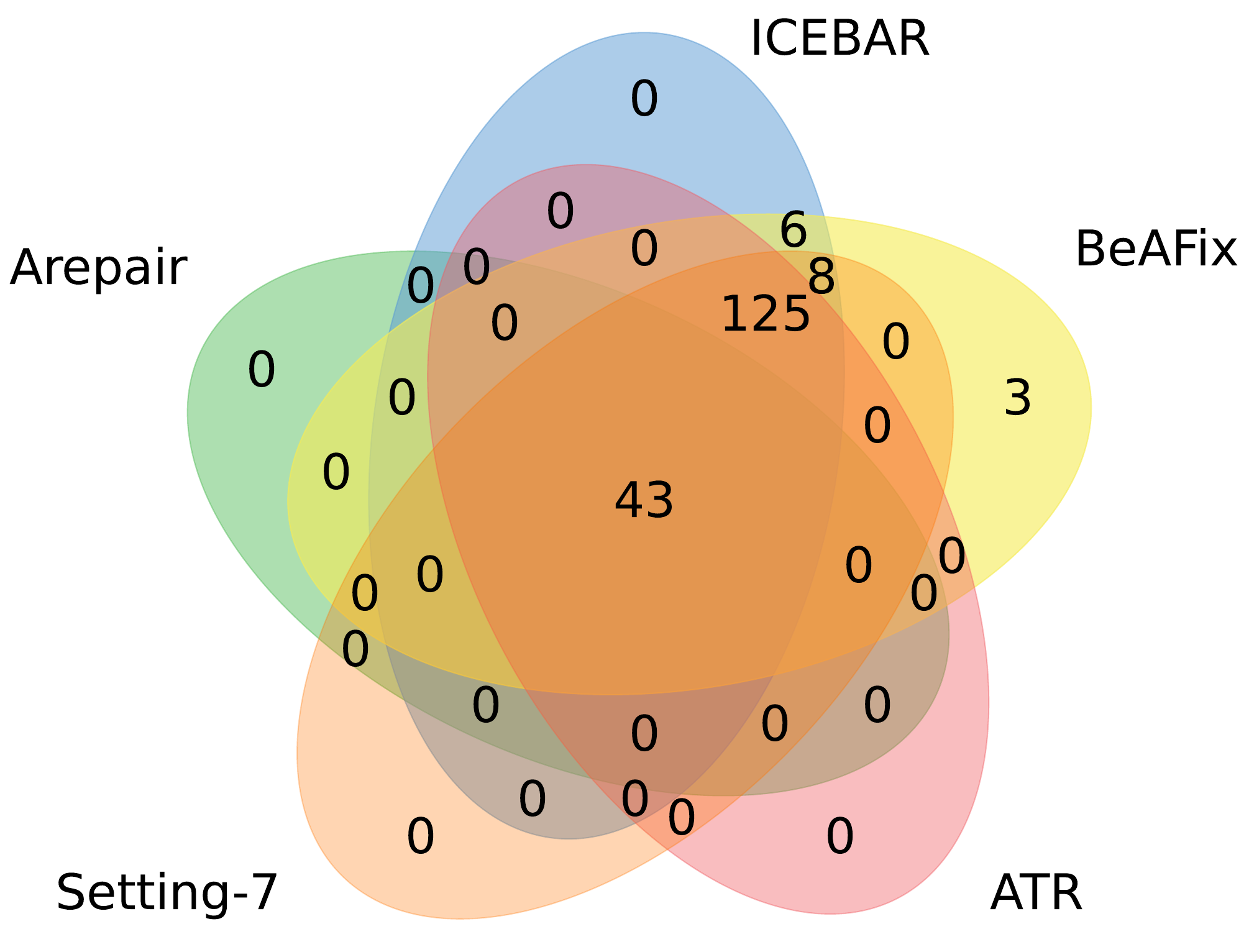}
    \end{minipage}
    \hfill
    \begin{minipage}{0.32\textwidth}
        \includegraphics[width=\linewidth]{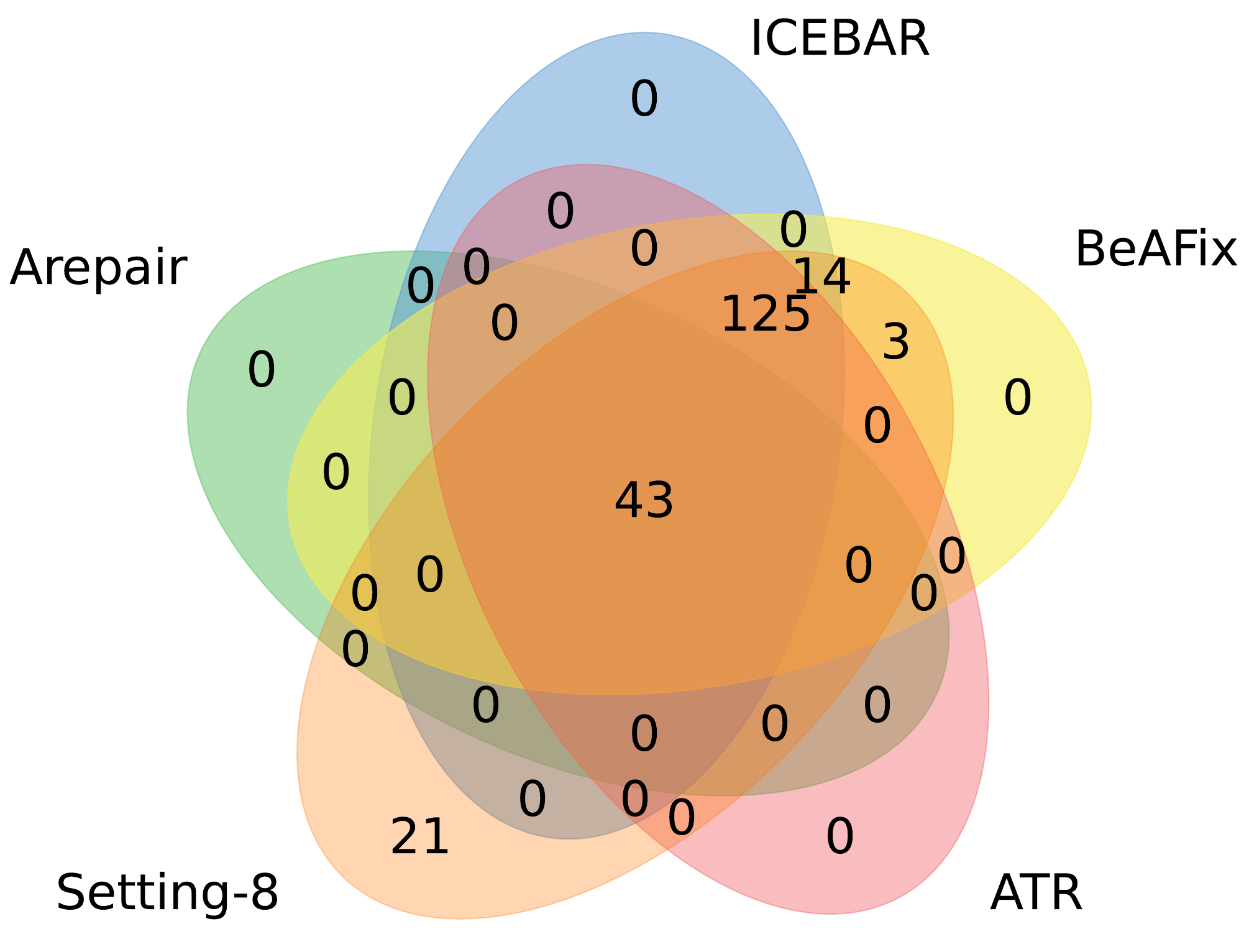}
    \end{minipage}
    \hfill
    \begin{minipage}{0.32\textwidth}
        \includegraphics[width=\linewidth]{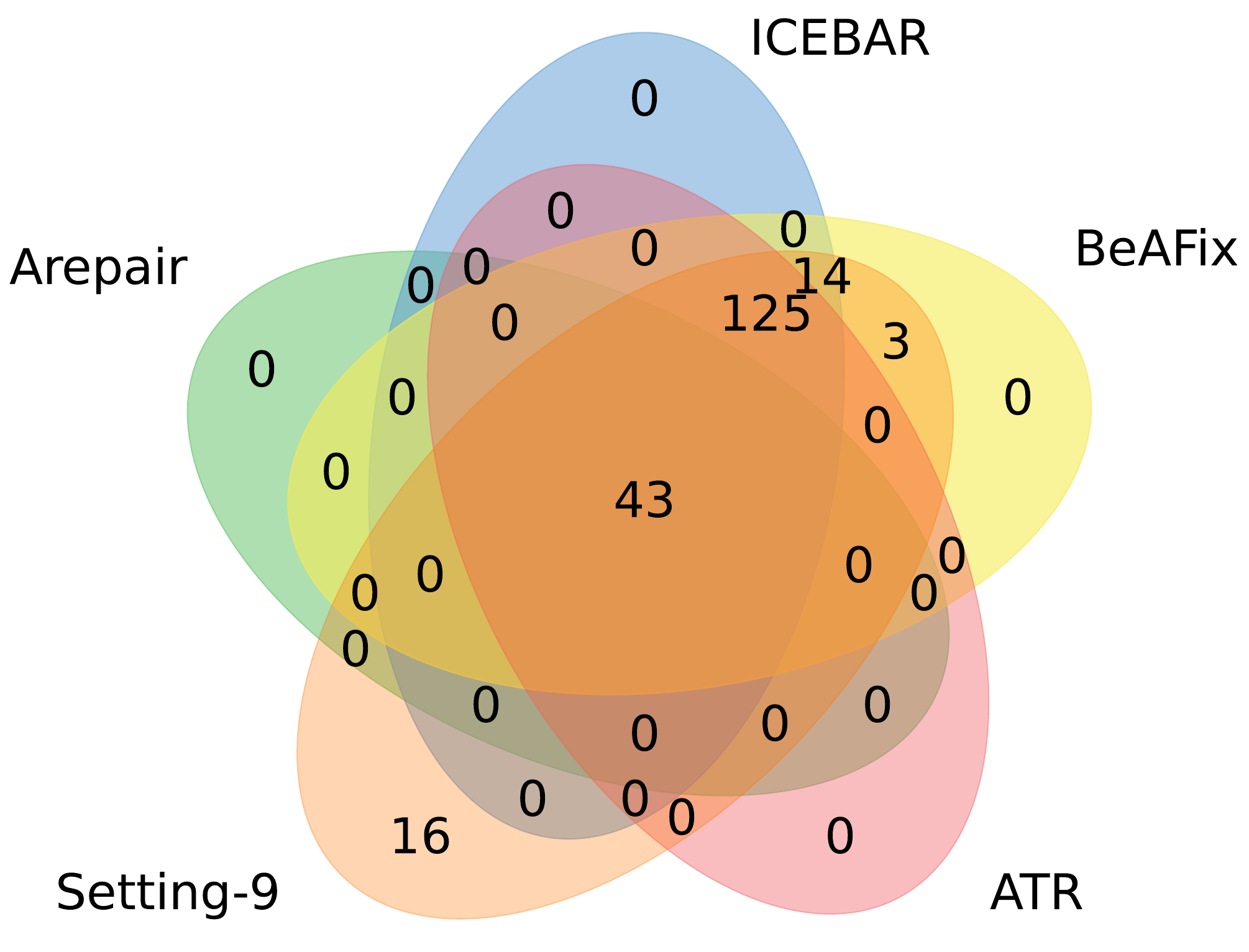}
    \end{minipage}
    
    \vspace{0.5cm} 
    
    \begin{minipage}{0.32\textwidth}
        \includegraphics[width=\linewidth]{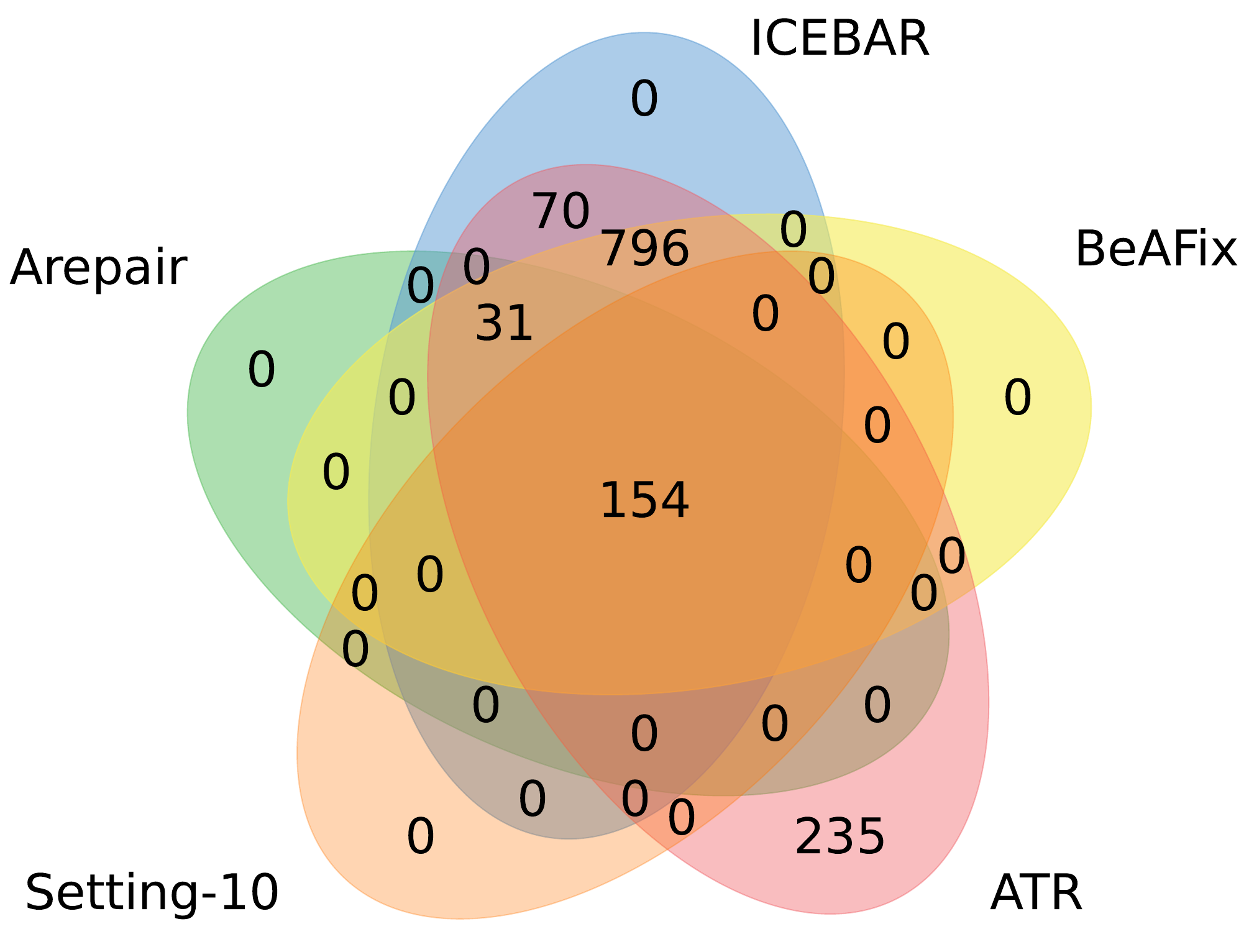}
    \end{minipage}
    \hfill
    \begin{minipage}{0.32\textwidth}
        \includegraphics[width=\linewidth]{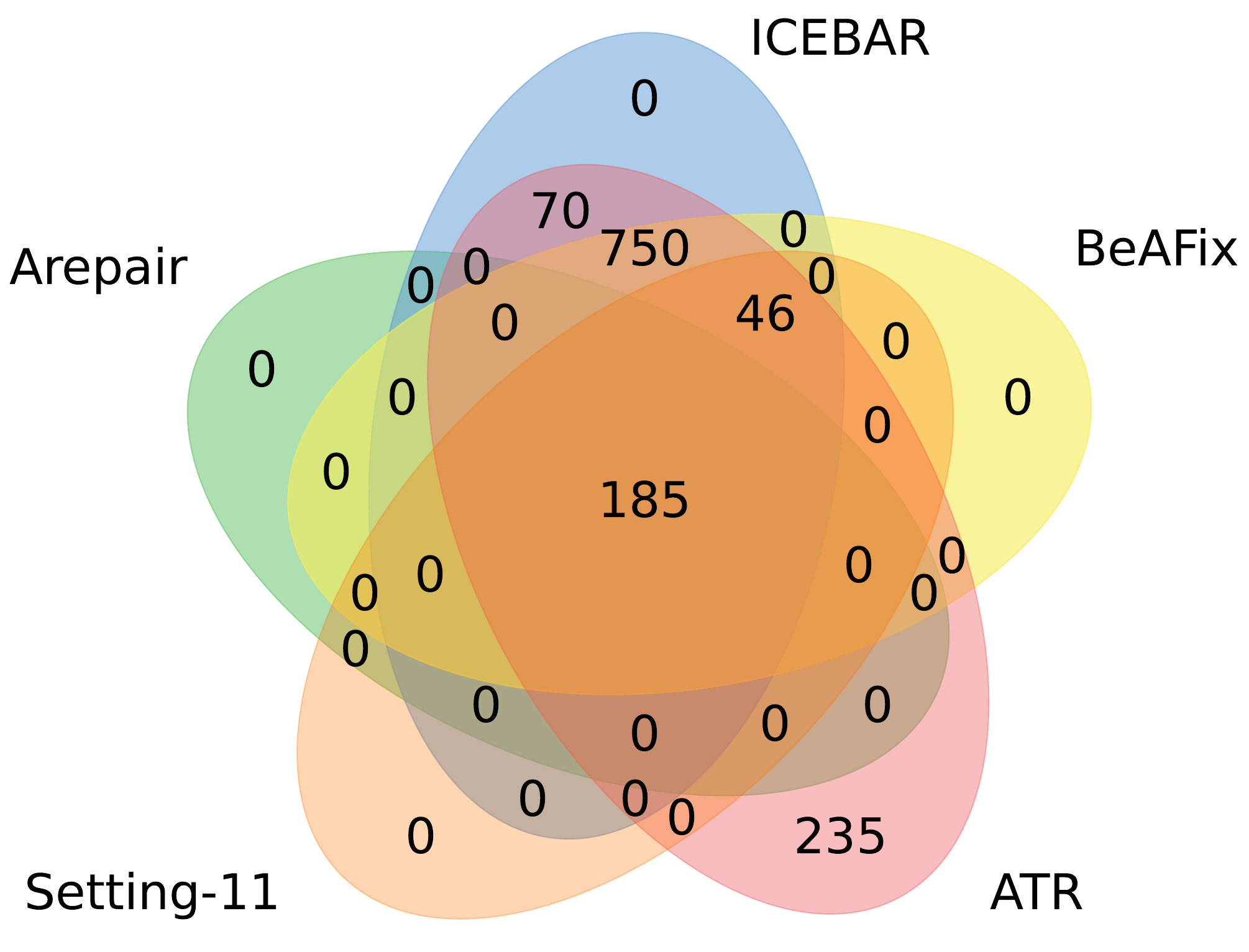}
    \end{minipage}
    \hfill
    \begin{minipage}{0.32\textwidth}
        \includegraphics[width=\linewidth]{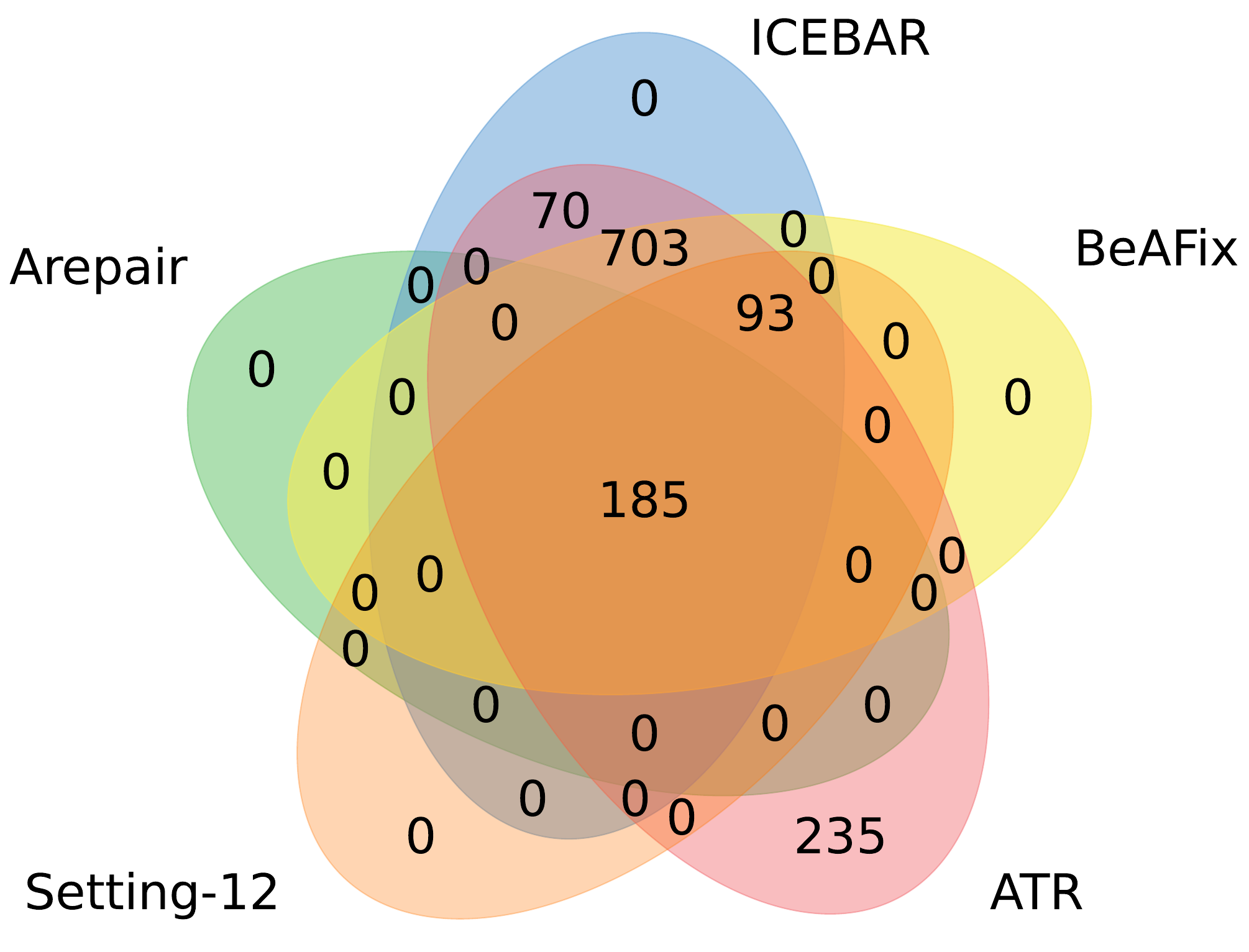}
    \end{minipage}
    
    \caption{\moh{Venn diagrams showing the exclusive and overlapping successful repairs for the Alloy4Fun benchmark achieved by different repair methods. These illustrate the complementary and unique capabilities of the APR pipeline compared to state-of-the-art (SoTA) repair tools. Settings 7-9 are based on the sampled 357 models, while settings 10-12 are based on the entire set of models in the benchmark. Accordingly, the results are reported for the SoTA tools.}}    
    \label{fig:venn_diagrams_A4F}
\end{figure*}